\documentclass[aps,prx,preprint,superscriptaddress]{revtex4-2}
\usepackage[T1]{fontenc}
\usepackage[utf8]{inputenc}
\usepackage{hyperref}
\usepackage{amsmath,amssymb,graphicx}

\begin{document}

\title{Polariton Laser in the Bardeen-Cooper-Schrieffer Regime}

\author{Jiaqi Hu}
 \affiliation{Applied Physics Program, University of Michigan, Ann Arbor, Michigan 48109, USA}
\author{Zhaorong Wang}
 \affiliation{Department of Electrical Engineering, University of Michigan, Ann Arbor, Michigan 48109, USA}
\author{Seonghoon Kim}
 \affiliation{Department of Electrical Engineering, University of Michigan, Ann Arbor, Michigan 48109, USA}
\author{Hui Deng}
 \email[dengh@umich.edu]{}
 \affiliation{Applied Physics Program, University of Michigan, Ann Arbor, Michigan 48109, USA}
 \affiliation{Department of Physics, University of Michigan, Ann Arbor, Michigan 48109, USA}
\author{Sebastian Brodbeck}
 \affiliation{Technische Physik, Universit\"{a}t W\"urzburg, Am Hubland, W\"urzburg 97074, Germany}
\author{Christian Schneider}
 \affiliation{Technische Physik, Universit\"{a}t W\"urzburg, Am Hubland, W\"urzburg 97074, Germany}
 \affiliation{Institute of Physics, University of Oldenburg, 26129 Oldenburg, Germany}
\author{Sven H\"ofling}
 \affiliation{Technische Physik, Universit\"{a}t W\"urzburg, Am Hubland, W\"urzburg 97074, Germany}
 \affiliation{SUPA, School of Physics and Astronomy, University of St Andrews, St Andrews KY16 9SS, United Kingdom}
\author{Nai H. Kwong}
 \affiliation{College of Optical Sciences, University of Arizona, Tucson, Arizona 85721, USA}
\author{Rolf Binder}
 \email[binder@optics.arizona.edu]{}
 \affiliation{College of Optical Sciences, University of Arizona, Tucson, Arizona 85721, USA}
 \affiliation{Department of Physics, University of Arizona, Tucson, Arizona 85721, USA}

\date{\today}

\begin{abstract}
Microcavity exciton polariton systems can have a wide range of macroscopic quantum effects that may be turned into better photonic technologies. 
Polariton Bose-Einstein condensation and photon lasing have been widely accepted in the limits of low and high carrier densities, 
but
identification of the expected Bardeen-Cooper-Schrieffer (BCS) state 
at intermediate densities remains elusive, as the optical-gain mechanism cannot be directly inferred from existing experiments. Here, using a microcavity with strong polarization selectivity, we gain direct experimental access to the reservoir absorption in the presence of polariton condensation and lasing, which reveals a fermionic gain mechanism underlying the polariton laser. A microscopic many-particle theory shows that this polariton lasing state is consistent with an open-dissipative-pumped system analog of a polaritonic BCS state.
\end{abstract}

\maketitle

\section{Introduction}

\begin{figure}[tp!]
\begin{center}
\includegraphics[angle=0, scale=0.5, trim= 0cm 0cm 0cm 0cm]{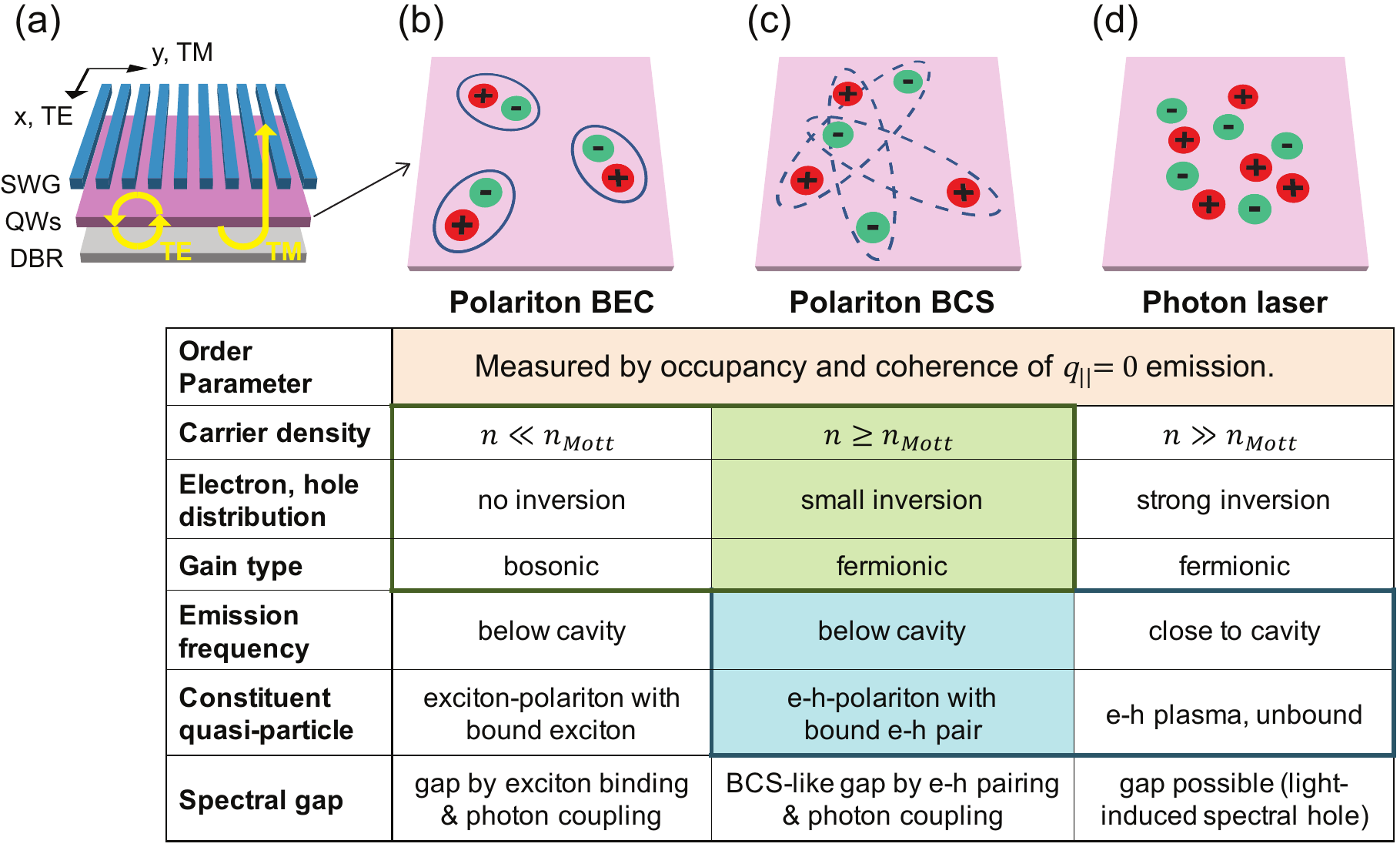}
\caption{\label{sketch-lasing.fig}\label{fig:sample}
{\bf Comparison of the polariton BEC, polariton BCS and photon laser.}
(a) A schematic of the subwavelength grating (SWG) based microcavity that allows access to the electronic reservoir of a polariton system. Quantum wells (QWs) are embedded in the cavity formed by a SWG and a distributed Bragg reflector (DBR). Fields polarized parallel to the bars (TE) are well confined in the cavity while the perpendicularly polarized fields (TM) transmit through the SWG. (b)-(d) Illustrations of three possible many-body states in the system with different quasiparticles in the electronic medium.
(b) Polariton BEC with bosonlike excitons (bound electron-hole, $e$-$h$, pairs at low density),
 (c) polariton BCS (bound $e$-$h$ pairs at elevated density),
 and (d) a conventional photon laser with plasmas of unbound $e$-$h$ pairs.
 Polaritons are formed in (b) and (c) via strong photon coupling.
 The table summarizes key properties of the three many-body states.
Thick green (blue) border lines indicate properties that distinguish a polariton BCS from a polariton BEC (photon laser). The properties confirmed for our system are highlighted correspondingly in green and blue.
}
\end{center}
\end{figure}

Collective quantum effects, formed through spontaneous symmetry breaking, have been intensively studied in a variety of physical system, such as Bose-Einstein condensation (BEC) \cite{dalfovo_theory_1999,leggett_bose-einstein_2001} in ultracold atomic gasses and the Bardeen-Cooper-Schrieffer (BCS) state in superconductors \cite{bardeen_theory_1957}.
While the original concepts of such quantum phases were developed for closed systems in thermal equilibrium, generalization to open quantum systems has been a fruitful path for studying a wider range of effects in physical systems. 
One such generalization is
quantum phases induced by an external driving field, such as light-induced superconductivity \cite{kaiser.2017} and, in the context of semiconductors, an early prediction where the coherence of an external light field  results in electron-hole ($e$-$h$) pairs exhibiting features similar to Cooper pairs in BCS states \cite{galitskii-etal.70}.
Another generalization is the concept of quasithermal equilibrium. For example, excitons (bound electron-hole pairs) in semiconductors do not exist in thermal equilibrium, but after creation might live long enough to assume quasithermal equilibrium and during that time exhibit exciton BEC \cite{keldysh-kozlov.68} or BCS \cite{comte-nozieres.82}, similar to BEC or BCS in closed systems.
When the excitons are strongly coupled to a cavity photon mode, polaritons are formed, which is the focus of this publication.
Polaritons have been discussed
to exhibit a range of many-body phases that bridge phenomena of both matter and light systems and involve both fermionic and bosonic quantum statistics \cite{keeling_collective_2007,carusotto_quantum_2013}. Two of the most studied phase transitions in the polariton system include, below the Mott density, the transitions from a thermal gas to a polariton BEC, accompanied by the emission of coherent light, or polariton lasing \cite{imamoglu_quantum_1996} [Fig.~\ref{sketch-lasing.fig}(b)], and, far above the Mott density, a conventional photon laser transition \cite{szymanska-etal.03} [Fig.~\ref{sketch-lasing.fig}(d)]. In between the two, a transition to a BCS-like polariton laser has been postulated theoretically \cite{keeling-etal.05, kamide_what_2010,byrnes_bcs_2010} [Fig.~\ref{sketch-lasing.fig}(c)], with only a few previous experimental attempts at demonstrating a polariton BCS state \cite{horikiri_highenergy_2016}.  As we
illustrate
in Fig.~\ref{sketch-lasing.fig}, while all three lead to coherent emission, they have distinct microscopic characteristics such as gain mechanism, quasiparticle type, and $e$-$h$ distribution functions.

Exciton-polariton BEC takes place at carrier densities $n$ much below the Mott transition density $n_\mathrm{Mott}$, $n \ll n_\mathrm{Mott}$, where electron-hole pairs are tightly bound by the Coulomb interaction to form excitons, which satisfy bosonic commutation properties [Fig.~\ref{sketch-lasing.fig}(b)] \cite{deng_exciton-polariton_2010}. The electronic distribution functions are far below Fermi degeneracy,
i.e. their value is far below 1 (unity),
and correspond to the exciton wave function. Coherence is formed via bosonic final state stimulation into the polariton ground state.
In contrast, conventional photon lasing takes place in the limit of high densities $n\gg n_\mathrm{Mott}$, where Coulomb interaction is screened, leading to uncorrelated, fermionic electron and hole plasmas. The electronic distributions are Fermi degenerate. Coherence is formed via stimulated scattering into a cavity mode when fermionic gain is provided by population inversion between the conduction and valence bands.  In between the polariton BEC and photon laser regimes, $n$ is high enough to disallow tightly bound exciton states, but low enough to allow sufficient electron-hole Coulomb correlations to form overlapping Cooper-pair like $e$-$h$ pairs.  In this intermediate regime,
absent a cavity, an excitonic
BCS state, with a coherent population of degenerate and weakly Coulomb-bound e-h pairs, has been predicted since the 1960s \cite{keldysh-kozlov.68,comte-nozieres.82,littlewood_models_2004,kremp-etal.2008,combescot-shiau.15}, though not yet demonstrated. In a strongly coupled microcavity, different types of polariton BCS states have been proposed, with Coulomb or photon-induced electron-hole pairing \cite{keeling_collective_2007,kamide_what_2010,byrnes_bcs_2010}.

Despite the distinct microscopic characteristics of the three phases, there is no additional symmetry breaking between them \footnote{It has recently been shown that non-hermitian phase transitions can occur in a system with broken U(1) symmetry {\protect\cite{hanai-etal.2019}},
but a robust relation between those
phase transitions and existing polariton condensation experiments has not yet been established.}.
This poses a challenge to identify the different phases experimentally.  Polariton condensation or lasing has been identified when the emitted light transforms from thermal to coherent \cite{deng_condensation_2002a,kasprzak_bose-einstein_2006} while maintaining a polaritonic dispersion with nonlinear interactions; the nonlinear interactions manifest themselves in frequency shift \cite{bajoni_polariton_2008} and coherence properties \cite{kim_coherent_2016} of the emission, thereby distinguishing the state from a photon laser.  Whether such a polariton laser originates from a BEC- or BCS-like state, however, is unclear---both a polariton BEC and a polariton BCS-like state would emit coherent light and show strong nonlinearities. The essential difference lies in the electronic media, which have been difficult to access experimentally,
except for the exciton component of a polariton BEC \cite{menard_revealing_2014}.

In this work, we gain direct access to the electronic component in the presence of lasing, by using an unconventional polarization-selective cavity and a special measurement technique, combining off-resonant continuous wave pump with a resonant probe in time-resolved spectroscopy. Through measurement of both the absorption and emission spectra of both the polariton lasing mode and its electronic reservoir, we identify fermionic gain above the Mott transition in the polariton laser, which distinguishes it from a polariton BEC. At the same time, bound electronic states persist, characteristic of a polaritonic BCS state but different from a conventional photon laser.
 To clarify the nature of the observed polariton laser, we developed a microscopic many-particle theory for the Coulomb correlated electron, hole and photon system. The theory correctly predicts the experimentally observed emission frequency and absorption spectra, and reveals distribution functions and interband polarizations resembling those of an ideal ($T=0\,\mathrm{K}$) polariton BCS state \cite{byrnes_bcs_2010,kamide_what_2010}, but modified by cavity dissipation and thermal dephasing.

\section{Experimental method}
To access the electronic media in the presence of polariton lasing, we use a cavity that simultaneously supports weakly and strongly coupled modes which, importantly, are in orthogonal linear polarizations.
As shown schematically in Fig.~\ref{fig:sample}(a) and detailed in Appendix~\ref{app:sample},
the top mirror of the $\lambda/2$ cavity is made of an $\mathrm{Al_{0.15}Ga_{0.85}As}$ high-contrast subwavelength grating.
The grating has very high reflectivity for transverse-electric- (TE) polarized light, but low for transverse-magnetic- (TM) polarized light \cite{sm}.
As a result, TE polaritons are formed
while TM excitons remain in the weak-coupling regime \cite{zhang_zerodimensional_2014}
.
Using such a cavity, we can access the reservoir (or bath) of the TE modes through the TM modes if they share a common electron and hole reservoir. This is readily achieved under nonresonant pumping. The observation of equal TE and TM emission outside the grating independent of the pump polarization confirms efficient scattering between exciton modes of different polarization \cite{sm}.
Moreover, within the grating area, the weakly and strongly coupled modes differ only over a small range of in-plane wave vector $q_{\parallel}$ near $q_{\parallel}=0$ compared to the large range of $q_{\parallel}$ occupied by all excitons or polaritons.
This ensures that the large-$q_{\parallel}$ modes, unmodified by strong coupling, do not have a significant difference of population between TE and TM polarizations 
despite very different dispersions and decay rates of the TE and TM modes near $q_{\parallel}=0$. The large-$q_{\parallel}$ modes account for most of the total carrier population, as the TE ground state polariton density is estimated to be $3.6\times 10^{-5}$ of the total carrier density below threshold and $1.2\times 10^{-3}$ even at the highest power used.
 Hence, the TE and TM modes share the same electron and hole reservoir, and via polarization-resolved spectroscopy, we can access simultaneously the optical properties of the polariton system and its underlying electronic media.

To identify the gain mechanism in the TE-mode lasing regime, we monitor the TM-mode absorption spectrum of the system via a separate, weak probe laser. To cleanly separate the photoluminescence (PL) and absorption, we use a two-color time-resolved spectroscopy. A continuous-wave pump is used, at 784 nm, about 30 meV above the exciton resonance, focused to a Gaussian spot of about $3~\mathrm{\mu m}$ in diameter at the center of the square grating. A resonant pulsed laser of about 150~fs pulse width is used as a probe, overlapping with the pump spatially.
We first measure the PL without the probe using spectrally resolved real-space and Fourier space imaging through a $4f$ confocal relay into a spectrometer. We then measure, in the presence of the PL, the probe absorption, using a streak camera. The reflected probe appears as a resolution-limited sharp peak in time in the streak image whereas the PL, produced under cw pumping, has uniform intensity in time. Hence, the PL is readily subtracted from the reflection \footnote{We integrate over the whole pulse to obtain the total intensity of reflected pulse, which is then divided by the reflection spectra from a reference gold mirror to produce the reflection spectra of the microcavity system.}. 

\section{Polariton lasing versus photon lasing}
\begin{figure}[tp!]
\begin{center}
\begin{minipage}{0.4\textwidth}
\includegraphics[angle=0, scale=0.45, trim= 6cm 0cm 0cm 0cm]{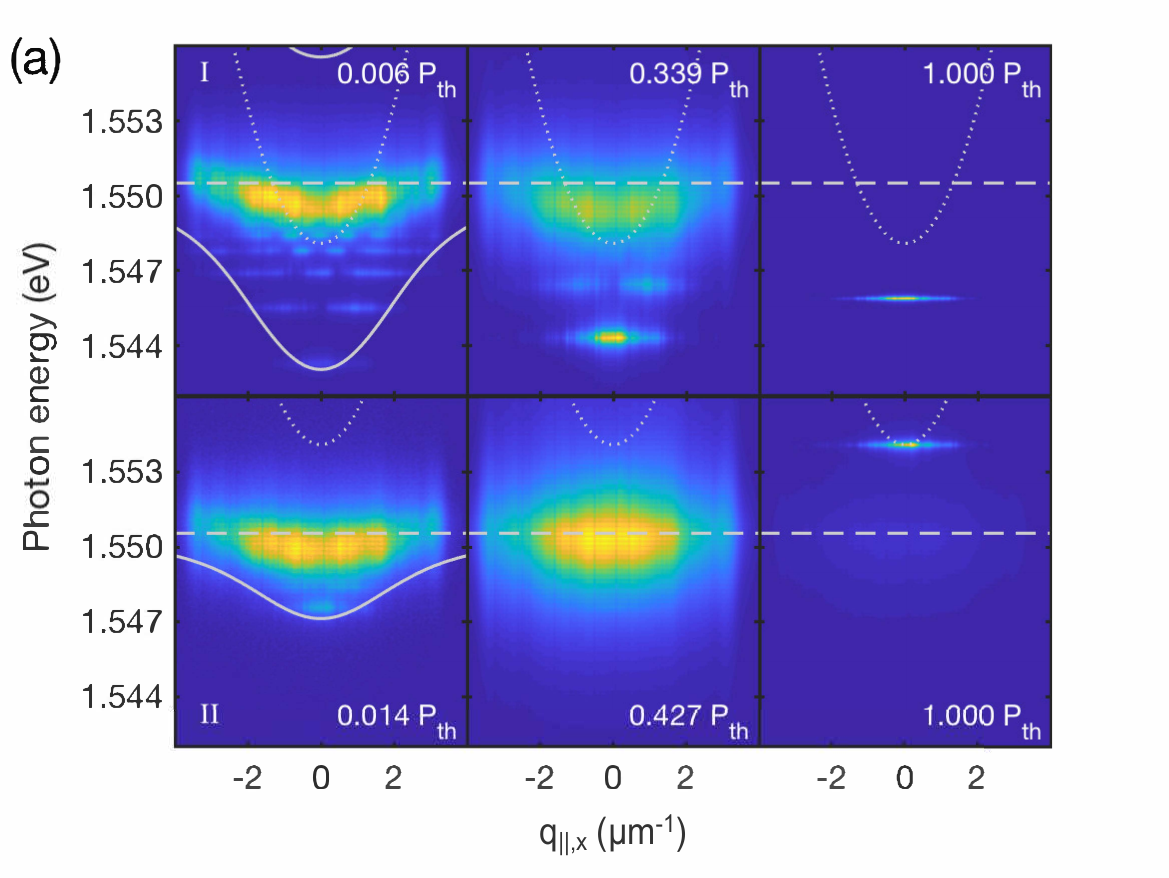}
\end{minipage}
\begin{minipage}{0.2\textwidth}
\includegraphics[angle=0, scale=0.45, trim= 1.8cm 0cm 0cm 0cm]{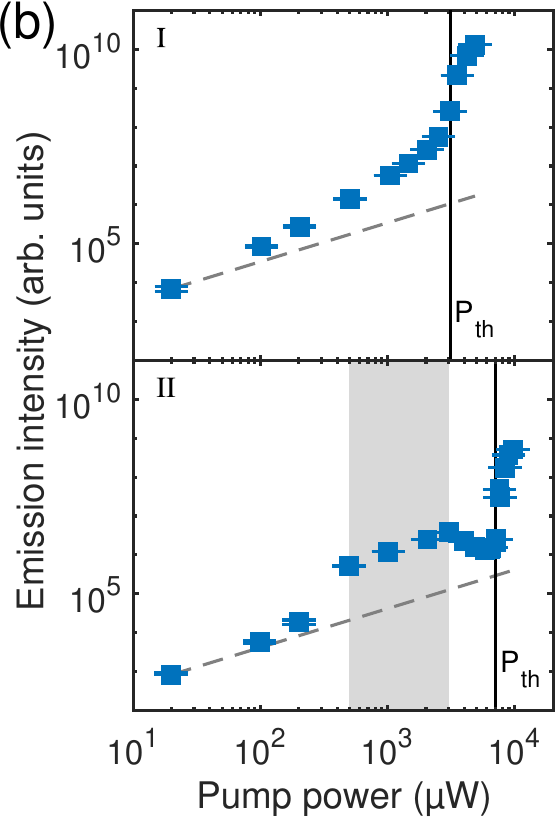}
\end{minipage}
\begin{minipage}{0.2\textwidth}
\includegraphics[angle=0, scale=0.45, trim= -1.2cm 0cm 0cm 0cm]{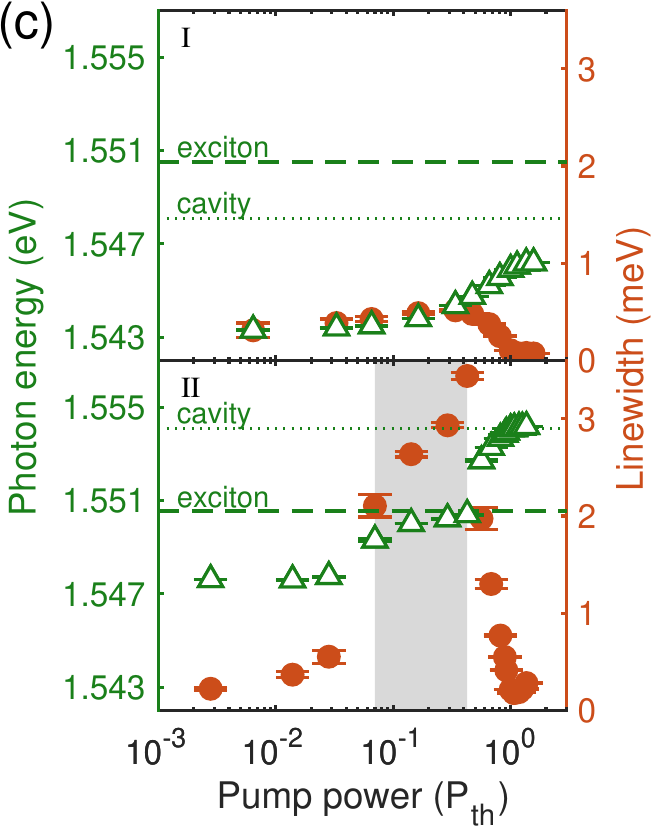}
\end{minipage}
\caption{\label{fig:lasing}
{\bf Emission properties of a BCS-like polariton laser (top row, detuning of –2.4~meV) and a photon laser (bottom row, detuning of 3.5~meV).}
(a) Fourier-space spectral images of TE-polarized emission from below to at the lasing threshold at $q_{\parallel,y}\sim 0$ at pump powers as labeled.
The top middle panel is at the pump power where the $q_{\parallel}=0$ mode has the largest linewidth. The bottom middle panel is at the pump power where the the $q_{\parallel}=0$ mode is much weaker than the peak at TM exciton energy and cannot be discerned.
The dashed lines mark the exciton resonances measured outside the grating region by reflection without a pump. The dotted (solid) curves are the calculated empty cavity (polariton) dispersions assuming no in-plane confinement; they are based on the measured exciton energy and the lower and upper polariton energies measured from TE reflection without pump (Appendix~\ref{app:TE_spectra}).
(b) The emission intensity of the TE $q_{\parallel}=0$ mode versus the pump power.
The dashed line is a reference of linear dependence.
(c) Pump power dependence of the emission photon energy (open green squares) and linewidth (filled orange circles) of the TE $q_{\parallel}=0$ modes.
The empty cavity at $q_{\parallel}=0$ (dotted lines) and exciton (dashed lines) resonances are also shown.
In (b) and (c), the gray region corresponds to where the TE ground state is much weaker than the peak at TM exciton energy and cannot be discerned; in (b), the vertical solid lines mark the lasing threshold.
}
\end{center}
\end{figure}

We measure two distinct types of lasing transition for the TE-polarized modes in devices with different cavity-exciton detunings.
In devices with negative to small positive detunings, polariton lasing was measured with typical features that have been so far identified with those of a polariton BEC. In more blue-detuned devices, we observe clear features of a photon laser. We show an example of each type in the top row and bottom row of Fig.~\ref{fig:lasing}, respectively. Additional examples of other detunings are shown in Ref.~\cite{sm}. The corresponding TM spectra and animations of detailed evolution of the spectra with power for the polariton and photon lasers are shown in Ref.~\cite{sm}.

In a polariton laser [top row in Fig.~2], the ground state remains distinct throughout the measured densities, with only slight line broadening below threshold and resolution-limited linewidth above threshold. Its frequency blueshifts continuously with power and remains well below the cavity or exciton resonances [Figs.~\ref{fig:lasing}(a) and \ref{fig:lasing}(c), top). These features suggest the ground state remains as a coupled state between electron-hole pair and photon; they are the same as widely reported for polariton BECs in the literature \cite{deng_condensation_2002a,kasprzak_bose-einstein_2006,balili_bose-einstein_2007}.

In stark contrast, in a photon laser [bottom row in Fig.~2], the lower polariton modes can no longer be discerned as the pumping density increases [Fig.~\ref{fig:lasing}(a), bottom middle panel; also see real-space spectra
in Ref.~\cite{sm}]. A mode with a rather broad linewidth emerges near the cavity resonance frequency near the threshold and becomes pinned at the cavity frequency above threshold [Fig.~\ref{fig:lasing}(c), bottom]. These features are fully consistent with the dissociation of bound states, both tightly bound exciton and $e$-$h$ pairs at the many-body level, before transition to photon lasing. The result suggests that electron-hole plasma is formed before a phase transition threshold is reached. This may be because of both a heavier effective mass of the lower polaritons that leads to a higher threshold density and a larger exciton fraction that leads to stronger scattering-induced dephasing.

Similar photon lasing transitions in polariton cavities have also been studied in earlier works \cite{bajoni_photon_2007,bajoni_polariton_2008,balili_role_2009}. Like in our observations, the photon lasing transitions were accompanied by significant linewidth broadening, many times more than that of a polariton lasing transition \cite{bajoni_polariton_2008,balili_role_2009}, as is consistent with the disappearance (retaining) of a bound state for photon (polariton) lasing transition.
In some of these works, the photon lasing frequency was reported to be below the original cavity resonance when the detuning is within about $\pm 10$~meV, which was attributed to a redshift of the cavity resonance at high carrier densities \cite{bajoni_photon_2007,balili_role_2009}. However no quantitative models were given to explain such a shift; and the carrier densities are likely much higher than that used in this study. The detuning of our polariton and photon lasing devices are similar in magnitude, –2.4~meV and +3.5~meV, respectively, for the examples shown in Fig.~2. The photon lasing we observed takes place close to the original cavity resonance despite a higher threshold density than that of polariton lasing. This confirms that there is negligible shift of the cavity resonance in our devices at the carrier densities we used.

\section{Gain mechanism of the polariton laser}

\begin{figure}[tp!]
\begin{center}
\begin{minipage}{0.32\textwidth}
\includegraphics[angle=0, scale=0.36, trim= 2.5cm 0cm 0cm 0.5cm]{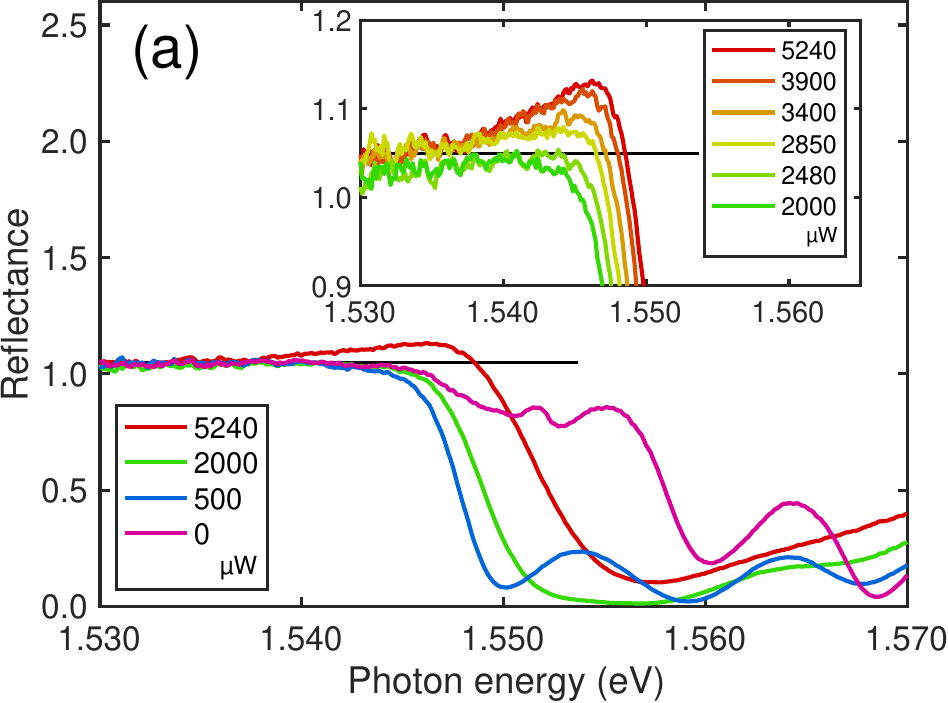}
\end{minipage}
\begin{minipage}{0.32\textwidth}
\includegraphics[angle=0, scale=0.36, trim= 0.5cm 0cm 0cm 0cm]{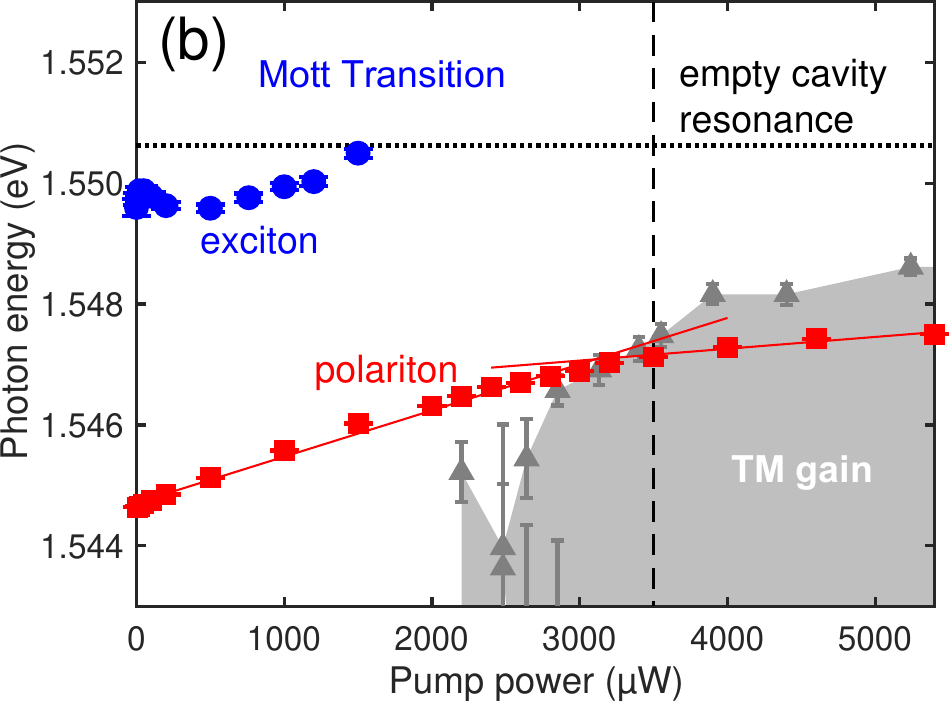}
\end{minipage}
\begin{minipage}{0.32\textwidth}
\includegraphics[angle=0, scale=0.62, trim= -0.6cm 0cm 0cm 0cm]{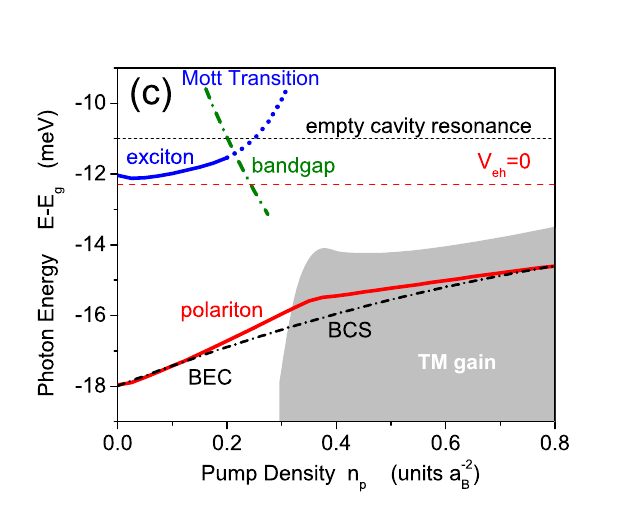}
\end{minipage}
\caption{\label{fig:R-spectra}\label{r771_vsDens.fig}
{\bf Measured and calculated gain and energetic positions of a polariton laser.}
(a) TM reflection spectra for a BCS-like polariton laser at different pump powers. The horizontal black lines mark reflectivity of 1 (see Appendix~\ref{app:gain} for calibration of the reflectivity). Inset: enlargement near the gain.
(b) Measured pump power dependence of the energetic positions of the TE $q_{\parallel}=0$ mode (red squares), TM exciton (blue circles), and the spectral bounds of gain (gray triangles). Error bars on the bounds are obtained by dividing the standard deviation of the reference of unity reflectivity by the local slope at the boundary. The empty cavity resonance is marked by the black dotted line. The
polariton lasing threshold,
marked by the black dashed line, coincides with the onset of the fermionic gain.
Red lines are linear fitting to TE polariton energy below and above threshold.
The exciton mode, extracted from TM reflection spectra (Appendix~\ref{app:exciton_energy}), is no longer resolvable above $1500~\mathrm{\mu W}$, due to dissociation of the excitons above the Mott transition.
(c) Calculated pump density ($n_p$) dependence of the energetic positions of the TE ground state (red solid line), TM exciton (blue solid line), gain region (gray area, determined from TM response spectra given in the Supplemental Maerial \cite{sm}),
ideal polariton BEC/BCS quasichemical potential assuming density $n=n_p$ (black dash-dotted line), the TE emission without $e$-$h$ interaction corresponding to an ideal photon laser (red dashed line),
renormalized band gap (green dash-dotted line),  and empty cavity resonance (black dotted line).
$a_B=14$~nm and $E_g$
are the exciton Bohr radius and quantum well band gap energy, respectively.
Effective electron  temperature $T[K]=40+50 n_p a_B^{2}$.
}
\end{center}
\end{figure}

While the observed polariton laser shows spectral features distinct from a photon laser and similar to a polariton BEC, the absorption spectra of the electronic media reveal a many-body phase different from BEC.
As described above, with linear polarization selectivity of the cavity, we are able to probe the electronic gain of the strongly coupled TE mode via the weakly coupled TM mode using time-resolved reflectance of a pulsed probe laser. 
We show examples of the TM reflection spectra of a polariton laser in Fig.~\ref{fig:R-spectra}(a).
Without pump, we measure TM absorption at the exciton resonance below the band continuum and above the TE polariton energy. With increasing pump power the exciton absorption becomes stronger. At even higher pump power but still below threshold, the discrete exciton resonance is no longer resolvable and the spectra have the shape of an absorption edge of the band continuum. This implies that there is strong screening and band gap renormalization to lower energies, and that the carrier density is already near the Mott density. Close to threshold, a peak above unity emerges; its height and width increase with further increase of the pump power [Fig.~\ref{fig:R-spectra}(a), inset]. This shows optical gain due to population inversion.
Corresponding TE emission and reflection spectra are provided in Appendix~\ref{app:TE_spectra}.

A condensed version of the experimental data, including the evolution of the resonances and gain with the pump power, is shown in Fig.~\ref{r771_vsDens.fig}(b).
The TM exciton resonance blueshifts with increasing pump power but quickly broadens and becomes nonresolvable, corresponding to the Mott transition at moderate pump power well below threshold. In contrast, the discrete TE resonance continues to exist despite the Mott transition and the presence of gain in TM; it blueshifts continuously with increasing carrier density throughout the pump powers used. Lasing takes place as the polariton ground state frequency enters the gain region and the frequency stays within the gain region at higher pump power, showing clearly a lasing transition driven by fermionic population inversion.

The total carrier density per quantum well (QW) at threshold is estimated to be $n_\mathrm{th} \sim 4\times10^{11}~\mathrm{cm}^{-2} > n_\mathrm{Mott}$ (see Appendix~\ref{app:density} for details), consistent with the onset of population inversion and fermionic gain.
We note that density estimates in polariton systems typically carry large uncertainties;  therefore they should be used not for identification of the many-body phases but rather as a consistency check.
In our experiment, regardless of the exact value of the density, the carrier reservoir that screens the electron-hole interaction in the
 TM exciton also screens the electron-hole interaction in the TE polariton; hence the direct observation of the Mott transition in TM confirms the fermionic nature of the gain medium.

We summarize our experimental observations as follows. The spontaneous phase transition is evidenced in the characteristic superlinear increase of the emission intensity at zero in-plane wave number, accompanied by sharp linewidth narrowing [Figs.~\ref{fig:lasing}(b) and \ref{fig:lasing}(c)]. These are common for the formation of all three possible phases in the system. However, distinct from a photon laser, we observe a well-defined single mode across the transition threshold, with an emission linewidth that remains narrow and emission frequency well below the cavity or exciton resonance frequencies without sudden shifts [Figs.~\ref{fig:lasing}(a) and \ref{fig:lasing}(c)]. This shows the emission originates from a bound state of electron, hole and photon. Furthermore, we distinguish our system from a BEC-like state by measurement of fermionic gain in its electronic reservoir due to population inversion, at a carrier density above the Mott density [Figs.~\ref{fig:R-spectra}(a) and \ref{fig:R-spectra}(b)].
Therefore, these observations are suggesting the possibility of a polariton BCS state: we highlight in the table in Fig.~\ref{fig:sample} the properties consistent with polariton BCS, and outline with thick green and blue borders which of these properties distinguish it from a polariton BEC in the low-density regime or a photon laser in the high-density regime, respectively.

While we have observed a BCS-like polariton laser and a photon laser in our devices, a transition in the BEC regime was not observed, possibly due to limited polariton lifetime. The transition from the BCS regime to photon lasing is expected with increasing excitation. However, we presently can only increase the power to about twice the BCS threshold before sample deformation takes place.  Within twice the BCS threshold, we did not observe a large shift of the lasing mode, which suggests the system remains in the BCS regime.

\section{Theoretical model of the system}

To analyze the microscopic picture underlying the experimental observation, we develop a
 theory based on a nonequilibrium Green's function approach to treat self-consistently the entire fermionic system as an open dissipative and pumped system.
The theory needs to correctly predict both the frequency shift due to density-dependent Coulomb interaction effects and the onset of fermionic gain even at the presence of Coulomb interactions.
The highest degree of agreement between theoretical results and experimental data
is obtained if low-frequency artifacts in the gain spectra \cite{indik-etal.96,girndt-etal.97} are avoided,
 which we achieve by including electronic correlations due to screening and the resulting partial cancellation of self-energy and $e$-$h$ vertex contributions (cf.\ Ref.~\cite{semkat-etal.09}).
Early theories that laid the foundation of the concept of polariton BCS states used localized two-level states represented as fermions to model the electronic system \cite{keeling-etal.05}. Later work considered a realistic electronic band structure and two-dimensional Coulomb interaction in the Hartree-Fock approximation \cite{byrnes_bcs_2010,kamide_what_2010}, but was limited to a closed, quasiequilibrium, $T=0\,\mathrm{K}$ polariton system, which we call an ``ideal polariton BCS state.'' A more recent work considered an open, dissipative and pumped system, but utilized a contact potential rather than a two-dimensional Coulomb potential \cite{yamaguchi_second_2013}.
Since BCS states are characterized by strong correlations among electrons and holes widely distributed in the configuration space, we use in our theory the two-dimensional Coulomb potential and electronic correlations and thereby extend the description of the electronic system in Refs.~\cite{byrnes_bcs_2010,kamide_what_2010,yamaguchi_second_2013}
 beyond the Hartree-Fock approximation. We also treat the two bright
  linear-polarization components (TE and TM)
  together with the carrier distributions functions
  fully self-consistently
  \footnote{In the anisotropic cavity, the role of dark coherences
differs from that in isotropic systems {\protect\cite{combescot-etal.07prl}} and are neglected.
}.
The incoherent pump is parametrized by the pump density $n_p$.

In the following, we summarize our theory,
 which is
 based on a nonequilibrium Green's function approach to treat self-consistently the entire fermionic system as an open dissipative and pumped system. In this section we present the equations of motion that we use in the numerical simulation.

 Since all two-time Green's functions entering
 the dynamically screened Hartree-Fock 
 theory are evaluated in the equal-time limit,
our theoretical approach can be viewed as an extended version of the semiconductor Bloch
equations \cite{binder-koch.95,haug-koch.04}
for the interband polarizations with zero center-of-mass
wave vector and as a function of the relative electron-hole wave vector $%
\mathbf{k}$.
Given in terms of electron and hole annihilation operators, we have
$P_{\pm}(\mathbf{k})=\langle a_{h, \pm 3/2, -\mathbf{k}}
a_{e, \mp 1/2 , \mathbf{k}}\rangle e^{i\omega _{0}t}$
where the subscripts $+$ ($-$) refer to circularly polarized bright interband transitions in the spin basis
from the 3/2 (–3/2) heavy-hole  (hh) band to the –1/2 (1/2) conduction band, where
 $\hbar \omega
_{0}\simeq E_{g}$ , with $E_{g}$ \ denoting the band gap energy, to make $P(%
\mathbf{k})$ slowly varying in time.
The Coulomb interactions entering the semiconductor Bloch equations are diagrammatically shown in
Fig.~1(a) of Ref. \cite{kwong-binder.00}.
 We omit electron-hole exchange effects, since in GaAs the exchange splitting is on the order of $\mu$eV \cite{malinowski-etal.99}, while all effects that we are observing are on the order of meV.

\begin{figure}[tp!]
\begin{center}
\centerline{\includegraphics[angle=0, scale=0.75, trim= 0cm 9cm 0cm 2cm]{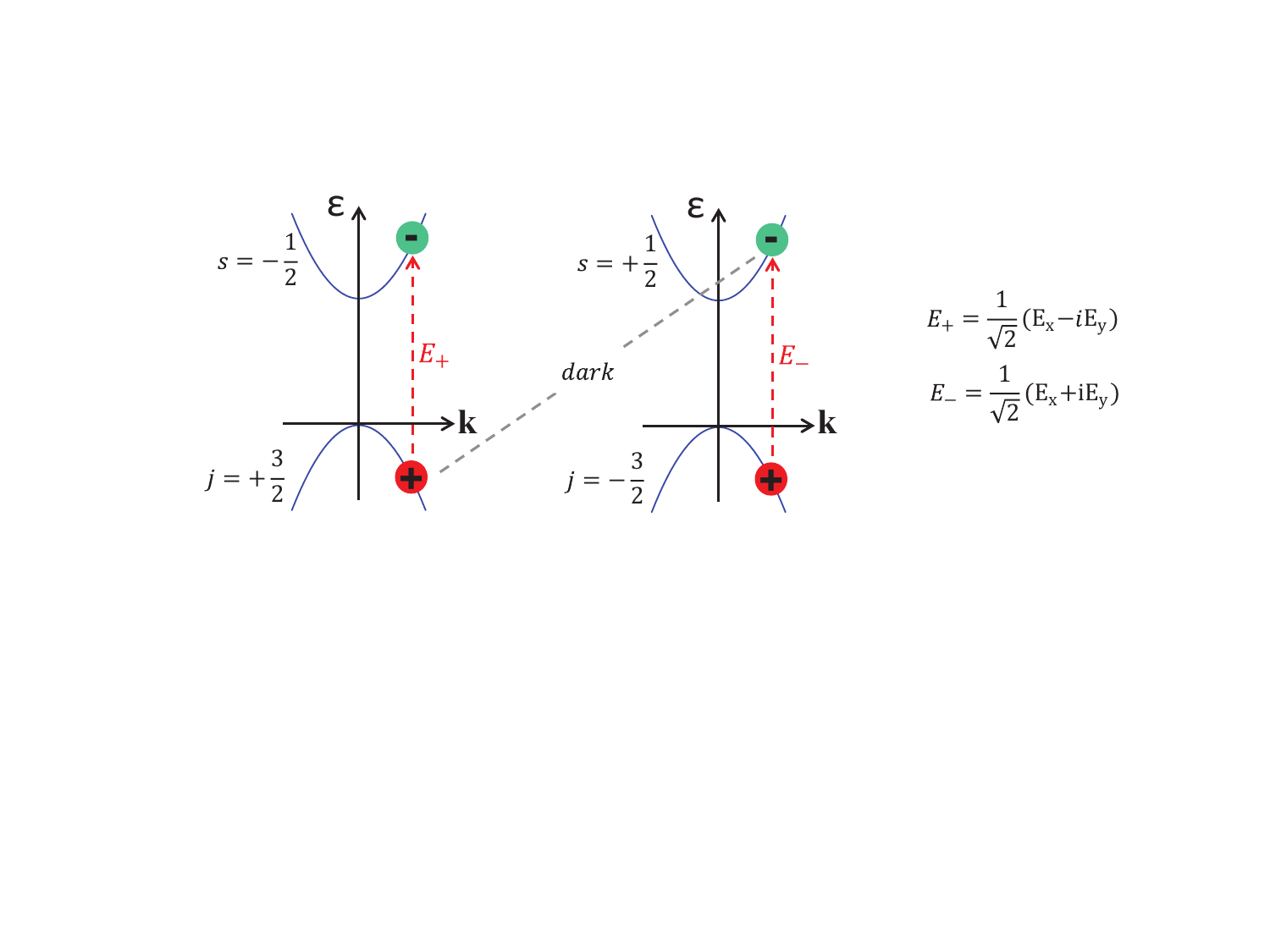}}
\caption{\label{sketch-selection-rules.fig}
{\bf Sketch of selection rules.}
 Selection rules involving heavy-hole and conduction bands in a thin GaAs quantum well. Both $x$- and $y$-polarized light fields
  induce optical transitions and interband polarizations, $P_x$ and $P_y$, respectively
  (red dashed lines) that
  provide equal amounts of population (or recombination in case of population inversion) in the two spin states ($\pm 1/2$ for electrons and $\pm 3/2$ for holes). For further explanation, see text.
}
\end{center}
\end{figure}

Since our cavity has a linear-polarization anisotropy, we transform the interband polarizations from the spin basis to the linear polarization basis:
\begin{eqnarray}
\label{equ:Px=Pp+pm}
P_{x}(
\mathbf{k}) &=& \frac{1}{\sqrt{2}}[P_{+}(\mathbf{k})+P_{-}(\mathbf{k})], \\
\label{equ:Py=Pp-pm}
 P_{y}(
\mathbf{k})&=&\frac{i}{\sqrt{2}}[P_{+}(\mathbf{k})-P_{-}(\mathbf{k})] .
\end{eqnarray}
The optical selection rules are shown in
Fig.~\ref{sketch-selection-rules.fig}.
We assume the effective electron $e$ and hole $h$
masses to be equal, $m_{e}=m_{h}$ (an approximation that is reasonable for
heavy-hole  states in
GaAs quantum wells).
As indicated in the figure,
both $x$- and $y$-polarized light fields
  induce optical transitions and interband polarizations, $P_x$ and $P_y$, respectively
  (red dashed lines) that
  provide equal amounts of population (or recombination in case of population inversion) in the two spin states ($\pm 1/2$ for electrons and $\pm 3/2$ for holes).
 This is the reason why, in the Hartree-Fock approximation, both excitonic interband polarizations,
  $P_x$ and $P_y$, are affected by a shared electron reservoir (in other words, both $P_x$ and $P_y$ experience the same phase-space filling
  from charge carrier populations in both spin states).

We assume that there is no spontaneously formed dark interband
coherence or dark condensate, e.g.
$  P^{(2)} (\mathbf{k}) = \langle a_{h,  3/2, -\mathbf{k}} a_{e,  1/2 , \mathbf{k}}\rangle$,
shown as dashed gray line in Fig.~\ref{sketch-selection-rules.fig},
where the superscript ``2'' refers to the angular momentum transfer (in units of $\hbar$) required for the transitions (dipole allowed transitions have
angular momentum transfer of $\pm 1$).
A coupling between $P_x$ and $P_y$ due to phase-space filling requires
  nonzero electron-spin (or hole-spin) coherences or spin imbalances (optically induced magnetism).
We  neglect electron-spin coherences
such as
$ f_e^{-1/2,1/2} (\mathbf{k}) =  \langle a^{\dag}_{e,  1/2, \mathbf{k}} a_{e,  - 1/2 , \mathbf{k}}\rangle$
and
hole-spin coherences
such as
$f_h^{-3/2,3/2} (\mathbf{k}) = \langle a^{\dag}_{h,  3/2, \mathbf{k}} a_{h,  - 3/2 , \mathbf{k}}\rangle$,
since the optical creation of these spin coherences requires a combination of bright and dark interband coherences.
Finally, we neglect spin imbalance, being defined, e.g., for the conduction band as
$ \Delta f_e (\mathbf{k}) =
\langle a^{\dag}_{e,  1/2, \mathbf{k}} a_{e,  1/2 , \mathbf{k}}\rangle
-
\langle a^{\dag}_{e,  - 1/2, \mathbf{k}} a_{e, - 1/2 , \mathbf{k}}\rangle
$, 
since the optical generation of spin imbalance cannot be achieved by a linearly polarized light field.
Only a superposition of cross-linearly polarized fields could optically generate a spin imbalance; however
in our case the pump beam is linearly polarized
and above threshold we also have only linearly polarized ($x$-polarized) lasing light.
We further clarify potential sources for spin coherences and spin imbalances at the end of this section.

 In the $x$-$y$ basis the  polarization equation in the
 screened Hartree-Fock approximation  reads
[compare, for example, Eqs.\ (12.19) and (15.2)-(15.4) in \cite%
{haug-koch.04}]
\begin{equation}
i\hbar \frac{\partial }{\partial t}P_{\ell }(\mathbf{k})=\left[ \varepsilon
_\mathrm{tot}(\mathbf{k})-\hbar \omega _{0}-i\gamma \right] P_{\ell }(\mathbf{k})-%
\left[ 1-2f(\mathbf{k})\right] \Omega _\mathrm{eff}^{\ell }(\mathbf{k})+\left.
i\hbar \frac{\partial }{\partial t}P_{\ell }(\mathbf{k})\right\vert _\mathrm{corr},
\label{equ:SBE-P-ell-dot}
\end{equation}%
where $\ell =x,y$, $\gamma $ is a phenomenological dephasing constant, the band-to-band renormalized
transition energy
\begin{equation}
\varepsilon _\mathrm{tot}(\mathbf{k})=\frac{\hbar ^{2}k^{2}}{2m_{r}}+E_{g}+2\Sigma ^\mathrm{HF} (%
\mathbf{k}),
\end{equation}%
where $m_{r}$ is the reduced $e$-$h$ mass, $m_{r}^{-1}=m_{e}^{-1}+m_{h}^{-1}$,
and $\Sigma ^\mathrm{HF} (\mathbf{k})$ is the unscreened Hartree-Fock self-energy
\begin{equation}
\label{equ:HF-selfenergy}
\Sigma ^\mathrm{HF}(\mathbf{k})=-\frac{1}{A}\sum\limits_{\mathbf{k}^{\prime }}V_{%
\mathbf{k-k}^{\prime }}^{c}f(\mathbf{k}^{\prime }),
\end{equation}%
where $V_{\mathbf{k-k}^{\prime }}^{c}$ is the unscreened two-dimension
Coulomb potential
\begin{equation}
V_{q}^{c}=\frac{2\pi e^{2}}{\varepsilon _{b}q},
\end{equation}%
with $e$ being the free-electron charge, $\varepsilon _{b}$ the background
dielectric constant, and $A$ the cross-sectional area of the system. The
correlation term in Eq.\ (\ref{equ:SBE-P-ell-dot}) consists of the electron
and hole correlation self-energies and the electron-hole vertex contribution,
\begin{equation}
\left. i\hbar \frac{\partial }{\partial t}P_{\ell }(\mathbf{k})\right\vert
_\mathrm{corr}=\left[ \Sigma _{e}^\mathrm{corr}(\mathbf{k)+}\Sigma _{h}^\mathrm{corr}(\mathbf{k)}%
\right] P_{\ell }(\mathbf{k})+\left. i\hbar \frac{\partial }{\partial t}%
P_{\ell }(\mathbf{k})\right\vert _\mathrm{corr}^\mathrm{vertex}.
\end{equation}%
The correlation self-energy is
\begin{equation}
\label{equ:corr-selfenergy}
\Sigma _{a}^\mathrm{corr}(\mathbf{k)=}\frac{1}{A}\sum_{\mathbf{k}^{\prime
}}\int\limits_{-\infty }^{\infty }\frac{d\hbar \omega }{\pi }\frac{%
[g_{B}(-\omega )+f(\mathbf{k}^{\prime })]
{\rm Im}V^\mathrm{scr}(\mathbf{k-k}%
^{\prime },\omega )}{\varepsilon _{a}(\mathbf{k)-}\varepsilon _{a}(\mathbf{k}%
^{\prime })-\hbar \omega +i\gamma_\mathrm{pl}} .
\end{equation}%
Here, $\gamma_\mathrm{pl}$ is an effective damping constant (subscript pl for plasma), which, for simplicity,
is taken as a parameter instead of calculating it self-consistently with the
self-energy, $\varepsilon _{a}(\mathbf{k)}=\frac{\hbar ^{2}k^{2}}{2m_{a}}$
are the unrenormalized band energies ($a=e,h$), and $g_{B}(\omega )$ the
Bose function:
\begin{equation}
g_{B}(\omega )=\frac{1}{e^{\hbar \omega /k_{B}T}-1} .
\end{equation}
 This form of $\Sigma _{a}^\mathrm{corr}(\mathbf{k)}$ is strictly valid in
quasithermal equilibrium. If the system is not in quasithermal
equilibrium, the present form is an approximation that assumes the plasmons
to be in quasithermal equilibrium, described by the Bose function, while
the carrier distributions $f(\mathbf{k})$ can be arbitrary, i.e. not
restricted to quasithermal equilibrium. We call this the equilibrium
plasmon approximation.

The dynamically screened Coulomb potential is given in terms of the
inverse plasma dielectric function
\begin{equation}
V^\mathrm{scr}(\mathbf{q,}\omega )=\varepsilon _\mathrm{pl}^{-1}(\mathbf{q,}\omega )V_{%
\mathbf{q}}^{c} .
\end{equation}%
We use a plasmon-pole model for the screening (see, for example  \cite{haug-schmittrink.84}),
\begin{equation}
\varepsilon _\mathrm{pl}^{-1}(q,\omega )=\varepsilon _{b}^{-1}\left( 1+\frac{\omega
_\mathrm{pl}^{2}}{2\omega _{q}}\left[ \frac{1}{\omega +i\delta -\omega _{q}}-\frac{1%
}{\omega +i\delta +\omega _{q}}\right] \right),
\end{equation}%
where $\delta $ is an infinitesimally small positive constant, the
two-component effective plasmon pole dispersion is given by
\begin{equation}
\hbar ^{2}\omega _{q}^{2}=\hbar ^{2}\omega _\mathrm{pl}^{2}\left( 1+\frac{q}{\kappa
}\right) +C_\mathrm{pl}\left( \frac{\hbar ^{2}q^{2}}{2m_{r}}\right) ^{2},
\end{equation}%
and the $q$-dependent squared plasma frequency is
\begin{equation}
\omega _\mathrm{pl}^{2}=\omega _\mathrm{pl}^{2}(q)=\frac{2\pi e^{2}q}{\varepsilon _{b}}%
\left( \frac{n^{e}}{m_{e}}+\frac{n^{h}}{m_{h}}\right).
\end{equation}%
Here, $n^{a}$ ($a=e,h$) denotes the density of the electrons and holes,
\begin{equation}
n^{a}=2\int \frac{d^{2}k}{(2\pi )^{2}}f^{a}(\mathbf{k}),
\end{equation}%
and, as mentioned above, we assume equal masses, $m_{e}=m_{h}\equiv m$, and thus
$n^{e}=n^{h}\equiv n$. The screening wave vector is $\kappa =\kappa
_{e}+\kappa _{h}$ with
\begin{equation}
\kappa _{a}=\frac{2m_{a}e^{2}}{\varepsilon _{b}\hbar ^{2}}f^{a}(k=0).
\end{equation}%
The $e$-$h$ vertex contribution is taken to be the one derived in Ref.\
\cite{semkat-etal.09}, with the additional simplification of making an effective
quasiparticle approximation that replaces the frequency dependence by
unrenormalized energy differences, thus neglecting quantum memory effects,
\begin{equation}
\label{equ:Pdot-corr-eh-vertex}
\left. i\hbar \frac{\partial }{\partial t}P_{\ell }(\mathbf{k})\right\vert
_\mathrm{corr}^\mathrm{vertex}=\frac{1}{A}\sum\limits_{\mathbf{k}^{\prime }}\Delta W_{%
\mathbf{k,k}^{\prime }}P_{\ell }(\mathbf{k}^{\prime }),
\end{equation}%
with
\begin{equation}
\Delta W_{\mathbf{k,k}^{\prime }}=2\int\limits_{-\infty }^{\infty }\frac{%
d\hbar \omega }{\pi }\frac{[g_{B}(-\omega )+f(\mathbf{k}^{\prime })]
{\rm Im}%
V^\mathrm{scr}(\mathbf{k-k}^{\prime },\omega )}{\varepsilon _{a}(\mathbf{k}^{\prime
}\mathbf{)-}\varepsilon _{a}(\mathbf{k})-\hbar \omega +i\gamma_\mathrm{pl}} .
\end{equation}%
Furthermore, in Eq.\ (\ref{equ:SBE-P-ell-dot}) the effective Rabi frequency is
\begin{equation}
\label{equ:eff-Rabi-frequ}
\Omega _\mathrm{eff}^{\ell }(\mathbf{k})\ =\left( a_{\mathbf{k}}^{c}E_{\ell }+\frac{%
1}{A}\sum\limits_{\mathbf{k}^{\prime }}V_{\mathbf{k-k}^{\prime }}^{c}P_{\ell
}(\mathbf{k}^{\prime })\right) .
\end{equation}%
The coupling coefficient,
\begin{equation}
a_{\mathbf{k}}^{c}=\frac{1}{2}d_{\text{cv}}(\mathbf{k})\Psi _\mathrm{cav}(z_\mathrm{QW})%
\sqrt{(8\pi \hbar \omega _\mathrm{cav})/\varepsilon _{b}},
\end{equation}%
is given in terms of the interband dipole matrix element between the conduction band (subscript c) and the valence band (subscript v) $d_{\text{cv}}(%
\mathbf{k})$, or more precisely, a matrix element of the $k$-gradient operator
(for more details about interband matrix elements in semiconductor models
with periodic boundary conditions, see Ref.~\cite{gu-etal.13}), the light mode
function $\Psi _\mathrm{cav}(z_\mathrm{QW})$ evaluated at the position of the quantum
well, and the background dielectric function $\varepsilon _{b}$.
We  assume $d_{\text{cv}}(\mathbf{k})$ to be independent of $\mathbf{k}$
up to a certain cutoff value $k_{max}$; in other words,
$a_{\mathbf{k}}^{c} =  a^{c}$  for $| \mathbf{k}| < k_\mathrm{max}$ ,
and  zero for $| \mathbf{k}| > k_\mathrm{max}$.
In the low-density limit, the
coupling coefficient can be related to the conventional polariton splitting $%
2\Omega $
 when the polarization is
written as $P(\mathbf{k,t})=\Phi (\mathbf{k})p(t)$ with a normalized
wave function $\Phi (\mathbf{k})$, which in the low-density limit is
the 1$s$ exciton wave function
\begin{equation}
\Phi_{1s}(\mathbf{k})=\frac{\sqrt{2\pi }a_{B}}{\left[ 1+\left( \frac{ka_{B}%
}{2}\right) ^{2}\right] ^{3/2}},
\end{equation}%
via%
\begin{equation}
\Omega =  \sqrt{N_\mathrm{QW}}  a^{c\ast} \Phi_{1s} (\mathbf{r=0}),
\end{equation}
where $N_\mathrm{QW}$ is the number of
quantum wells in the cavity.

The equation for the distribution function reads
\begin{equation}
\hbar \frac{\partial }{\partial t}f(\mathbf{k})=
{\rm Im}
\left[ \Omega
_\mathrm{eff}^{x}(\mathbf{k})^{\ast }P_{x}(\mathbf{k})+\Omega _\mathrm{eff}^{y}(\mathbf{k}%
)^{\ast }P_{y}(\mathbf{k})\right] +\hbar \left. \frac{\partial }{\partial t}%
f(\mathbf{k})\right\vert _\mathrm{relax}+\hbar \left. \frac{\partial }{\partial t}f(%
\mathbf{k})\right\vert _\mathrm{pump} , \label{equ:fk-dot-with-relax-and-pump}
\end{equation}%
with a relaxation term%
\begin{equation}
\hbar \left. \frac{\partial }{\partial t}f(\mathbf{k})\right\vert
_\mathrm{relax}=-\gamma_{F}\left[ f(\mathbf{k})-f_{F}(\mathbf{k})\right]
- \gamma_\mathrm{nr} f(\mathbf{k}),
\end{equation}%
where $f_{F}(\mathbf{k})$ is the Fermi function
\begin{equation}
f_{F}(\mathbf{k})=\frac{1}{e^{[\varepsilon _{a}(\mathbf{k)-\mu }%
_{a}]/k_{B}T}+1} ,
\end{equation}%
$\gamma_{F}$ an effective intraband relaxation rate and
$\gamma_\mathrm{nr}$ the nonradiative decay rate.
At each time instance, the Fermi function is
normalized to the same density as $f(\mathbf{k})$. We model the pump process
by assuming the relaxation into a Fermi function to be sufficiently fast so
that the pump term can be taken as
\begin{equation}
\hbar \left. \frac{\partial }{\partial t}f(\mathbf{k})\right\vert
_\mathrm{pump}=-\gamma_{p}\left[ f(\mathbf{k})-f_{p}(\mathbf{k})\right],
\end{equation}%
where $f_{p}(\mathbf{k})$ is taken to be a Fermi function normalized as
\begin{equation}
n_{p}=2\int \frac{d^{2}k}{(2\pi )^{2}}f_{p}(\mathbf{k}) .
\end{equation}%
We use the pump density $n_{p}$ as an input parameter to our theory.
For simplicity we assume here that all thermal functions entering our model, including
the bath distribution function  $f_{p}(\mathbf{k})$, are at the same effective temperature $T$.
This temperature can generally be different from the lattice temperature, since it accounts
for the dynamical equilibrium between the creation of carriers high in the bands and the electron-hole
recombination (and other loss) processes.

For comparison with a system without $e$-$h$ interaction, we omit the Coulomb interaction in Eq.\ (\ref{equ:eff-Rabi-frequ}) and
the $e$-$h$ correlation (\ref{equ:Pdot-corr-eh-vertex}). A complete omission of the Coulomb interaction, however, would imply that there is no
band gap reduction and hence the system cannot lase at the cavity frequency, which in our case is below the band gap. Therefore, we do not
completely omit the self-energy. However, the $e$-$h$ correlation (\ref{equ:Pdot-corr-eh-vertex}) effectively cancels that large imaginary part of the
correlation self-energy (\ref{equ:corr-selfenergy}). Hence, for the purpose of this study we omit both the
correlation self-energy (\ref{equ:corr-selfenergy}) and $e$-$h$ correlation (\ref{equ:Pdot-corr-eh-vertex}), and keep only the Hartree-Fock self-energy
(\ref{equ:HF-selfenergy}).

 Finally, the light field amplitudes are determined as follows. For
the $y$-polarization (TM) we take $E_{y}$ to be a weak external perturbation field, $%
E_{y}= E_\mathrm{pert}$ , for example a spectrally wide short femtosecond pulse. To
model the TE field, for which we have a high-$Q$ cavity, we use the quasimode
equation
\begin{equation}
i\hbar \frac{\partial }{\partial t}E_{x}=\left( \hbar \omega _\mathrm{cav}-\hbar
\omega _{0}-i\gamma_\mathrm{cav}\right) E_{x}-\frac{N_\mathrm{QW}}{A}\sum\limits_{\mathbf{%
k}}a_{\mathbf{k}}^{c\ast }P_{x}(\mathbf{k}) , \label{equ:quasi-mode-E-dot}
\end{equation}%
where $\gamma_\mathrm{cav}$ is the cavity decay rate.

The numerical values for the parameters used in the calculations are given in  the Supplemental Material \cite{sm}.

To briefly summarize, we solve numerically the equations for the wave-vector-dependent $X$ and $Y$ interband polarizations,  Eq. (\ref{equ:SBE-P-ell-dot}),
the wave-vector-dependent carrier distribution, Eq. (\ref{equ:fk-dot-with-relax-and-pump}), and the light field amplitude at the position of the quantum wells,
Eq. (\ref{equ:quasi-mode-E-dot}), fully  self-consistently as time-differential equations until the steady state is reached.
The solution of these equations yields the theoretical results shown below, notably, the spectral position of the polariton emission,
cf. Fig.~\ref{r771_vsDens.fig}(c),
 and the $\mathbf{k}$-dependent distribution and interband polarization functions, cf. Figs.~\ref{r955-fk.fig}(a) and \ref{r955-fk.fig}(b)
 We limit
ourselves to $s$-wave solutions, meaning that all $\mathbf{k}$-dependent
functions depend only on the magnitude of the wave vector, $k=|\mathbf{k}|$,
and $k$ ranges between 0 and $k_{\max }.$
 After numerically Fourier transforming the results to frequency, we find the TE emission frequency from the maximum of the spectral intensity of the TE  light field,
 $|E_x(\omega)|^2$.
 Examples of these spectra, for the parameters used in Fig.~\ref{r771_vsDens.fig}(c),
  are shown in the
 Supplemental Material \cite{sm}.
  To obtain  the TM exciton resonance and gain spectra,
 we  add a small ultrafast
(femtosecond) external light field $E_\mathrm{pert}$ to probe the system.
Specifically, we replace $E_y$ by $E_\mathrm{pert}$ in
Eqs. (\ref{equ:SBE-P-ell-dot}) and (\ref{equ:fk-dot-with-relax-and-pump}). The TM spectra are then given by the susceptibility
$
\chi (\omega )=\delta P_{y}^\mathrm{tot}(\omega ) / E_\mathrm{pert}(\omega ) $ where $\delta P_{y}^\mathrm{tot}(\omega )$ is the perturbation-field-induced  total TM polarization.  For the interpretation of the experimental results, we use only exciton frequency and the spectral location of the gain, not the gain line shape. However, figures of the gain line shape, together with additional details of the theory and parameter values used in the calculations, are given in Ref.~\cite{sm}. The gain lineshape that we obtain is free of the artificial absorption below the gain, which one obtains in an unscreened Hartree-Fock theory, as was discussed, e.g., in Refs. \cite{indik-etal.96,girndt-etal.97}. It is important to note, however, that we verified that none of our qualitative conclusions would change if we would use an unscreened Hartree-Fock theory. Only the quantitative agreement between theory and experiment shown below [cf. Figs.~\ref{r771_vsDens.fig}(b) and \ref{r771_vsDens.fig}(c)] would be slightly less good.

Before presenting our numerical results, we want to briefly come back to the issue of spin coherences and spin imbalances mentioned above.
Within the screened Hartree-Fock theory and the Coulomb interactions shown
in Fig. 1(a) of Ref. \cite{kwong-binder.00},
all source terms for spin coherence
$ f_e^{-1/2,1/2} (\mathbf{k})$ involve polarization with angular momentum transfer 2, for example,
$
\frac{%
1}{A}\sum\limits_{\mathbf{k}^{\prime }}V_{\mathbf{k-k}^{\prime }}^{c}P^{\ast}_{+
}(\mathbf{k}^{\prime })
P^{(2)} (\mathbf{k})$.
Hence, unless we would find experimental indications of a  spontaneous dark coherence such as $P^{(2)} (\mathbf{k})$,
which does not couple directly to the cavity mode and therefore does not exhibit polariton effects similar to $P_x$, and
which would still need to be strong enough to give observable effects on the meV scale, we do not expect, and hence neglect, optically induced spontaneous spin coherence.
The reason to neglect optically induced spontaneous spin imbalance, e.g., $\Delta f_e (\mathbf{k})$,
lies in the fact that its source term is proportional to products of TE and TM fields, such as
$\Omega_\mathrm{eff}^{x}(\mathbf{k})^{\ast } P_{y} (\mathbf{k}) $. Since in steady state there is no TM field (a small TM field exists only transiently when we probe the system's TM response), these source terms are negligible in our experimental configuration.

\section{Theoretical results and discussion}

Figure~\ref{r771_vsDens.fig}(c) shows a representative theoretical result, with parameters chosen to correspond to the experiment (cavity-exciton detuning 1 meV, cavity decay rate 0.2 meV, exciton binding energy 12 meV, normal-mode splitting 12 meV).
We use a pump-dependent effective electron  temperature to account for pump-induced heating, $T[ {\rm K} ]=40+50 n_p a_B^{2}$ ($T$ models the dynamic equilibrium between high-energy pump and $e$-$h$ pair recombination).
The results are rather insensitive to the exact input parameters;
qualitatively similar results for fixed temperature are shown in Ref.~\cite{sm}.

The results reproduce all observed experimental features with excellent agreement.
For the TM exciton, we obtain a blueshift accompanied by strong band gap renormalization. At $n_p\sim 0.1 n_\mathrm{Mott}$, the exciton merges with the renormalized band gap and dissociates, which corresponds well to the observed disappearance of the exciton resonance into band-edge absorption. Further increasing density, we obtain gain due to population inversion, in agreement with the measured gain in TM field.

Gain in the TM field implies direct gain in the TE field.
This is because the equations for the TE and TM interband polarization differ only in that they contain different light field amplitudes. The Coulomb interaction terms as well as the distribution functions that enter the equations, for example as the inversion or Pauli blocking factor, are the same. Therefore, TE lasing is expected as the polariton resonance enters the gain region, as is observed experimentally [Fig.~\ref{r771_vsDens.fig}(b)].

The Mott transition and fermionic gain reproduced by the theory confirm that the observed lasing transition is not an exciton-polariton BEC.
At the same time, Coulomb effects continue to play a substantial role in the laser, rendering characteristics distinct from the photon laser but similar to the ideal polariton BCS state. A first manifestation is the spectral shift of the lasing state. As shown in Fig.~\ref{r771_vsDens.fig}(c) and in agreement with experimental results,
the lower polariton (LP) experiences a blueshift of about 2meV before the TE laser threshold is reached,
at which point it
levels off (i.e., its slope decreases), and across the threshold the blue-shift is continuous,
remaining significantly below that of the photon laser but tracking closely the ideal BCS state. In comparison, with Coulomb interactions turned off, we obtain photon lasing near the cavity frequency
similar to the observed photon lasing.

\begin{figure}[tp!]
\begin{center}
\includegraphics[angle=0, scale=1.8, trim= .5cm .4cm 0cm 0cm]{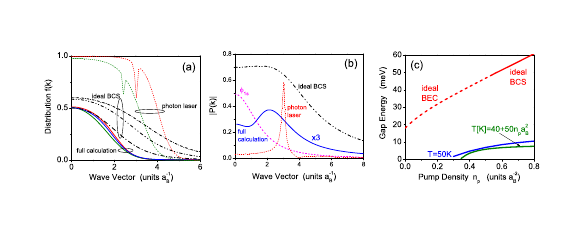}
\caption{\label{r955-fk.fig}
{\bf Carrier distributions and wave functions.}
(a) Distribution functions at $T=50$K.
 Solid lines: full calculation at pump densities $n_{p}a_B^{2}=$ 0.3, 0.45 and 0.6
(threshold pump density $n^\mathrm{th}_{p} \approx 0.3 a_B^{-2}$), $\gamma = 1.5$meV.
Short dashed lines: without $e$-$h$ interaction
at $n_{p}a_B^{2}=$ 1.6 (green) and 2.6 (red),
$n^\mathrm{th}_{p} \approx 1 a_B^{-2}$,
$\gamma = 0.15$ meV.
Black lines: ideal polariton BCS at densities $n a_B^{2}=$ 0.6 (long dash-dotted line), 1.2 (dash-dot-dot line) and 1.8 (short dash-dotted line), showing that at large densities $f_\mathrm{BCS}(k=0)$ saturates at about 0.6.
(b) Corresponding magnitudes of polarizations (effective wave functions); same line styles as in (a).
The dashed magenta line shows the 1s exciton wave function normalized to a density of
 $4 \times 10^{10}$cm$^{-2}$.
 (c) Pair gap energy of ideal system (red line) and phenomenological estimate of the pair gap energy of the experimental system (green and blue lines).
}
\end{center}
\end{figure}

With the theory reproducing  the spectral properties measured in the experiment, we furthermore use the theory to examine the reciprocal-space carrier distribution functions $f(k)$ and the electron-hole interband polarization $|P(k)|$, which characterize the microscopic mechanism of the lasing transition. As shown in Fig.~\ref{r955-fk.fig}, both $f(k)$ and $|P(k)|$ of the polariton laser are qualitatively similar to those of an ideal BCS state but different from those of a photon laser.

As seen in Fig.~\ref{r955-fk.fig}(a),  $f(k)$ of both the ideal BCS state and the polariton laser saturate
only slightly above 0.5 (the minimum value required for fermionic gain), for example the value of the red solid line close to $k=0$ is approximately
0.51, and in additional calculations using the unscreened HF approximation (not shown), the corresponding value is about 0.56,
while $f(k)$ of the photon laser approaches unity above threshold.
Sharp kinetic holes can develop in the distribution functions of the photon laser, shown here with a small dephasing of 0.15 meV for clarity, but not in the polariton laser or the ideal BCS state.
These results for the polariton laser and the ideal BCS state are largely insensitive to the value of the dephasing.

Equally important is the interband polarizations $|P(k)|$, corresponding to the order parameter [Fig.~\ref{r955-fk.fig}(b)]. For a photon laser,
it is sharply peaked below---and zero at---the transparency wave vector, which corresponds to the quasichemical potential.
In contrast, $|P(k)|$ of both the BCS polariton laser and the BCS state (cf.\ Ref.~\cite{byrnes_bcs_2010})
 do not vanish at the transparency wave vector and are
instead broadly distributed in the reciprocal space, indicating bound states
  even at densities $n$ where the mean distance between polaritons becomes comparable to the Bohr radius $a_B$, i.e.,\ $n a^2_B \approx 1$.
The magnitude of $|P(k)|$ in our polariton laser is smaller than that of the ideal BCS due to cavity dissipation and dephasing.

Lastly, we estimate the BCS gap corresponding to the minimum pair-breaking excitation (superscript xc) energy $2 \min_{k} E^\mathrm{xc}(\mathbf{k})$,
which is discussed in detail in the Supplemental Material \cite{sm}.
The phenomenological estimate uses the same formal expression of the excitation energies of electrons and hole in BEC/BCS state $E^\mathrm{xc}(\mathbf{k})$ and a minimization with respect to
$\mathbf{k}$. As shown in Fig.~\ref{r955-fk.fig}(c), the pair gap opens at a density corresponding to the BCS regime [see Fig.~\ref{fig:R-spectra}(c)] and increases to about 7~meV at twice threshold.
From Fig.~\ref{r955-fk.fig}(c) we see that
the gap is substantially smaller than that in the ideal system, which is expected due to cavity dissipation and dephasing and is consistent with the reduced order parameter $|P(k)|$. Such a reduction of the gap due to losses was predicted in Ref.~\cite{szymanska-etal.03}. Further discussion and results for the dependence of our phenomenological gap estimate on the dephasing rate are given in Ref.~\cite{sm}.

Our theory correctly reproduces not only the shift of the TE emission, including the size of the overall shift between zero
pumping and threshold, and the leveling off of the shift above threshold, but also the behavior (shift and vanishing)
of the TM exciton and the occurrence of TM gain, which sets in at a pump density (or pump power) at which the TE threshold
 occurs. The concurrence of these various theory-experiment agreements with our presently used theory is an important
 indication that the agreement is not coincidental or an artifact of fitting parameters. In addition to the theory
 presented here, we have developed several other theories that did not reproduce the various agreements just mentioned,
  in spite of the fact that some of those theories had fitting parameters with substantial effect on the predicted
  polariton shift.

\section{Conclusion}
In conclusion, we demonstrate
a system that exhibits spontaneous symmetry breaking
(polariton condensate or laser)
 which, although having spectral features commonly identified in a polariton BEC, is shown to take place above the Mott transition, with physical characteristics consistent with an open-dissipative-pumped system analog of a
 polaritonic BCS state \cite{footnote_ShimanoPRL}.

We summarize our results in the table in Fig.~\ref{fig:sample}, where we highlight in green the properties we used to distinguish our
polariton state
from a polariton BEC in the low-density regime, and in blue from a photon laser in the high-density regime.
Distinct from a photon laser, our polariton laser is formed by a bound state of electron, hole, and photon, manifested in spectral features including an emission linewidth that remains narrow and emission frequency well below the cavity or exciton resonance frequencies without sudden shifts. These spectral features closely resemble a polariton BEC. However, we show that spectral features of the laser emission alone is insufficient for identifying it to be in the BEC or BCS regime. Fermionic gain in the electronic reservoir was measured in our system, suggesting a BCS-like state.

The experimental observations are described by a fermionic many-particle theory that extends the ideal BCS theory to an open, pumped and dissipative system at nonzero temperature. The theory, validated by the experiments, furthermore reveals at the microscopic level that our polariton laser crucially involves the electron-hole interactions typical for the formation of a polaritonic BCS state, and has electron distributions and interband coherences that are qualitatively similar to the zero-temperature polariton BCS states.
The pair-breaking excitation gap, which can neither be ruled out nor confirmed with our present experimental setup, is found to be substantially reduced from that of an ideal polariton BCS state due to the elevated electron temperature, dephasing, and cavity dissipation.

Future theoretical work may
include extensions of the theory that would  account for bosonic effects within a fermionic theory. For example, the theory should account for
(bosonic) populations of normal-state TE polaritons and TM excitons, i.e., incoherent polaritons and excitons that form a bath. One possible way to achieve this
 could be the extension of the self-energy to the level of the $T$-matrix approximation where sidebands in the single-particle spectral function can be associated with excitonic populations
\cite{%
schaefer-etal.88b,
haussmann.93,%
schmielau-etal.00,%
pieri-etal.04,%
kwong-etal.09%
},
but now generalized to include polariton effects in the TM channel.
Furthermore, the treatment of screening should be improved by accounting for excitonic screening in the low-density regime. This could be achieved
through a generalization of the screening function developed in Ref. \cite{ropke-der.79}. A further extension would account, at a microscopic level
and ideally without phenomenological parameters, for bosonic scattering in the fermionic framework in order to describe consistently an open and pumped system with bosonic gain via polariton scattering from an incoherent bath and fermionic gain.

Future experiments to directly probe the BCS gap using intraband terahertz spectroscopy \cite{menard_revealing_2014} could yield important information \cite{binder-arxiv.2020}. 
It will be interesting to examine the possibility of a crossovers among polariton BEC, polariton BCS, and photon lasing state,
determine whether a non-Hermitian phase transition may take place \cite{hanai-etal.2019}, and  explore conditions to realize other possible phases of the electron-hole-photon coupled system.

\appendix

\section{\label{app:sample}Microcavity sample}
The samples have three sets of four 12-nm-wide GaAs quantum wells with 4~nm AlAs barriers embedded at the three central antinodes of a $\lambda/2$ $\mathrm{AlAs}$ cavity. The bottom mirror is a distributed Bragg reflector (DBR) with 30 pairs of $\mathrm{Al_{0.15}Ga_{0.85}As/AlAs}$ layers. The top mirror consists of 2.5 pairs of DBR and an $\mathrm{Al_{0.15}Ga_{0.85}As}$ subwavelength grating suspending over the DBR. The grating is about 80~nm thick, with a 40\% duty cycle and a grating period of 520~nm. The lower polariton (LP) and upper polariton (UP) resonances of each device are measured by reflection and low-power photoluminescence spectroscopy [see examples in Fig.~\ref{fig:lasing}(a) and Appendix~\ref{app:TE_spectra}].
 The exciton resonances are measured from the unetched part next to the device. Using the measured LP, UP and exciton resonances, we estimate the cavity resonance frequency, detuning, and normal-mode splitting for each grating device. The size of the gratings is $7.5 \times 7.5~\mathrm{\mu m^2}$. The different reflectance under the grating and outside the grating leads to laterally confined, fully discrete TE-cavity and TE-polariton  modes as evident from Fig.~\ref{fig:lasing}(a). The sample is kept at $10~\mathrm{K}$ in a Janis ST-500 or Montana CR-509 cryostat for all the measurements.

\section{\label{app:gain}Measuring the bandwidth of the optical gain}
To determine the gain bandwidth requires measurement of the absolute reflectivity, which we obtain by a two-step calibration process. We first calibrate the reflectivity by normalizing the reflected pulse laser intensity on the sample to that on a gold mirror mounted in the cryostat next to the sample. There is still error due to imperfect reflection from the gold mirror as well as laser fluctuation and slight changes in the optical path when moving between the sample and the mirror. We then estimate and correct for the error using a spectral regime well below any cavity resonances but still inside the DBR stop band, where the reflectivity should be unity at no or low pump powers. As shown in Fig.~\ref{fig:R-spectra}(a), such a spectra region corresponds to $1.53-1.54~\mathrm{eV}$, where the reflectivity is close to 1 and varies within about $2.4\%$.
Hence we use the mean of the reflectivity in this spectral region between $0$ and $500~\mathrm{\mu W}$ pump power as a reference of unity. This allows us to accurately determine deviation from unity in the reflection spectra.

Gain is identified where a local maximum of the reflectivity between $1.54$ and $1.55~\mathrm{eV}$ is above unity as determined from above. The uncertainty of the gain boundary is estimated by dividing the standard deviation $\sigma$ of the reference reflectivity by the local slope at the boundary.

\section{\label{app:TE_spectra}Angle-integrated TE spectra of a BCS-like polariton laser}
\begin{figure}[tp!]
\begin{center}
\begin{minipage}{0.6\textwidth}
\begin{center}
\includegraphics[angle=0, scale=0.5, trim= -1cm 0cm 0cm 0cm]{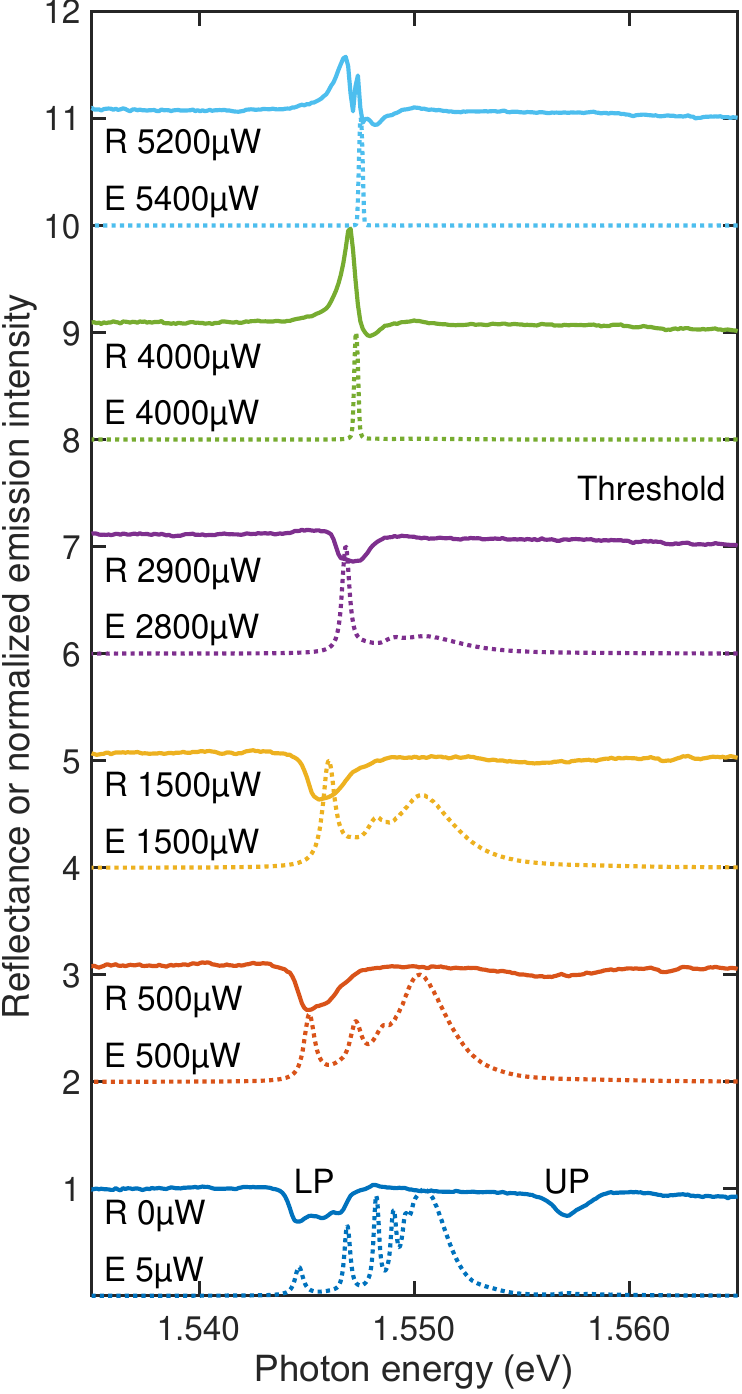}
\end{center}
\end{minipage}
\caption{\label{fig:FigS_spectra}
{\bf Measured TE reflection and emission spectra of a BCS-like polariton laser.}
The spectra at different pump laser powers are displaced vertically for clarity. Because of the finite in-plane size of the grating, the polariton modes are fully discrete, shown as multiple discrete peaks or dips in the spectra. The reflection (solid lines) is integrated in both $q_x$ and $q_y$, where all the modes are detectable; the emission intensity (dashed lines) is only integrated along $q_x$ at $q_y \sim 0$; therefore, only the modes with nonzero intensity along $q_y=0$ are detectable \cite{zhang_zerodimensional_2014}. For example, at lowest pump laser power ($0~\mathrm{\mu W}$ for reflection and $5~\mathrm{\mu W}$ for emission), the lowest energy peak in both reflection and emission spectra corresponds to the LP ground state. The second lowest energy peak in emission spectrum is the second excited state that has a node at $q_x = 0$, and it also shows up in the reflection. In between these two peaks, there is another peak in the reflection spectrum corresponding to the first excited state that has a node at $q_y=0$. In the reflection spectrum, only one UP mode can be resolved. The UP mode has low emission intensity and appears only as a very small peak in the emission spectrum. All the detected modes match closely in frequency between the reflection and emission spectra.}
\end{center}
\end{figure}

In Fig.~\ref{fig:FigS_spectra} we show the 
TE-polarized reflection and emission spectra of the device shown in Figs.~\ref{fig:R-spectra}(a) and \ref{fig:R-spectra}(b) of the main text at different pump laser powers integrated over $q$. 
LP and UP resonances are clearly seen in the reflection spectra at low pump powers. The multiple discrete LP modes due to the finite in-plane size of the grating are also clearly resolved at low pump powers. As the pump power increases, the emission intensity from LP ground state increases sharply, much faster than other modes, signaling lasing. At the highest power, the oscillations near the lasing peak in the reflection spectra are artifacts due to the subtraction of the strong lasing emission, as discussed in the main text. To get the energies of LP and UP modes at zero pump, we do curve fitting for TE reflection spectra with the model,
\begin{eqnarray}
R_\mathrm{TE}(E)=b-\sum_i c_i L_i(E, E_{0i}, \Gamma_i), \\
L_i(E, E_{0i}, \Gamma_i) = \frac{(\Gamma_i/2)^2}{(E-E_{0i})^2+(\Gamma_i/2)^2},
\end{eqnarray}
where $b$, $c_i$, $E_{0i}$, and $\Gamma_i$ are the fitting parameters. $L_i(E, E_{0i}, \Gamma_i)$ represent the resonances with energies $E_{0i}$ and FWHM $\Gamma_i$. The number of resonances used in the model depends on the number of resonances visible in the spectra. We use the energy of the lowest one in the LP (UP) regions as LP (UP) ground state energy.
In all our data, the emission frequencies of the lowest polariton or photon mode are obtained from TE emission spectra by curve fitting with the model
\begin{equation}
I_\mathrm{TE}(E)=\sum_i c_i f_i(E, E_{0i}, \Gamma_i),
\end{equation}
where $c_i$, $E_{0i}$, and $\Gamma_i$ are the fitting parameters. Each of the peaks $f_i(E, E_{0i}, \Gamma_i)$ is either Lorentzian
\begin{equation}\label{eq:Lorentzian}
f_i = \frac{(\Gamma_i/2)^2}{(E-E_{0i})^2+(\Gamma_i/2)^2},
\end{equation}
typically at low pump power, or Gaussian,
\begin{equation}\label{eq:Gaussian}
f_i = e^{-(E-E_{0i})^2/\{2[\Gamma_i/(2\sqrt{2\ln{2}})]^2\}},
\end{equation}
typically at high pump power, depending on which model returns a smaller mean squared error.

\section{\label{app:exciton_energy}Examples of exciton energy estimate}
\begin{figure}[tp!]
\begin{center}
\includegraphics[angle=0, scale=0.5, trim= 0cm 0cm 0cm -1cm]{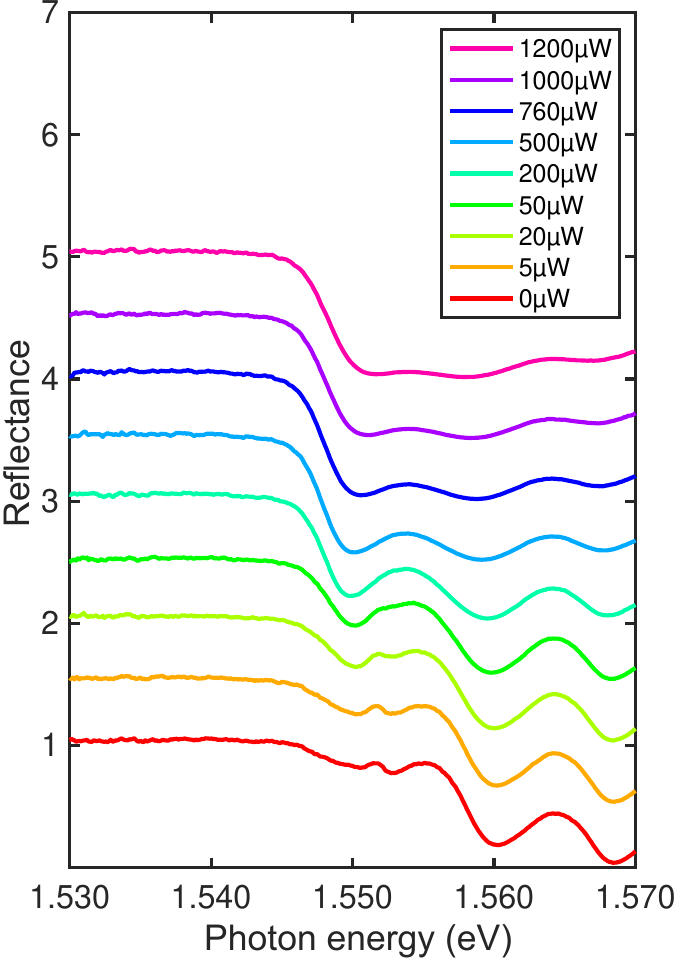}
\caption{\label{fig:FigS_TM_R_low_power}
{\bf TM-polarized reflection spectra at low pump power.} These spectra are used for the exciton energy estimation in Fig.~\ref{fig:R-spectra}(b) of the main text. The spectra at different pump laser powers are displaced vertically for clarity.
}
\end{center}
\end{figure}

\begin{figure}[tp!]
\begin{center}
\begin{minipage}{0.4\textwidth}
\begin{center}
\includegraphics[angle=0, scale=0.5, trim= 0cm 0cm -1cm 0cm]{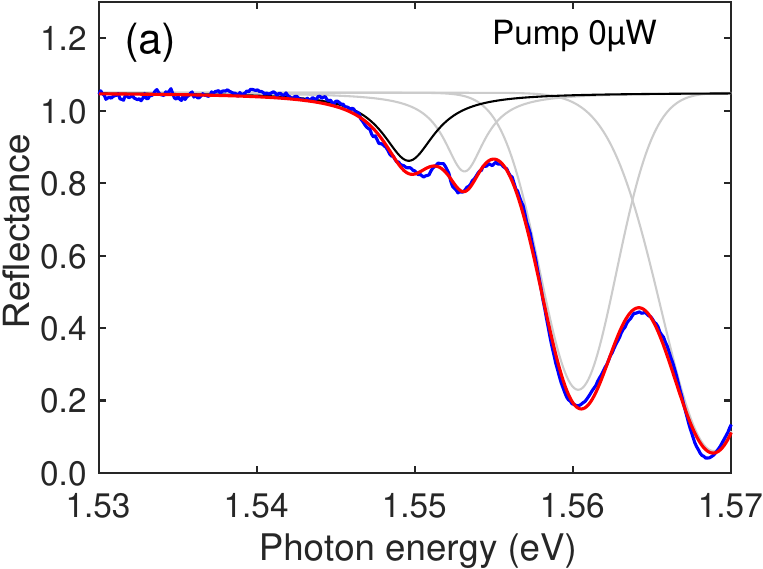}
\end{center}
\end{minipage}%
\begin{minipage}{0.4\textwidth}
\begin{center}
\includegraphics[angle=0, scale=0.5, trim= -1cm 0cm 0cm 0cm]{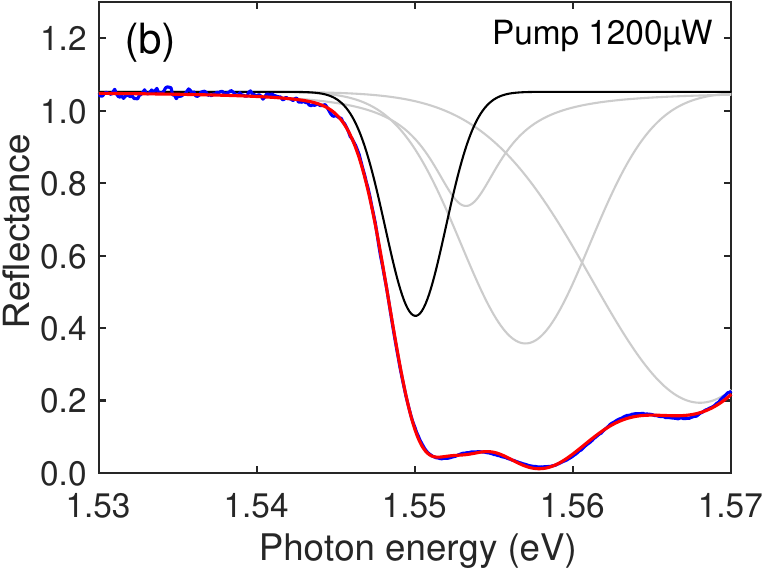}
\end{center}
\end{minipage}
\caption{\label{fig:FigS_exciton_energy_fitting}
{\bf Examples of exciton energy estimation from TM reflection spectra at pump powers $0~\mathrm{\mu W}$ (a) and $2000~\mathrm{\mu W}$ (b).}
The blue curve is the experimental spectra and the red curve is the fitted spectra. The black curve is the lowest energy peak used in the fitting, whose position is taken to be the exciton energy, and the gray curves are the other three peaks described in the text.
}
\end{center}
\end{figure}

We fit the TM reflection spectra to estimate the exciton energy at low pump powers, with the model
\begin{equation}
R_\mathrm{TM}(E)=b - \sum_i c_i f_i(E, E_{0i}, \Gamma_i).
\end{equation}
The four peaks $f_i(E, E_{0i}, \Gamma_i)$ used in the fitting are, from high energy to low energy, a broad TM cavity mode, light-hole exciton, and two heavy-hole exciton peaks possibly due to the inhomogeneity among the quantum wells. The energy of the lowest peak is used as the exciton energy. The line shapes of each peak are either Lorentzian or Gaussian [see Eqs.~\ref{eq:Lorentzian} and \ref{eq:Gaussian}] depending on which combination gives the smallest mean squared error. We show the complete series of data used for Fig.~\ref{fig:R-spectra}(b) of the main text in Fig.~\ref{fig:FigS_TM_R_low_power} and examples of curve fitting at two different pump powers in Fig.~\ref{fig:FigS_exciton_energy_fitting}.
At even higher pump powers we expect a redshift of the band gap and observe that the discrete peaks are no longer resolvable [see Figs.~\ref{fig:R-spectra}(a) and \ref{fig:R-spectra}(b) of the main text] and therefore this energy estimation is not applied to those pump powers.

\section{\label{app:density}Estimating the carrier density in the experiments}
We estimate an experimental pump density, corresponding to the pump density $n_p$ in the theory, based on the absorption of the pump laser and the PL decay time of the TM-polarized emission:
\begin{align}\label{eq:npump}
\Phi_\mathrm{pump} = & P \frac{\lambda}{hc} \eta \alpha,\\
n_\mathrm{pump} = & \Phi_\mathrm{pump} \tau \frac{1}{A} \frac{1}{N_\mathrm{QW}}.
\end{align}
Here $\Phi_\mathrm{pump}$ is the population of carriers excited by the pump in unit time. $P$ is the pump laser power measured before the objective lens. $\eta = 0.7$ is the transmission of the objective lens, $\alpha$ is the total absorption of the pump, ${hc}/{\lambda}$ is the photon energy at the excitation wavelength, and $\tau$ is the average lifetime of carriers. $A$ is the diffusion area of the carriers measured from real-space PL (example and pump power dependence given in the Supplemental Material \cite{sm}, and $N_\mathrm{QW}$ is the number of QWs in the cavity.
To determine $\alpha$, we measure independently the reflectivity of a TM-polarized laser at the excitation wavelength of $784~\mathrm{nm}$. At this wavelength, the top mirror has low reflectivity;
the bottom mirror has very high reflectivity and there is negligible transmission through the bottom mirror and therefore negligible absorption in the substrate. The materials of the DBR and grating layers also have larger band gap than the QWs and their absorption should be very small and is ignored. We assume the pump laser is either reflected or absorbed by the QWs, and obtain $\alpha=1-R$.
$\alpha$ and $A$ vary among different devices. For the device shown in Figs.~\ref{fig:R-spectra}(a) and \ref{fig:R-spectra}(b) of the main text, $\alpha \sim 0.3$ and $A \sim 32~\mathrm{\mu m^2}$.
We determine $\tau$ from the decay time of TM emission, measured by time-resolved PL of the TM emission with a pulsed excitation laser centered at the same wavelength as the cw pump laser used in the main measurement. It is found to be around $600~\mathrm{ps}$ with the excitation power ranging from $1$ to $300~\mathrm{\mu W}$. Nonradiative processes are ignored.

Below threshold, in the linear regime, the steady-state carrier density $n_\mathrm{carrier}$ is equal to the pump density $n_\mathrm{pump}$.
Above threshold, it is necessary to account for the strong, superlinear emission from the lasing mode. We assume that all decay processes other than the laser emission are still linear and proportional to the carrier density. Therefore, we subtract the emission rate of the laser $\Phi_\mathrm{laser}$ from the pump to get an estimate of the carrier density:
\begin{align}
\Phi_\mathrm{laser} & = N_{c} \frac{1}{\eta_\mathrm{col}} \frac{A_{q_\mathrm{mode}}}{A_{q_\mathrm{col}}},\\
n_\mathrm{carrier} & = (\Phi_\mathrm{pump}-\Phi_\mathrm{laser}) \tau \frac{1}{A} \frac{1}{N_\mathrm{QW}}.
\end{align}
Here $N_c$ is the collected photon count from the emission of the lasing mode, $\eta_\mathrm{col} = 2.85\times10^{-3}$ is the independently calibrated total collection efficiency, and ${A_{q_\mathrm{mode}}}/{A_{q_\mathrm{col}}}\sim 7.5$ is the ratio of the area of the lasing mode to the area of collection in Fourier space, which also varies among devices.

To estimate ground state polariton density, we use
\begin{align}
n_\mathrm{LP} = \Phi_\mathrm{LP} \tau_\mathrm{LP} \frac{1}{A_\mathrm{LP}},
\end{align}
where $\Phi_\mathrm{LP}$ is the emission rate of the polariton ground state, calculated similarly as $\Phi_\mathrm{laser}$ but with the emission photon count and the mode area of the polariton ground state. Above threshold, the lasing state (BCS-like polariton lasing or photon lasing) is used in this estimation, $\Phi_\mathrm{LP}=\Phi_\mathrm{laser}$. $\tau_\mathrm{LP}=1.5~\mathrm{ps}$ is the polariton lifetime and $A_\mathrm{LP}\sim 5.3~\mathrm{\mu m^2}$ is the real-space area of the polariton mode.

In this density estimation, quantities $P$, $\lambda$, $N_\mathrm{QW}$, $\eta$, $\eta_\mathrm{col}$, $R$, $\tau$, and $N_c$ are directly measured. $\alpha$ is estimated from $R$. $A$, $A_{q_\mathrm{mode}}$, $A_{q_\mathrm{col}}$, and $A_\mathrm{LP}$ are estimated based on measured 1D spatial profiles of the emission along $x$ and $y$ directions. $\tau_\mathrm{LP}$ is estimated from measured emission linewidth. This density estimation aims to provide an upper bound of density but should be within small uncertainty.

\section{\label{app:earlier-theories}Summary of alternative theoretical approaches}

In this appendix, we briefly summarize important features of alternative theories that we developed and that
failed to explain the experimental data. Common to the alternative models we tried, is the use of spin-dependent exciton-exciton interactions
  accounting for fermionic exchange that are commonly used in bosonic models, and that we quantitatively
  developed in a nonperturbative (in the Coulomb interaction) fashion leading to spin-dependent $T$ matrices;
  see Ref. \cite{takayama-etal.02}.
  Using the excitonic $T$ matrices, it is straightforward to calculate the resulting polariton shift
  as a function of polariton density \cite{schumacher-etal.07prb}. In order to understand the fact that
  the shift, as a function of density, levels off at a certain density (or pump power in the experiment),
  we extended the first version of the theory and included  spin-dependent (fermionic) phase-space filling.
  However, the effect of phase-space filling on the polariton shift is opposite to the
  observation: the shift increase nonlinearly as a function of density, rather than leveling off as in the experimental observation.

We further refined the model, accounting for each polarization
  (TE and TM) to have both coherent (zero center-of-mass momentum) and incoherent (nonzero
  center-of-mass momentum) exciton densities, as well as an electron-hole plasma reservoir,
  given by an adjustable ionization ratio (fitting parameter, in practice not chosen to be the same as
   the conventional Saha equation predicts). The condensation fraction, i.e., the fraction of the
    excitons in the TE zero center-of-mass momentum state, is also a fitting parameter.
    The model also accounts for interactions (Coulomb and phase-space filling)
    between bright and dark excitons, and the dark exciton fraction is a fitting parameter.
    The screening of the excitonic $T$-matrix interactions due to the plasma reservoir is included in the model.
    The model ensures, in a phenomenological fashion, that the single-particle distribution function, which is a sum of
    exciton contributions (proportional to the squared
    1$s$-exciton wave functions) and steplike (Fermi function) contributions, does not exceed unity.

   Within this model, the only way we found to obtain the experimental feature of the LP shift
     leveling off at threshold was to adjust the condensation fraction as a function of density.
     This is reasonable, because we know that the condensation fraction is zero below threshold,
     and nonzero above. Increasing the condensation fraction above threshold yields indeed the
     leveling-off of the LP shift. However, we found that the condensation fraction required
     to yield an appreciable effect (similar to the experiment) was unrealistically large.
     Only if the condensate density was comparable to or larger than the reservoir density
     did we obtain the leveling-off in a way similar to the experiment. Analyzing our experiment
     clearly excludes the possibility that the condensate density is as large as the reservoir density.
     Hence, the earlier model was unsuccessful in modeling our experiment. In contrast, the present model
     reproduces all salient features of the experiment shown Figs.~\ref{fig:R-spectra}(b) and \ref{fig:R-spectra}(c).

\begin{acknowledgments} J.H., Z.W., and H.D. acknowledge financial support from the U.S. Air Force Office of Scientific Research under Grant No. FA9550-15-1-0240 and the U.S. National Science Foundation (NSF) under Grant No. DMR 1150593.
     R.B.  acknowledges financial support from the U.S. National Science Foundation (NSF) under Grant No. DMR 1839570 and CPU time at HPC (University of Arizona).
    The W\"urzburg group gratefully acknowledges support by the state of Bavaria.
    J.H. and Z.W. performed the experiments and data analysis. Z.W. and H.D. designed the experiments. S.K. fabricated the devices on the wafer. S.B., C.S., and S.H. grew the wafer. R.B. and N.H.K.  performed the theoretical analysis and simulations. J.H., H.D., and R.B. wrote the manuscript. H.D. and R.B. supervised the project. All authors discussed the results and the manuscript.

The authors declare that they have no competing financial interests.

\end{acknowledgments}

\nocite{fetter-walecka.71}
\nocite{fetter-walecka.71}
\nocite{galitskii-etal.70}
\nocite{jerome-etal.67,
keldysh-kozlov.68,
hanamura-haug.77,
comte-nozieres.82}
\nocite{jahnke-henneberger.92,byrnes_bcs_2010,kamide_what_2010}
\nocite{fetter-walecka.71}
\nocite{comte-nozieres.82}
\nocite{comte-nozieres.82}
\nocite{keeling-etal.05,byrnes_bcs_2010,kamide_what_2010}
\nocite{byrnes_bcs_2010,kamide_what_2010}
\nocite{galitskii-etal.70}
\nocite{szymanska-etal.03}
\nocite{szymanska-etal.03}

\bibliography{references}

\begin{thebibliography}{58}%
\makeatletter
\providecommand \@ifxundefined [1]{%
 \@ifx{#1\undefined}
}%
\providecommand \@ifnum [1]{%
 \ifnum #1\expandafter \@firstoftwo
 \else \expandafter \@secondoftwo
 \fi
}%
\providecommand \@ifx [1]{%
 \ifx #1\expandafter \@firstoftwo
 \else \expandafter \@secondoftwo
 \fi
}%
\providecommand \natexlab [1]{#1}%
\providecommand \enquote  [1]{``#1''}%
\providecommand \bibnamefont  [1]{#1}%
\providecommand \bibfnamefont [1]{#1}%
\providecommand \citenamefont [1]{#1}%
\providecommand \href@noop [0]{\@secondoftwo}%
\providecommand \href [0]{\begingroup \@sanitize@url \@href}%
\providecommand \@href[1]{\@@startlink{#1}\@@href}%
\providecommand \@@href[1]{\endgroup#1\@@endlink}%
\providecommand \@sanitize@url [0]{\catcode `\\12\catcode `\$12\catcode
  `\&12\catcode `\#12\catcode `\^12\catcode `\_12\catcode `\%12\relax}%
\providecommand \@@startlink[1]{}%
\providecommand \@@endlink[0]{}%
\providecommand \url  [0]{\begingroup\@sanitize@url \@url }%
\providecommand \@url [1]{\endgroup\@href {#1}{\urlprefix }}%
\providecommand \urlprefix  [0]{URL }%
\providecommand \Eprint [0]{\href }%
\providecommand \doibase [0]{https://doi.org/}%
\providecommand \selectlanguage [0]{\@gobble}%
\providecommand \bibinfo  [0]{\@secondoftwo}%
\providecommand \bibfield  [0]{\@secondoftwo}%
\providecommand \translation [1]{[#1]}%
\providecommand \BibitemOpen [0]{}%
\providecommand \bibitemStop [0]{}%
\providecommand \bibitemNoStop [0]{.\EOS\space}%
\providecommand \EOS [0]{\spacefactor3000\relax}%
\providecommand \BibitemShut  [1]{\csname bibitem#1\endcsname}%
\let\auto@bib@innerbib\@empty
\bibitem [{\citenamefont {Dalfovo}\ \emph {et~al.}(1999)\citenamefont
  {Dalfovo}, \citenamefont {Giorgini}, \citenamefont {Pitaevskii},\ and\
  \citenamefont {Stringari}}]{dalfovo_theory_1999}%
  \BibitemOpen
  \bibfield  {author} {\bibinfo {author} {\bibfnamefont {F.}~\bibnamefont
  {Dalfovo}}, \bibinfo {author} {\bibfnamefont {S.}~\bibnamefont {Giorgini}},
  \bibinfo {author} {\bibfnamefont {L.~P.}\ \bibnamefont {Pitaevskii}},\ and\
  \bibinfo {author} {\bibfnamefont {S.}~\bibnamefont {Stringari}},\ }\bibfield
  {title} {\bibinfo {title} {Theory of {{Bose}}-{{Einstein}} condensation in
  trapped gases},\ }\href@noop {} {\bibfield  {journal} {\bibinfo  {journal}
  {Reviews of Modern Physics}\ }\textbf {\bibinfo {volume} {71}},\ \bibinfo
  {pages} {463} (\bibinfo {year} {1999})}\BibitemShut {NoStop}%
\bibitem [{\citenamefont {Leggett}(2001)}]{leggett_bose-einstein_2001}%
  \BibitemOpen
  \bibfield  {author} {\bibinfo {author} {\bibfnamefont {A.~J.}\ \bibnamefont
  {Leggett}},\ }\bibfield  {title} {\bibinfo {title} {Bose-{{Einstein}}
  condensation in the alkali gases: {{Some}} fundamental concepts},\
  }\href@noop {} {\bibfield  {journal} {\bibinfo  {journal} {Reviews of Modern
  Physics}\ }\textbf {\bibinfo {volume} {73}},\ \bibinfo {pages} {307}
  (\bibinfo {year} {2001})}\BibitemShut {NoStop}%
\bibitem [{\citenamefont {Bardeen}\ \emph {et~al.}(1957)\citenamefont
  {Bardeen}, \citenamefont {Cooper},\ and\ \citenamefont
  {Schrieffer}}]{bardeen_theory_1957}%
  \BibitemOpen
  \bibfield  {author} {\bibinfo {author} {\bibfnamefont {J.}~\bibnamefont
  {Bardeen}}, \bibinfo {author} {\bibfnamefont {L.~N.}\ \bibnamefont
  {Cooper}},\ and\ \bibinfo {author} {\bibfnamefont {J.~R.}\ \bibnamefont
  {Schrieffer}},\ }\bibfield  {title} {\bibinfo {title} {Theory of
  {{Superconductivity}}},\ }\href {https://doi.org/10.1103/PhysRev.108.1175}
  {\bibfield  {journal} {\bibinfo  {journal} {Physical Review}\ }\textbf
  {\bibinfo {volume} {108}},\ \bibinfo {pages} {1175} (\bibinfo {year}
  {1957})}\BibitemShut {NoStop}%
\bibitem [{\citenamefont {Kaiser}(2017)}]{kaiser.2017}%
  \BibitemOpen
  \bibfield  {author} {\bibinfo {author} {\bibfnamefont {S.}~\bibnamefont
  {Kaiser}},\ }\bibfield  {title} {\bibinfo {title} {{Light-induced
  superconductivity in high-Tc cuprates}},\ }\href@noop {} {\bibfield
  {journal} {\bibinfo  {journal} {Phys. Scr.}\ }\textbf {\bibinfo {volume}
  {92}},\ \bibinfo {pages} {1 } (\bibinfo {year} {2017})}\BibitemShut {NoStop}%
\bibitem [{\citenamefont {Galitskii}\ \emph {et~al.}(1970)\citenamefont
  {Galitskii}, \citenamefont {Goreslavskii},\ and\ \citenamefont
  {Elesin}}]{galitskii-etal.70}%
  \BibitemOpen
  \bibfield  {author} {\bibinfo {author} {\bibfnamefont {V.~M.}\ \bibnamefont
  {Galitskii}}, \bibinfo {author} {\bibfnamefont {S.~P.}\ \bibnamefont
  {Goreslavskii}},\ and\ \bibinfo {author} {\bibfnamefont {V.~F.}\ \bibnamefont
  {Elesin}},\ }\bibfield  {title} {\bibinfo {title} {{Electric and Magnetic
  Properties of a Semiconductor in the Field of a Strong Electromagnetic
  Wave}},\ }\href@noop {} {\bibfield  {journal} {\bibinfo  {journal} {Soviet
  Physics JETP}\ }\textbf {\bibinfo {volume} {30}},\ \bibinfo {pages} {117}
  (\bibinfo {year} {1970})}\BibitemShut {NoStop}%
\bibitem [{\citenamefont {Keldysh}\ and\ \citenamefont
  {Kozlov}(1968)}]{keldysh-kozlov.68}%
  \BibitemOpen
  \bibfield  {author} {\bibinfo {author} {\bibfnamefont {L.}~\bibnamefont
  {Keldysh}}\ and\ \bibinfo {author} {\bibfnamefont {A.}~\bibnamefont
  {Kozlov}},\ }\bibfield  {title} {\bibinfo {title} {Collective properties of
  excitons in semiconductors},\ }\href@noop {} {\bibfield  {journal} {\bibinfo
  {journal} {Sov. Phys. JETP}\ }\textbf {\bibinfo {volume} {27}},\ \bibinfo
  {pages} {521} (\bibinfo {year} {1968})}\BibitemShut {NoStop}%
\bibitem [{\citenamefont {Comte}\ and\ \citenamefont
  {Nozieres}(1982)}]{comte-nozieres.82}%
  \BibitemOpen
  \bibfield  {author} {\bibinfo {author} {\bibfnamefont {C.}~\bibnamefont
  {Comte}}\ and\ \bibinfo {author} {\bibfnamefont {P.}~\bibnamefont
  {Nozieres}},\ }\bibfield  {title} {\bibinfo {title} {Exciton {Bose}
  condensation: the ground state of an electron-hole gas i. mean field
  description of a simplified model},\ }\href@noop {} {\bibfield  {journal}
  {\bibinfo  {journal} {J. Physique}\ }\textbf {\bibinfo {volume} {43}},\
  \bibinfo {pages} {1069} (\bibinfo {year} {1982})}\BibitemShut {NoStop}%
\bibitem [{\citenamefont {Keeling}\ \emph {et~al.}(2007)\citenamefont
  {Keeling}, \citenamefont {Marchetti}, \citenamefont {Szyma{\'n}ska},\ and\
  \citenamefont {Littlewood}}]{keeling_collective_2007}%
  \BibitemOpen
  \bibfield  {author} {\bibinfo {author} {\bibfnamefont {J.}~\bibnamefont
  {Keeling}}, \bibinfo {author} {\bibfnamefont {F.~M.}\ \bibnamefont
  {Marchetti}}, \bibinfo {author} {\bibfnamefont {M.~H.}\ \bibnamefont
  {Szyma{\'n}ska}},\ and\ \bibinfo {author} {\bibfnamefont {P.~B.}\
  \bibnamefont {Littlewood}},\ }\bibfield  {title} {\bibinfo {title}
  {Collective coherence in planar semiconductor microcavities},\ }\href
  {https://doi.org/10.1088/0268-1242/22/5/R01} {\bibfield  {journal} {\bibinfo
  {journal} {Semiconductor Science and Technology}\ }\textbf {\bibinfo {volume}
  {22}},\ \bibinfo {pages} {R1} (\bibinfo {year} {2007})}\BibitemShut {NoStop}%
\bibitem [{\citenamefont {Carusotto}\ and\ \citenamefont
  {Ciuti}(2013)}]{carusotto_quantum_2013}%
  \BibitemOpen
  \bibfield  {author} {\bibinfo {author} {\bibfnamefont {I.}~\bibnamefont
  {Carusotto}}\ and\ \bibinfo {author} {\bibfnamefont {C.}~\bibnamefont
  {Ciuti}},\ }\bibfield  {title} {\bibinfo {title} {Quantum fluids of light},\
  }\href {https://doi.org/10.1103/RevModPhys.85.299} {\bibfield  {journal}
  {\bibinfo  {journal} {Reviews of Modern Physics}\ }\textbf {\bibinfo {volume}
  {85}},\ \bibinfo {pages} {299} (\bibinfo {year} {2013})}\BibitemShut
  {NoStop}%
\bibitem [{\citenamefont {Imamoglu}\ and\ \citenamefont
  {Ram}(1996)}]{imamoglu_quantum_1996}%
  \BibitemOpen
  \bibfield  {author} {\bibinfo {author} {\bibfnamefont {A.}~\bibnamefont
  {Imamoglu}}\ and\ \bibinfo {author} {\bibfnamefont {R.~J.}\ \bibnamefont
  {Ram}},\ }\bibfield  {title} {\bibinfo {title} {Quantum dynamics of exciton
  lasers},\ }\href@noop {} {\bibfield  {journal} {\bibinfo  {journal} {Phys.
  Lett. A}\ }\textbf {\bibinfo {volume} {214}},\ \bibinfo {pages} {193}
  (\bibinfo {year} {1996})}\BibitemShut {NoStop}%
\bibitem [{\citenamefont {Szymanska}\ \emph {et~al.}(2003)\citenamefont
  {Szymanska}, \citenamefont {Littlewood},\ and\ \citenamefont
  {Simons}}]{szymanska-etal.03}%
  \BibitemOpen
  \bibfield  {author} {\bibinfo {author} {\bibfnamefont {M.}~\bibnamefont
  {Szymanska}}, \bibinfo {author} {\bibfnamefont {P.}~\bibnamefont
  {Littlewood}},\ and\ \bibinfo {author} {\bibfnamefont {B.}~\bibnamefont
  {Simons}},\ }\bibfield  {title} {\bibinfo {title} {{Polariton condensation
  and lasing in optical microcavities: The decoherence-driven crossover}},\
  }\href@noop {} {\bibfield  {journal} {\bibinfo  {journal} {Phys. Rev. A}\
  }\textbf {\bibinfo {volume} {68}},\ \bibinfo {pages} {013818} (\bibinfo
  {year} {2003})}\BibitemShut {NoStop}%
\bibitem [{\citenamefont {Keeling}\ \emph {et~al.}(2005)\citenamefont
  {Keeling}, \citenamefont {Eastham}, \citenamefont {Szymanska},\ and\
  \citenamefont {Littlewood}}]{keeling-etal.05}%
  \BibitemOpen
  \bibfield  {author} {\bibinfo {author} {\bibfnamefont {J.}~\bibnamefont
  {Keeling}}, \bibinfo {author} {\bibfnamefont {P.~R.}\ \bibnamefont
  {Eastham}}, \bibinfo {author} {\bibfnamefont {M.~H.}\ \bibnamefont
  {Szymanska}},\ and\ \bibinfo {author} {\bibfnamefont {P.~B.}\ \bibnamefont
  {Littlewood}},\ }\bibfield  {title} {\bibinfo {title} {{{BCS}}-{{BEC}}
  crossover in a system of microcavity polaritons},\ }\href
  {https://doi.org/10.1103/PhysRevB.72.115320} {\bibfield  {journal} {\bibinfo
  {journal} {Physical Review B}\ }\textbf {\bibinfo {volume} {72}},\ \bibinfo
  {pages} {115320} (\bibinfo {year} {2005})}\BibitemShut {NoStop}%
\bibitem [{\citenamefont {Kamide}\ and\ \citenamefont
  {Ogawa}(2010)}]{kamide_what_2010}%
  \BibitemOpen
  \bibfield  {author} {\bibinfo {author} {\bibfnamefont {K.}~\bibnamefont
  {Kamide}}\ and\ \bibinfo {author} {\bibfnamefont {T.}~\bibnamefont {Ogawa}},\
  }\bibfield  {title} {\bibinfo {title} {What {{Determines}} the {{Wave
  Function}} of {{Electron}}-{{Hole Pairs}} in {{Polariton Condensates}}?},\
  }\href {https://doi.org/10.1103/PhysRevLett.105.056401} {\bibfield  {journal}
  {\bibinfo  {journal} {Physical Review Letters}\ }\textbf {\bibinfo {volume}
  {105}},\ \bibinfo {pages} {056401} (\bibinfo {year} {2010})}\BibitemShut
  {NoStop}%
\bibitem [{\citenamefont {Byrnes}\ \emph {et~al.}(2010)\citenamefont {Byrnes},
  \citenamefont {Horikiri}, \citenamefont {Ishida},\ and\ \citenamefont
  {Yamamoto}}]{byrnes_bcs_2010}%
  \BibitemOpen
  \bibfield  {author} {\bibinfo {author} {\bibfnamefont {T.}~\bibnamefont
  {Byrnes}}, \bibinfo {author} {\bibfnamefont {T.}~\bibnamefont {Horikiri}},
  \bibinfo {author} {\bibfnamefont {N.}~\bibnamefont {Ishida}},\ and\ \bibinfo
  {author} {\bibfnamefont {Y.}~\bibnamefont {Yamamoto}},\ }\bibfield  {title}
  {\bibinfo {title} {{{BCS Wave}}-{{Function Approach}} to the {{BEC}}-{{BCS
  Crossover}} of {{Exciton}}-{{Polariton Condensates}}},\ }\href
  {https://doi.org/10.1103/PhysRevLett.105.186402} {\bibfield  {journal}
  {\bibinfo  {journal} {Physical Review Letters}\ }\textbf {\bibinfo {volume}
  {105}},\ \bibinfo {pages} {186402} (\bibinfo {year} {2010})}\BibitemShut
  {NoStop}%
\bibitem [{\citenamefont {Horikiri}\ \emph {et~al.}(2016)\citenamefont
  {Horikiri}, \citenamefont {Yamaguchi}, \citenamefont {Kamide}, \citenamefont
  {Matsuo}, \citenamefont {Byrnes}, \citenamefont {Ishida}, \citenamefont
  {L{\"o}ffler}, \citenamefont {H{\"o}fling}, \citenamefont {Shikano},
  \citenamefont {Ogawa}, \citenamefont {Forchel},\ and\ \citenamefont
  {Yamamoto}}]{horikiri_highenergy_2016}%
  \BibitemOpen
  \bibfield  {author} {\bibinfo {author} {\bibfnamefont {T.}~\bibnamefont
  {Horikiri}}, \bibinfo {author} {\bibfnamefont {M.}~\bibnamefont {Yamaguchi}},
  \bibinfo {author} {\bibfnamefont {K.}~\bibnamefont {Kamide}}, \bibinfo
  {author} {\bibfnamefont {Y.}~\bibnamefont {Matsuo}}, \bibinfo {author}
  {\bibfnamefont {T.}~\bibnamefont {Byrnes}}, \bibinfo {author} {\bibfnamefont
  {N.}~\bibnamefont {Ishida}}, \bibinfo {author} {\bibfnamefont
  {A.}~\bibnamefont {L{\"o}ffler}}, \bibinfo {author} {\bibfnamefont
  {S.}~\bibnamefont {H{\"o}fling}}, \bibinfo {author} {\bibfnamefont
  {Y.}~\bibnamefont {Shikano}}, \bibinfo {author} {\bibfnamefont
  {T.}~\bibnamefont {Ogawa}}, \bibinfo {author} {\bibfnamefont
  {A.}~\bibnamefont {Forchel}},\ and\ \bibinfo {author} {\bibfnamefont
  {Y.}~\bibnamefont {Yamamoto}},\ }\bibfield  {title} {\bibinfo {title}
  {High-energy side-peak emission of exciton-polariton condensates in high
  density regime},\ }\href {https://doi.org/10.1038/srep25655} {\bibfield
  {journal} {\bibinfo  {journal} {Scientific Reports}\ }\textbf {\bibinfo
  {volume} {6}},\ \bibinfo {pages} {25655} (\bibinfo {year}
  {2016})}\BibitemShut {NoStop}%
\bibitem [{\citenamefont {Deng}\ \emph {et~al.}(2010)\citenamefont {Deng},
  \citenamefont {Haug},\ and\ \citenamefont
  {Yamamoto}}]{deng_exciton-polariton_2010}%
  \BibitemOpen
  \bibfield  {author} {\bibinfo {author} {\bibfnamefont {H.}~\bibnamefont
  {Deng}}, \bibinfo {author} {\bibfnamefont {H.}~\bibnamefont {Haug}},\ and\
  \bibinfo {author} {\bibfnamefont {Y.}~\bibnamefont {Yamamoto}},\ }\bibfield
  {title} {\bibinfo {title} {Exciton-polariton {Bose}-{Einstein}
  condensation},\ }\href {https://doi.org/10.1103/RevModPhys.82.1489}
  {\bibfield  {journal} {\bibinfo  {journal} {Reviews of Modern Physics}\
  }\textbf {\bibinfo {volume} {82}},\ \bibinfo {pages} {1489} (\bibinfo {year}
  {2010})}\BibitemShut {NoStop}%
\bibitem [{\citenamefont {Littlewood}\ \emph {et~al.}(2004)\citenamefont
  {Littlewood}, \citenamefont {Eastham}, \citenamefont {Keeling}, \citenamefont
  {Marchetti}, \citenamefont {Simons},\ and\ \citenamefont
  {Szymanska}}]{littlewood_models_2004}%
  \BibitemOpen
  \bibfield  {author} {\bibinfo {author} {\bibfnamefont {P.~B.}\ \bibnamefont
  {Littlewood}}, \bibinfo {author} {\bibfnamefont {P.~R.}\ \bibnamefont
  {Eastham}}, \bibinfo {author} {\bibfnamefont {J.~M.~J.}\ \bibnamefont
  {Keeling}}, \bibinfo {author} {\bibfnamefont {F.~M.}\ \bibnamefont
  {Marchetti}}, \bibinfo {author} {\bibfnamefont {B.~D.}\ \bibnamefont
  {Simons}},\ and\ \bibinfo {author} {\bibfnamefont {M.~H.}\ \bibnamefont
  {Szymanska}},\ }\bibfield  {title} {\bibinfo {title} {Models of coherent
  exciton condensation},\ }\href {https://doi.org/10.1088/0953-8984/16/35/003}
  {\bibfield  {journal} {\bibinfo  {journal} {Journal of Physics: Condensed
  Matter}\ }\textbf {\bibinfo {volume} {16}},\ \bibinfo {pages} {S3597}
  (\bibinfo {year} {2004})}\BibitemShut {NoStop}%
\bibitem [{\citenamefont {Kremp}\ \emph {et~al.}(2008)\citenamefont {Kremp},
  \citenamefont {Semkat},\ and\ \citenamefont {Henneberger}}]{kremp-etal.2008}%
  \BibitemOpen
  \bibfield  {author} {\bibinfo {author} {\bibfnamefont {D.}~\bibnamefont
  {Kremp}}, \bibinfo {author} {\bibfnamefont {D.}~\bibnamefont {Semkat}},\ and\
  \bibinfo {author} {\bibfnamefont {K.}~\bibnamefont {Henneberger}},\
  }\bibfield  {title} {\bibinfo {title} {{Quantum condensation in electron-hole
  plasmas}},\ }\href@noop {} {\bibfield  {journal} {\bibinfo  {journal} {Phys.
  Rev. B}\ }\textbf {\bibinfo {volume} {78}},\ \bibinfo {pages} {125315}
  (\bibinfo {year} {2008})}\BibitemShut {NoStop}%
\bibitem [{\citenamefont {Shiau}\ and\ \citenamefont
  {Combescot}(2016)}]{combescot-shiau.15}%
  \BibitemOpen
  \bibfield  {author} {\bibinfo {author} {\bibfnamefont {S.}~\bibnamefont
  {Shiau}}\ and\ \bibinfo {author} {\bibfnamefont {M.}~\bibnamefont
  {Combescot}},\ }\href@noop {} {\emph {\bibinfo {title} {{Excitons and Cooper
  Pairs: Two Composite Bosons in Many-Body Physics}}}}\ (\bibinfo  {publisher}
  {Oxford University Press},\ \bibinfo {address} {Oxford, England},\ \bibinfo
  {year} {2016})\BibitemShut {NoStop}%
\bibitem [{Note1()}]{Note1}%
  \BibitemOpen
  \bibinfo {note} {It has recently been shown that non-hermitian phase
  transitions can occur in a system with broken U(1) symmetry {\protect \cite
  {hanai-etal.2019}}, but a robust relation between those phase transitions and
  existing polariton condensation experiments has not yet been
  established.}\BibitemShut {Stop}%
\bibitem [{\citenamefont {Deng}\ \emph {et~al.}(2002)\citenamefont {Deng},
  \citenamefont {Weihs}, \citenamefont {Santori}, \citenamefont {Bloch},\ and\
  \citenamefont {Yamamoto}}]{deng_condensation_2002a}%
  \BibitemOpen
  \bibfield  {author} {\bibinfo {author} {\bibfnamefont {H.}~\bibnamefont
  {Deng}}, \bibinfo {author} {\bibfnamefont {G.}~\bibnamefont {Weihs}},
  \bibinfo {author} {\bibfnamefont {C.}~\bibnamefont {Santori}}, \bibinfo
  {author} {\bibfnamefont {J.}~\bibnamefont {Bloch}},\ and\ \bibinfo {author}
  {\bibfnamefont {Y.}~\bibnamefont {Yamamoto}},\ }\bibfield  {title} {\bibinfo
  {title} {Condensation of {{Semiconductor Microcavity Exciton Polaritons}}},\
  }\href {https://doi.org/10.1126/science.1074464} {\bibfield  {journal}
  {\bibinfo  {journal} {Science}\ }\textbf {\bibinfo {volume} {298}},\ \bibinfo
  {pages} {199} (\bibinfo {year} {2002})}\BibitemShut {NoStop}%
\bibitem [{\citenamefont {Kasprzak}\ \emph {et~al.}(2006)\citenamefont
  {Kasprzak}, \citenamefont {Richard}, \citenamefont {Kundermann},
  \citenamefont {Baas}, \citenamefont {Jeambrun}, \citenamefont {Keeling},
  \citenamefont {Marchetti}, \citenamefont {Szyma{\'n}ska}, \citenamefont
  {Andr{\'e}}, \citenamefont {Staehli}, \citenamefont {Savona}, \citenamefont
  {Littlewood}, \citenamefont {Deveaud},\ and\ \citenamefont
  {Dang}}]{kasprzak_bose-einstein_2006}%
  \BibitemOpen
  \bibfield  {author} {\bibinfo {author} {\bibfnamefont {J.}~\bibnamefont
  {Kasprzak}}, \bibinfo {author} {\bibfnamefont {M.}~\bibnamefont {Richard}},
  \bibinfo {author} {\bibfnamefont {S.}~\bibnamefont {Kundermann}}, \bibinfo
  {author} {\bibfnamefont {A.}~\bibnamefont {Baas}}, \bibinfo {author}
  {\bibfnamefont {P.}~\bibnamefont {Jeambrun}}, \bibinfo {author}
  {\bibfnamefont {J.~M.~J.}\ \bibnamefont {Keeling}}, \bibinfo {author}
  {\bibfnamefont {F.~M.}\ \bibnamefont {Marchetti}}, \bibinfo {author}
  {\bibfnamefont {M.~H.}\ \bibnamefont {Szyma{\'n}ska}}, \bibinfo {author}
  {\bibfnamefont {R.}~\bibnamefont {Andr{\'e}}}, \bibinfo {author}
  {\bibfnamefont {J.~L.}\ \bibnamefont {Staehli}}, \bibinfo {author}
  {\bibfnamefont {V.}~\bibnamefont {Savona}}, \bibinfo {author} {\bibfnamefont
  {P.~B.}\ \bibnamefont {Littlewood}}, \bibinfo {author} {\bibfnamefont
  {B.}~\bibnamefont {Deveaud}},\ and\ \bibinfo {author} {\bibfnamefont {L.~S.}\
  \bibnamefont {Dang}},\ }\bibfield  {title} {\bibinfo {title}
  {Bose\textendash{}{{Einstein}} condensation of exciton polaritons},\ }\href
  {https://doi.org/10.1038/nature05131} {\bibfield  {journal} {\bibinfo
  {journal} {Nature}\ }\textbf {\bibinfo {volume} {443}},\ \bibinfo {pages}
  {409} (\bibinfo {year} {2006})}\BibitemShut {NoStop}%
\bibitem [{\citenamefont {Bajoni}\ \emph {et~al.}(2008)\citenamefont {Bajoni},
  \citenamefont {Senellart}, \citenamefont {Wertz}, \citenamefont {Sagnes},
  \citenamefont {Miard}, \citenamefont {Lema{\^\i}tre},\ and\ \citenamefont
  {Bloch}}]{bajoni_polariton_2008}%
  \BibitemOpen
  \bibfield  {author} {\bibinfo {author} {\bibfnamefont {D.}~\bibnamefont
  {Bajoni}}, \bibinfo {author} {\bibfnamefont {P.}~\bibnamefont {Senellart}},
  \bibinfo {author} {\bibfnamefont {E.}~\bibnamefont {Wertz}}, \bibinfo
  {author} {\bibfnamefont {I.}~\bibnamefont {Sagnes}}, \bibinfo {author}
  {\bibfnamefont {A.}~\bibnamefont {Miard}}, \bibinfo {author} {\bibfnamefont
  {A.}~\bibnamefont {Lema{\^\i}tre}},\ and\ \bibinfo {author} {\bibfnamefont
  {J.}~\bibnamefont {Bloch}},\ }\bibfield  {title} {\bibinfo {title} {Polariton
  laser using single micropillar {{GaAs-GaAlAs}} semiconductor cavities},\
  }\href {https://doi.org/10.1103/PhysRevLett.100.047401} {\bibfield  {journal}
  {\bibinfo  {journal} {Physical Review Letters}\ }\textbf {\bibinfo {volume}
  {100}},\ \bibinfo {pages} {047401} (\bibinfo {year} {2008})}\BibitemShut
  {NoStop}%
\bibitem [{\citenamefont {Kim}\ \emph {et~al.}(2016)\citenamefont {Kim},
  \citenamefont {Zhang}, \citenamefont {Wang}, \citenamefont {Fischer},
  \citenamefont {Brodbeck}, \citenamefont {Kamp}, \citenamefont {Schneider},
  \citenamefont {Höfling},\ and\ \citenamefont {Deng}}]{kim_coherent_2016}%
  \BibitemOpen
  \bibfield  {author} {\bibinfo {author} {\bibfnamefont {S.}~\bibnamefont
  {Kim}}, \bibinfo {author} {\bibfnamefont {B.}~\bibnamefont {Zhang}}, \bibinfo
  {author} {\bibfnamefont {Z.}~\bibnamefont {Wang}}, \bibinfo {author}
  {\bibfnamefont {J.}~\bibnamefont {Fischer}}, \bibinfo {author} {\bibfnamefont
  {S.}~\bibnamefont {Brodbeck}}, \bibinfo {author} {\bibfnamefont
  {M.}~\bibnamefont {Kamp}}, \bibinfo {author} {\bibfnamefont {C.}~\bibnamefont
  {Schneider}}, \bibinfo {author} {\bibfnamefont {S.}~\bibnamefont
  {Höfling}},\ and\ \bibinfo {author} {\bibfnamefont {H.}~\bibnamefont
  {Deng}},\ }\bibfield  {title} {\bibinfo {title} {Coherent {Polariton}
  {Laser}},\ }\href {https://doi.org/10.1103/PhysRevX.6.011026} {\bibfield
  {journal} {\bibinfo  {journal} {Phys. Rev. X}\ }\textbf {\bibinfo {volume}
  {6}},\ \bibinfo {pages} {011026} (\bibinfo {year} {2016})}\BibitemShut
  {NoStop}%
\bibitem [{\citenamefont {M{\'e}nard}\ \emph {et~al.}(2014)\citenamefont
  {M{\'e}nard}, \citenamefont {Poellmann}, \citenamefont {Porer}, \citenamefont
  {Leierseder}, \citenamefont {Galopin}, \citenamefont {Lemaître},
  \citenamefont {Amo}, \citenamefont {Bloch},\ and\ \citenamefont
  {Huber}}]{menard_revealing_2014}%
  \BibitemOpen
  \bibfield  {author} {\bibinfo {author} {\bibfnamefont {J.-M.}\ \bibnamefont
  {M{\'e}nard}}, \bibinfo {author} {\bibfnamefont {C.}~\bibnamefont
  {Poellmann}}, \bibinfo {author} {\bibfnamefont {M.}~\bibnamefont {Porer}},
  \bibinfo {author} {\bibfnamefont {U.}~\bibnamefont {Leierseder}}, \bibinfo
  {author} {\bibfnamefont {E.}~\bibnamefont {Galopin}}, \bibinfo {author}
  {\bibfnamefont {A.}~\bibnamefont {Lemaître}}, \bibinfo {author}
  {\bibfnamefont {A.}~\bibnamefont {Amo}}, \bibinfo {author} {\bibfnamefont
  {J.}~\bibnamefont {Bloch}},\ and\ \bibinfo {author} {\bibfnamefont
  {R.}~\bibnamefont {Huber}},\ }\bibfield  {title} {\bibinfo {title} {Revealing
  the dark side of a bright exciton–polariton condensate},\ }\href
  {https://doi.org/10.1038/ncomms5648} {\bibfield  {journal} {\bibinfo
  {journal} {Nature Communications}\ }\textbf {\bibinfo {volume} {5}},\
  \bibinfo {pages} {4648} (\bibinfo {year} {2014})}\BibitemShut {NoStop}%
\bibitem [{sm()}]{sm}%
  \BibitemOpen
  \href@noop {} {}\bibinfo {note} {See Supplemental Material at [URL will be
  inserted by publisher] for the simulated reflectivity of the grating and
  electric field distribution in the cavity, data and theoretical discussion
  for the common electron and hole reservoir of TE and TM modes, additional
  spectra for TE polariton and TM exciton, data for devices with other
  detunings, and animations of detailed evolution of the spectra with power, as
  well as numerical parameters used in the theoretical calculations and details
  about the theoretical estimates of the BCS-like gap.}\BibitemShut {Stop}%
\bibitem [{\citenamefont {Zhang}\ \emph {et~al.}(2014)\citenamefont {Zhang},
  \citenamefont {Wang}, \citenamefont {Brodbeck}, \citenamefont {Schneider},
  \citenamefont {Kamp}, \citenamefont {H{\"o}fling},\ and\ \citenamefont
  {Deng}}]{zhang_zerodimensional_2014}%
  \BibitemOpen
  \bibfield  {author} {\bibinfo {author} {\bibfnamefont {B.}~\bibnamefont
  {Zhang}}, \bibinfo {author} {\bibfnamefont {Z.}~\bibnamefont {Wang}},
  \bibinfo {author} {\bibfnamefont {S.}~\bibnamefont {Brodbeck}}, \bibinfo
  {author} {\bibfnamefont {C.}~\bibnamefont {Schneider}}, \bibinfo {author}
  {\bibfnamefont {M.}~\bibnamefont {Kamp}}, \bibinfo {author} {\bibfnamefont
  {S.}~\bibnamefont {H{\"o}fling}},\ and\ \bibinfo {author} {\bibfnamefont
  {H.}~\bibnamefont {Deng}},\ }\bibfield  {title} {\bibinfo {title}
  {Zero-dimensional polariton laser in a subwavelength grating-based vertical
  microcavity},\ }\href {https://doi.org/10.1038/lsa.2014.16} {\bibfield
  {journal} {\bibinfo  {journal} {Light: Science \& Applications}\ }\textbf
  {\bibinfo {volume} {3}},\ \bibinfo {pages} {e135} (\bibinfo {year}
  {2014})}\BibitemShut {NoStop}%
\bibitem [{Note2()}]{Note2}%
  \BibitemOpen
  \bibinfo {note} {We integrate over the whole pulse to obtain the total
  intensity of reflected pulse, which is then divided by the reflection spectra
  from a reference gold mirror to produce the reflection spectra of the
  microcavity system.}\BibitemShut {Stop}%
\bibitem [{\citenamefont {Balili}\ \emph {et~al.}(2007)\citenamefont {Balili},
  \citenamefont {Hartwell}, \citenamefont {Snoke}, \citenamefont {Pfeiffer},\
  and\ \citenamefont {West}}]{balili_bose-einstein_2007}%
  \BibitemOpen
  \bibfield  {author} {\bibinfo {author} {\bibfnamefont {R.}~\bibnamefont
  {Balili}}, \bibinfo {author} {\bibfnamefont {V.}~\bibnamefont {Hartwell}},
  \bibinfo {author} {\bibfnamefont {D.}~\bibnamefont {Snoke}}, \bibinfo
  {author} {\bibfnamefont {L.}~\bibnamefont {Pfeiffer}},\ and\ \bibinfo
  {author} {\bibfnamefont {K.}~\bibnamefont {West}},\ }\bibfield  {title}
  {\bibinfo {title} {Bose-{{Einstein Condensation}} of {{Microcavity
  Polaritons}} in a {{Trap}}},\ }\href
  {https://doi.org/10.1126/science.1140990} {\bibfield  {journal} {\bibinfo
  {journal} {Science}\ }\textbf {\bibinfo {volume} {316}},\ \bibinfo {pages}
  {1007} (\bibinfo {year} {2007})}\BibitemShut {NoStop}%
\bibitem [{\citenamefont {Bajoni}\ \emph {et~al.}(2007)\citenamefont {Bajoni},
  \citenamefont {Senellart}, \citenamefont {Lema{\^\i}tre},\ and\ \citenamefont
  {Bloch}}]{bajoni_photon_2007}%
  \BibitemOpen
  \bibfield  {author} {\bibinfo {author} {\bibfnamefont {D.}~\bibnamefont
  {Bajoni}}, \bibinfo {author} {\bibfnamefont {P.}~\bibnamefont {Senellart}},
  \bibinfo {author} {\bibfnamefont {A.}~\bibnamefont {Lema{\^\i}tre}},\ and\
  \bibinfo {author} {\bibfnamefont {J.}~\bibnamefont {Bloch}},\ }\bibfield
  {title} {\bibinfo {title} {Photon lasing in {{GaAs}} microcavity:
  Similarities with a polariton condensate},\ }\href
  {https://doi.org/10.1103/PhysRevB.76.201305} {\bibfield  {journal} {\bibinfo
  {journal} {Physical Review B}\ }\textbf {\bibinfo {volume} {76}},\ \bibinfo
  {pages} {201305R} (\bibinfo {year} {2007})}\BibitemShut {NoStop}%
\bibitem [{\citenamefont {Balili}\ \emph {et~al.}(2009)\citenamefont {Balili},
  \citenamefont {Nelsen}, \citenamefont {Snoke}, \citenamefont {Pfeiffer},\
  and\ \citenamefont {West}}]{balili_role_2009}%
  \BibitemOpen
  \bibfield  {author} {\bibinfo {author} {\bibfnamefont {R.}~\bibnamefont
  {Balili}}, \bibinfo {author} {\bibfnamefont {B.}~\bibnamefont {Nelsen}},
  \bibinfo {author} {\bibfnamefont {D.~W.}\ \bibnamefont {Snoke}}, \bibinfo
  {author} {\bibfnamefont {L.}~\bibnamefont {Pfeiffer}},\ and\ \bibinfo
  {author} {\bibfnamefont {K.}~\bibnamefont {West}},\ }\bibfield  {title}
  {\bibinfo {title} {Role of the stress trap in the polariton quasiequilibrium
  condensation in {{GaAs}} microcavities},\ }\href
  {https://doi.org/10.1103/PhysRevB.79.075319} {\bibfield  {journal} {\bibinfo
  {journal} {Physical Review B}\ }\textbf {\bibinfo {volume} {79}},\ \bibinfo
  {pages} {075319} (\bibinfo {year} {2009})}\BibitemShut {NoStop}%
\bibitem [{\citenamefont {Indik}\ \emph {et~al.}(1996)\citenamefont {Indik},
  \citenamefont {Binder}, \citenamefont {Mlejnek}, \citenamefont {Moloney},
  \citenamefont {Hughes}, \citenamefont {Knorr},\ and\ \citenamefont
  {Koch}}]{indik-etal.96}%
  \BibitemOpen
  \bibfield  {author} {\bibinfo {author} {\bibfnamefont {R.~A.}\ \bibnamefont
  {Indik}}, \bibinfo {author} {\bibfnamefont {R.}~\bibnamefont {Binder}},
  \bibinfo {author} {\bibfnamefont {M.}~\bibnamefont {Mlejnek}}, \bibinfo
  {author} {\bibfnamefont {J.~V.}\ \bibnamefont {Moloney}}, \bibinfo {author}
  {\bibfnamefont {S.}~\bibnamefont {Hughes}}, \bibinfo {author} {\bibfnamefont
  {A.}~\bibnamefont {Knorr}},\ and\ \bibinfo {author} {\bibfnamefont {S.~W.}\
  \bibnamefont {Koch}},\ }\bibfield  {title} {\bibinfo {title} {Role of plasma
  cooling, heating and memory effects in sub-picosecond pulse propagation in
  semiconductor amplifiers},\ }\href@noop {} {\bibfield  {journal} {\bibinfo
  {journal} {Phys. Rev. A}\ }\textbf {\bibinfo {volume} {53}},\ \bibinfo
  {pages} {3614} (\bibinfo {year} {1996})}\BibitemShut {NoStop}%
\bibitem [{\citenamefont {Girndt}\ \emph {et~al.}(1997)\citenamefont {Girndt},
  \citenamefont {Jahnke}, \citenamefont {Chow}, \citenamefont {Knorr},
  \citenamefont {Koch},\ and\ \citenamefont {Chow}}]{girndt-etal.97}%
  \BibitemOpen
  \bibfield  {author} {\bibinfo {author} {\bibfnamefont {A.}~\bibnamefont
  {Girndt}}, \bibinfo {author} {\bibfnamefont {F.}~\bibnamefont {Jahnke}},
  \bibinfo {author} {\bibfnamefont {W.~W.}\ \bibnamefont {Chow}}, \bibinfo
  {author} {\bibfnamefont {A.}~\bibnamefont {Knorr}}, \bibinfo {author}
  {\bibfnamefont {S.~W.}\ \bibnamefont {Koch}},\ and\ \bibinfo {author}
  {\bibfnamefont {W.~W.}\ \bibnamefont {Chow}},\ }\bibfield  {title} {\bibinfo
  {title} {Multi-band bloch equations and gain spectra of highly excited ii-vi
  semiconductor quantum wells},\ }\href@noop {} {\bibfield  {journal} {\bibinfo
   {journal} {Phys. Stat. Sol (b)}\ }\textbf {\bibinfo {volume} {202}},\
  \bibinfo {pages} {725 } (\bibinfo {year} {1997})}\BibitemShut {NoStop}%
\bibitem [{\citenamefont {Semkat}\ \emph {et~al.}(2009)\citenamefont {Semkat},
  \citenamefont {Richter}, \citenamefont {Kremp}, \citenamefont {Manzke},
  \citenamefont {Kraeft},\ and\ \citenamefont {Henneberger}}]{semkat-etal.09}%
  \BibitemOpen
  \bibfield  {author} {\bibinfo {author} {\bibfnamefont {D.}~\bibnamefont
  {Semkat}}, \bibinfo {author} {\bibfnamefont {F.}~\bibnamefont {Richter}},
  \bibinfo {author} {\bibfnamefont {D.}~\bibnamefont {Kremp}}, \bibinfo
  {author} {\bibfnamefont {G.}~\bibnamefont {Manzke}}, \bibinfo {author}
  {\bibfnamefont {W.~D.}\ \bibnamefont {Kraeft}},\ and\ \bibinfo {author}
  {\bibfnamefont {K.}~\bibnamefont {Henneberger}},\ }\bibfield  {title}
  {\bibinfo {title} {{Ionization equilibrium in an excited semiconductor: Mott
  transition versus Bose-Einstein condensation}},\ }\href@noop {} {\bibfield
  {journal} {\bibinfo  {journal} {Phys. Rev. B}\ }\textbf {\bibinfo {volume}
  {80}},\ \bibinfo {pages} {155201} (\bibinfo {year} {2009})}\BibitemShut
  {NoStop}%
\bibitem [{\citenamefont {Yamaguchi}\ \emph {et~al.}(2013)\citenamefont
  {Yamaguchi}, \citenamefont {Kamide}, \citenamefont {Nii}, \citenamefont
  {Ogawa},\ and\ \citenamefont {Yamamoto}}]{yamaguchi_second_2013}%
  \BibitemOpen
  \bibfield  {author} {\bibinfo {author} {\bibfnamefont {M.}~\bibnamefont
  {Yamaguchi}}, \bibinfo {author} {\bibfnamefont {K.}~\bibnamefont {Kamide}},
  \bibinfo {author} {\bibfnamefont {R.}~\bibnamefont {Nii}}, \bibinfo {author}
  {\bibfnamefont {T.}~\bibnamefont {Ogawa}},\ and\ \bibinfo {author}
  {\bibfnamefont {Y.}~\bibnamefont {Yamamoto}},\ }\bibfield  {title} {\bibinfo
  {title} {Second {{Thresholds}} in {{BEC}}-{{BCS}}-{{Laser Crossover}} of
  {{Exciton}}-{{Polariton Systems}}},\ }\href
  {https://doi.org/10.1103/PhysRevLett.111.026404} {\bibfield  {journal}
  {\bibinfo  {journal} {Physical Review Letters}\ }\textbf {\bibinfo {volume}
  {111}},\ \bibinfo {pages} {026404} (\bibinfo {year} {2013})}\BibitemShut
  {NoStop}%
\bibitem [{Note3()}]{Note3}%
  \BibitemOpen
  \bibinfo {note} {In the anisotropic cavity, the role of dark coherences
  differs from that in isotropic systems {\protect \cite
  {combescot-etal.07prl}} and are neglected.}\BibitemShut {Stop}%
\bibitem [{\citenamefont {Binder}\ and\ \citenamefont
  {Koch}(1995)}]{binder-koch.95}%
  \BibitemOpen
  \bibfield  {author} {\bibinfo {author} {\bibfnamefont {R.}~\bibnamefont
  {Binder}}\ and\ \bibinfo {author} {\bibfnamefont {S.}~\bibnamefont {Koch}},\
  }\bibfield  {title} {\bibinfo {title} {Nonequilibrium semiconductor
  dynamics},\ }\href@noop {} {\bibfield  {journal} {\bibinfo  {journal} {Prog.
  Quant. Electr.}\ }\textbf {\bibinfo {volume} {19}},\ \bibinfo {pages} {307}
  (\bibinfo {year} {1995})}\BibitemShut {NoStop}%
\bibitem [{\citenamefont {Haug}\ and\ \citenamefont
  {Koch}(2004)}]{haug-koch.04}%
  \BibitemOpen
  \bibfield  {author} {\bibinfo {author} {\bibfnamefont {H.}~\bibnamefont
  {Haug}}\ and\ \bibinfo {author} {\bibfnamefont {S.~W.}\ \bibnamefont
  {Koch}},\ }\href@noop {} {\emph {\bibinfo {title} {Quantum Theory of the
  Optical and Electronic Properties of Semiconductors}}},\ \bibinfo {edition}
  {4th}\ ed.\ (\bibinfo  {publisher} {World Scientific},\ \bibinfo {address}
  {Singapore},\ \bibinfo {year} {2004})\BibitemShut {NoStop}%
\bibitem [{\citenamefont {Kwong}\ and\ \citenamefont
  {Binder}(2000)}]{kwong-binder.00}%
  \BibitemOpen
  \bibfield  {author} {\bibinfo {author} {\bibfnamefont {N.~H.}\ \bibnamefont
  {Kwong}}\ and\ \bibinfo {author} {\bibfnamefont {R.}~\bibnamefont {Binder}},\
  }\bibfield  {title} {\bibinfo {title} {Green's function approach to the
  dynamics-controlled truncation formalism: Derivation of the {$\chi^{(3)}$}
  equations of motion},\ }\href@noop {} {\bibfield  {journal} {\bibinfo
  {journal} {Phys. Rev. B}\ }\textbf {\bibinfo {volume} {61}},\ \bibinfo
  {pages} {8341 } (\bibinfo {year} {2000})}\BibitemShut {NoStop}%
\bibitem [{\citenamefont {Malinowski}\ \emph {et~al.}(1999)\citenamefont
  {Malinowski}, \citenamefont {Guerrier}, \citenamefont {Traynor}, ,\ and\
  \citenamefont {Harley}}]{malinowski-etal.99}%
  \BibitemOpen
  \bibfield  {author} {\bibinfo {author} {\bibfnamefont {A.}~\bibnamefont
  {Malinowski}}, \bibinfo {author} {\bibfnamefont {D.~J.}\ \bibnamefont
  {Guerrier}}, \bibinfo {author} {\bibfnamefont {N.~J.}\ \bibnamefont
  {Traynor}}, ,\ and\ \bibinfo {author} {\bibfnamefont {R.~T.}\ \bibnamefont
  {Harley}},\ }\bibfield  {title} {\bibinfo {title} {{Larmor beats and
  conduction electron g factors in InGaAs/GaAs quantum wells}},\ }\href@noop {}
  {\bibfield  {journal} {\bibinfo  {journal} {Phys. Rev. B}\ }\textbf {\bibinfo
  {volume} {60}},\ \bibinfo {pages} {7728} (\bibinfo {year}
  {1999})}\BibitemShut {NoStop}%
\bibitem [{\citenamefont {Haug}\ and\ \citenamefont
  {Schmitt-Rink}(1984)}]{haug-schmittrink.84}%
  \BibitemOpen
  \bibfield  {author} {\bibinfo {author} {\bibfnamefont {H.}~\bibnamefont
  {Haug}}\ and\ \bibinfo {author} {\bibfnamefont {S.}~\bibnamefont
  {Schmitt-Rink}},\ }\bibfield  {title} {\bibinfo {title} {Electron theory of
  the optical properties of laser-excited semiconductors},\ }\href@noop {}
  {\bibfield  {journal} {\bibinfo  {journal} {Prog. Quantum Electron.}\
  }\textbf {\bibinfo {volume} {9}},\ \bibinfo {pages} {3} (\bibinfo {year}
  {1984})}\BibitemShut {NoStop}%
\bibitem [{\citenamefont {Gu}\ \emph {et~al.}(2013)\citenamefont {Gu},
  \citenamefont {Kwong},\ and\ \citenamefont {Binder}}]{gu-etal.13}%
  \BibitemOpen
  \bibfield  {author} {\bibinfo {author} {\bibfnamefont {B.}~\bibnamefont
  {Gu}}, \bibinfo {author} {\bibfnamefont {N.}~\bibnamefont {Kwong}},\ and\
  \bibinfo {author} {\bibfnamefont {R.}~\bibnamefont {Binder}},\ }\bibfield
  {title} {\bibinfo {title} {Relation between the interband dipole and momentum
  matrix elements in semiconductors},\ }\href@noop {} {\bibfield  {journal}
  {\bibinfo  {journal} {Phys. Rev. B}\ }\textbf {\bibinfo {volume} {87}},\
  \bibinfo {pages} {125301} (\bibinfo {year} {2013})}\BibitemShut {NoStop}%
\bibitem [{foo()}]{footnote_ShimanoPRL}%
  \BibitemOpen
  \href@noop {} {}\bibinfo {note} {After submission of our manuscript to the
  arXiv:1902.00142, the observation of a light-driven BCS-like state in
  semiconductors (in contrast to ours not formed through spontaneous symmetry
  breaking, but through an external coherent light source) has been reported,
  Murotani et al., PRL 123, 197401 (2019)}\BibitemShut {NoStop}%
\bibitem [{\citenamefont {Sch{\"{a}}fer}\ \emph {et~al.}(1988)\citenamefont
  {Sch{\"{a}}fer}, \citenamefont {Schuldt},\ and\ \citenamefont
  {Treusch}}]{schaefer-etal.88b}%
  \BibitemOpen
  \bibfield  {author} {\bibinfo {author} {\bibfnamefont {W.}~\bibnamefont
  {Sch{\"{a}}fer}}, \bibinfo {author} {\bibfnamefont {K.~H.}\ \bibnamefont
  {Schuldt}},\ and\ \bibinfo {author} {\bibfnamefont {J.}~\bibnamefont
  {Treusch}},\ }\bibfield  {title} {\bibinfo {title} {Many-body effects in
  resonantly excited dense exciton systems},\ }\href@noop {} {\bibfield
  {journal} {\bibinfo  {journal} {phys. stat. sol. (b)}\ }\textbf {\bibinfo
  {volume} {147}},\ \bibinfo {pages} {699} (\bibinfo {year}
  {1988})}\BibitemShut {NoStop}%
\bibitem [{\citenamefont {Haussmann}(1993)}]{haussmann.93}%
  \BibitemOpen
  \bibfield  {author} {\bibinfo {author} {\bibfnamefont {R.}~\bibnamefont
  {Haussmann}},\ }\bibfield  {title} {\bibinfo {title} {{Crossover from BCS
  superconductivity to Bose-Einstein condensation: a self-consistent theory}},\
  }\href@noop {} {\bibfield  {journal} {\bibinfo  {journal} {Zeitschrift
  f{\"{u}}r Physik B}\ }\textbf {\bibinfo {volume} {91}},\ \bibinfo {pages}
  {291} (\bibinfo {year} {1993})}\BibitemShut {NoStop}%
\bibitem [{\citenamefont {Schmielau}\ \emph {et~al.}(2000)\citenamefont
  {Schmielau}, \citenamefont {Manske}, \citenamefont {Tamme},\ and\
  \citenamefont {Henneberger}}]{schmielau-etal.00}%
  \BibitemOpen
  \bibfield  {author} {\bibinfo {author} {\bibfnamefont {T.}~\bibnamefont
  {Schmielau}}, \bibinfo {author} {\bibfnamefont {G.}~\bibnamefont {Manske}},
  \bibinfo {author} {\bibfnamefont {D.}~\bibnamefont {Tamme}},\ and\ \bibinfo
  {author} {\bibfnamefont {K.}~\bibnamefont {Henneberger}},\ }\bibfield
  {title} {\bibinfo {title} {T-matrix approach to the linear optical response
  of highly excited semiconductors},\ }\href@noop {} {\bibfield  {journal}
  {\bibinfo  {journal} {phys. stat. sol. (b)}\ }\textbf {\bibinfo {volume}
  {221}},\ \bibinfo {pages} {215 } (\bibinfo {year} {2000})}\BibitemShut
  {NoStop}%
\bibitem [{\citenamefont {Pieri}\ \emph {et~al.}(2004)\citenamefont {Pieri},
  \citenamefont {Pisani},\ and\ \citenamefont {Strinati}}]{pieri-etal.04}%
  \BibitemOpen
  \bibfield  {author} {\bibinfo {author} {\bibfnamefont {P.}~\bibnamefont
  {Pieri}}, \bibinfo {author} {\bibfnamefont {L.}~\bibnamefont {Pisani}},\ and\
  \bibinfo {author} {\bibfnamefont {G.~C.}\ \bibnamefont {Strinati}},\
  }\bibfield  {title} {\bibinfo {title} {{BCS-BEC crossover at finite
  temperature in the broken-symmetry phase}},\ }\href@noop {} {\bibfield
  {journal} {\bibinfo  {journal} {Phys. Rev. B}\ }\textbf {\bibinfo {volume}
  {70}},\ \bibinfo {pages} {094508 } (\bibinfo {year} {2004})}\BibitemShut
  {NoStop}%
\bibitem [{\citenamefont {Kwong}\ \emph {et~al.}(2009)\citenamefont {Kwong},
  \citenamefont {Rupper},\ and\ \citenamefont {Binder}}]{kwong-etal.09}%
  \BibitemOpen
  \bibfield  {author} {\bibinfo {author} {\bibfnamefont {N.~H.}\ \bibnamefont
  {Kwong}}, \bibinfo {author} {\bibfnamefont {G.}~\bibnamefont {Rupper}},\ and\
  \bibinfo {author} {\bibfnamefont {R.}~\bibnamefont {Binder}},\ }\bibfield
  {title} {\bibinfo {title} {{Self-consistent T-matrix theory of semiconductor
  light-absorption and luminescence}},\ }\href@noop {} {\bibfield  {journal}
  {\bibinfo  {journal} {Phys. Rev. B}\ }\textbf {\bibinfo {volume} {79}},\
  \bibinfo {pages} {155205} (\bibinfo {year} {2009})}\BibitemShut {NoStop}%
\bibitem [{\citenamefont {Ropke}\ and\ \citenamefont
  {Der}(1979)}]{ropke-der.79}%
  \BibitemOpen
  \bibfield  {author} {\bibinfo {author} {\bibfnamefont {G.}~\bibnamefont
  {Ropke}}\ and\ \bibinfo {author} {\bibfnamefont {R.}~\bibnamefont {Der}},\
  }\bibfield  {title} {\bibinfo {title} {{The influence of two-particle states
  (excitons) on the dielectic function of the electron-hole plasma}},\
  }\href@noop {} {\bibfield  {journal} {\bibinfo  {journal} {phys. stat. sol.
  (b)}\ }\textbf {\bibinfo {volume} {92}},\ \bibinfo {pages} {501 } (\bibinfo
  {year} {1979})}\BibitemShut {NoStop}%
\bibitem [{bin()}]{binder-arxiv.2020}%
  \BibitemOpen
  \href@noop {} {}\bibinfo {note} {After submission of our manuscript, we
  obtained further corroboration of our polaritonic BCS gap estimates, as well
  as a more detailed understanding of the fluctuation modes that could be
  probed with interband transitions and possibly THz spectroscopy, see
  arxiv.org/abs/2007.13253}\BibitemShut {NoStop}%
\bibitem [{\citenamefont {Hanai}\ \emph {et~al.}(2019)\citenamefont {Hanai},
  \citenamefont {Edelman}, \citenamefont {Ohashi},\ and\ \citenamefont
  {Littlewood}}]{hanai-etal.2019}%
  \BibitemOpen
  \bibfield  {author} {\bibinfo {author} {\bibfnamefont {R.}~\bibnamefont
  {Hanai}}, \bibinfo {author} {\bibfnamefont {A.}~\bibnamefont {Edelman}},
  \bibinfo {author} {\bibfnamefont {Y.}~\bibnamefont {Ohashi}},\ and\ \bibinfo
  {author} {\bibfnamefont {P.}~\bibnamefont {Littlewood}},\ }\bibfield  {title}
  {\bibinfo {title} {{Non-Hermitian phase transition from a polariton
  Bose-Einstein condensate to a photon laser}},\ }\href@noop {} {\bibfield
  {journal} {\bibinfo  {journal} {Phys. Rev. Lett.}\ }\textbf {\bibinfo
  {volume} {122}},\ \bibinfo {pages} {185301} (\bibinfo {year}
  {2019})}\BibitemShut {NoStop}%
\bibitem [{\citenamefont {Takayama}\ \emph {et~al.}(2002)\citenamefont
  {Takayama}, \citenamefont {Kwong}, \citenamefont {Rumyantsev}, \citenamefont
  {Kuwata-Gonokami},\ and\ \citenamefont {Binder}}]{takayama-etal.02}%
  \BibitemOpen
  \bibfield  {author} {\bibinfo {author} {\bibfnamefont {R.}~\bibnamefont
  {Takayama}}, \bibinfo {author} {\bibfnamefont {N.~H.}\ \bibnamefont {Kwong}},
  \bibinfo {author} {\bibfnamefont {I.}~\bibnamefont {Rumyantsev}}, \bibinfo
  {author} {\bibfnamefont {M.}~\bibnamefont {Kuwata-Gonokami}},\ and\ \bibinfo
  {author} {\bibfnamefont {R.}~\bibnamefont {Binder}},\ }\bibfield  {title}
  {\bibinfo {title} {T-matrix analysis of biexcitonic correlations in the
  nonlinear optical response of semiconductor quantum wells},\ }\href@noop {}
  {\bibfield  {journal} {\bibinfo  {journal} {Eur. Phys. J. B}\ }\textbf
  {\bibinfo {volume} {25}},\ \bibinfo {pages} {445} (\bibinfo {year}
  {2002})}\BibitemShut {NoStop}%
\bibitem [{\citenamefont {Schumacher}\ \emph {et~al.}(2007)\citenamefont
  {Schumacher}, \citenamefont {Kwong},\ and\ \citenamefont
  {Binder}}]{schumacher-etal.07prb}%
  \BibitemOpen
  \bibfield  {author} {\bibinfo {author} {\bibfnamefont {S.}~\bibnamefont
  {Schumacher}}, \bibinfo {author} {\bibfnamefont {N.~H.}\ \bibnamefont
  {Kwong}},\ and\ \bibinfo {author} {\bibfnamefont {R.}~\bibnamefont
  {Binder}},\ }\bibfield  {title} {\bibinfo {title} {Influence of
  exciton-exciton correlations on the polarization characteristics of polariton
  amplification in semiconductor microcavities},\ }\href@noop {} {\bibfield
  {journal} {\bibinfo  {journal} {Phys. Rev. B}\ }\textbf {\bibinfo {volume}
  {76}},\ \bibinfo {pages} {245324} (\bibinfo {year} {2007})}\BibitemShut
  {NoStop}%
\bibitem [{\citenamefont {Fetter}\ and\ \citenamefont
  {Walecka}(1971)}]{fetter-walecka.71}%
  \BibitemOpen
  \bibfield  {author} {\bibinfo {author} {\bibfnamefont {A.}~\bibnamefont
  {Fetter}}\ and\ \bibinfo {author} {\bibfnamefont {J.}~\bibnamefont
  {Walecka}},\ }\href@noop {} {\emph {\bibinfo {title} {Quantum Theory of
  Many-Particle Systems}}}\ (\bibinfo  {publisher} {McGraw Hill},\ \bibinfo
  {address} {New York},\ \bibinfo {year} {1971})\BibitemShut {NoStop}%
\bibitem [{\citenamefont {Jerome}\ \emph {et~al.}(1967)\citenamefont {Jerome},
  \citenamefont {Rice},\ and\ \citenamefont {Kohn}}]{jerome-etal.67}%
  \BibitemOpen
  \bibfield  {author} {\bibinfo {author} {\bibfnamefont {D.}~\bibnamefont
  {Jerome}}, \bibinfo {author} {\bibfnamefont {T.}~\bibnamefont {Rice}},\ and\
  \bibinfo {author} {\bibfnamefont {W.}~\bibnamefont {Kohn}},\ }\bibfield
  {title} {\bibinfo {title} {Excitonic insulator},\ }\href@noop {} {\bibfield
  {journal} {\bibinfo  {journal} {Phys. Rev.}\ }\textbf {\bibinfo {volume}
  {158}},\ \bibinfo {pages} {462} (\bibinfo {year} {1967})}\BibitemShut
  {NoStop}%
\bibitem [{\citenamefont {Hanamura}\ and\ \citenamefont
  {Haug}(1977)}]{hanamura-haug.77}%
  \BibitemOpen
  \bibfield  {author} {\bibinfo {author} {\bibfnamefont {E.}~\bibnamefont
  {Hanamura}}\ and\ \bibinfo {author} {\bibfnamefont {H.}~\bibnamefont
  {Haug}},\ }\bibfield  {title} {\bibinfo {title} {Condensation effects of
  excitons},\ }\href@noop {} {\bibfield  {journal} {\bibinfo  {journal}
  {Physics Reports}\ }\textbf {\bibinfo {volume} {33}},\ \bibinfo {pages} {209}
  (\bibinfo {year} {1977})}\BibitemShut {NoStop}%
\bibitem [{\citenamefont {Jahnke}\ and\ \citenamefont
  {Henneberger}(1992)}]{jahnke-henneberger.92}%
  \BibitemOpen
  \bibfield  {author} {\bibinfo {author} {\bibfnamefont {F.}~\bibnamefont
  {Jahnke}}\ and\ \bibinfo {author} {\bibfnamefont {K.}~\bibnamefont
  {Henneberger}},\ }\bibfield  {title} {\bibinfo {title} {{Light-induced
  effects in the interband absorption of semiconductors}},\ }\href@noop {}
  {\bibfield  {journal} {\bibinfo  {journal} {Phys. Rev. B}\ }\textbf {\bibinfo
  {volume} {45}},\ \bibinfo {pages} {4077} (\bibinfo {year}
  {1992})}\BibitemShut {NoStop}%
\bibitem [{\citenamefont {Combescot}\ \emph {et~al.}(2007)\citenamefont
  {Combescot}, \citenamefont {Betbeder-Matibet},\ and\ \citenamefont
  {Combescot}}]{combescot-etal.07prl}%
  \BibitemOpen
  \bibfield  {author} {\bibinfo {author} {\bibfnamefont {M.}~\bibnamefont
  {Combescot}}, \bibinfo {author} {\bibfnamefont {O.}~\bibnamefont
  {Betbeder-Matibet}},\ and\ \bibinfo {author} {\bibfnamefont {R.}~\bibnamefont
  {Combescot}},\ }\bibfield  {title} {\bibinfo {title} {{Bose-Einstein
  condensation in semiconductors: the key role of dark excitons}},\ }\href@noop
  {} {\bibfield  {journal} {\bibinfo  {journal} {Phys. Rev. Lett.}\ }\textbf
  {\bibinfo {volume} {99}},\ \bibinfo {pages} {176403} (\bibinfo {year}
  {2007})}\BibitemShut {NoStop}%
\end{thebibliography}%


\begin{thebibliography}{11}%
\makeatletter
\providecommand \@ifxundefined [1]{%
 \@ifx{#1\undefined}
}%
\providecommand \@ifnum [1]{%
 \ifnum #1\expandafter \@firstoftwo
 \else \expandafter \@secondoftwo
 \fi
}%
\providecommand \@ifx [1]{%
 \ifx #1\expandafter \@firstoftwo
 \else \expandafter \@secondoftwo
 \fi
}%
\providecommand \natexlab [1]{#1}%
\providecommand \enquote  [1]{``#1''}%
\providecommand \bibnamefont  [1]{#1}%
\providecommand \bibfnamefont [1]{#1}%
\providecommand \citenamefont [1]{#1}%
\providecommand \href@noop [0]{\@secondoftwo}%
\providecommand \href [0]{\begingroup \@sanitize@url \@href}%
\providecommand \@href[1]{\@@startlink{#1}\@@href}%
\providecommand \@@href[1]{\endgroup#1\@@endlink}%
\providecommand \@sanitize@url [0]{\catcode `\\12\catcode `\$12\catcode
  `\&12\catcode `\#12\catcode `\^12\catcode `\_12\catcode `\%12\relax}%
\providecommand \@@startlink[1]{}%
\providecommand \@@endlink[0]{}%
\providecommand \url  [0]{\begingroup\@sanitize@url \@url }%
\providecommand \@url [1]{\endgroup\@href {#1}{\urlprefix }}%
\providecommand \urlprefix  [0]{URL }%
\providecommand \Eprint [0]{\href }%
\providecommand \doibase [0]{http://dx.doi.org/}%
\providecommand \selectlanguage [0]{\@gobble}%
\providecommand \bibinfo  [0]{\@secondoftwo}%
\providecommand \bibfield  [0]{\@secondoftwo}%
\providecommand \translation [1]{[#1]}%
\providecommand \BibitemOpen [0]{}%
\providecommand \bibitemStop [0]{}%
\providecommand \bibitemNoStop [0]{.\EOS\space}%
\providecommand \EOS [0]{\spacefactor3000\relax}%
\providecommand \BibitemShut  [1]{\csname bibitem#1\endcsname}%
\let\auto@bib@innerbib\@empty
\bibitem [{\citenamefont {Fetter}\ and\ \citenamefont
  {Walecka}(1971)}]{fetter-walecka.71}%
  \BibitemOpen
  \bibfield  {author} {\bibinfo {author} {\bibfnamefont {Alexander}\
  \bibnamefont {Fetter}}\ and\ \bibinfo {author} {\bibfnamefont {John}\
  \bibnamefont {Walecka}},\ }\href@noop {} {\emph {\bibinfo {title} {Quantum
  Theory of Many-Particle Systems}}}\ (\bibinfo  {publisher} {McGraw Hill},\
  \bibinfo {address} {New York},\ \bibinfo {year} {1971})\BibitemShut {NoStop}%
\bibitem [{\citenamefont {Galitskii}\ \emph {et~al.}(1970)\citenamefont
  {Galitskii}, \citenamefont {Goreslavskii},\ and\ \citenamefont
  {Elesin}}]{galitskii-etal.70}%
  \BibitemOpen
  \bibfield  {author} {\bibinfo {author} {\bibfnamefont {V.~M.}\ \bibnamefont
  {Galitskii}}, \bibinfo {author} {\bibfnamefont {S.~P.}\ \bibnamefont
  {Goreslavskii}}, \ and\ \bibinfo {author} {\bibfnamefont {V.~F.}\
  \bibnamefont {Elesin}},\ }\bibfield  {title} {\enquote {\bibinfo {title}
  {{Electric and Magnetic Properties of a Semiconductor in the Field of a
  Strong Electromagnetic Wave}},}\ }\href@noop {} {\bibfield  {journal}
  {\bibinfo  {journal} {Soviet Physics JETP}\ }\textbf {\bibinfo {volume}
  {30}},\ \bibinfo {pages} {117--122} (\bibinfo {year} {1970})}\BibitemShut
  {NoStop}%
\bibitem [{\citenamefont {Jerome}\ \emph {et~al.}(1967)\citenamefont {Jerome},
  \citenamefont {Rice},\ and\ \citenamefont {Kohn}}]{jerome-etal.67}%
  \BibitemOpen
  \bibfield  {author} {\bibinfo {author} {\bibfnamefont {D.}~\bibnamefont
  {Jerome}}, \bibinfo {author} {\bibfnamefont {T.M.}\ \bibnamefont {Rice}}, \
  and\ \bibinfo {author} {\bibfnamefont {W.}~\bibnamefont {Kohn}},\ }\bibfield
  {title} {\enquote {\bibinfo {title} {Excitonic insulator},}\ }\href@noop {}
  {\bibfield  {journal} {\bibinfo  {journal} {Phys. Rev.}\ }\textbf {\bibinfo
  {volume} {158}},\ \bibinfo {pages} {462--475} (\bibinfo {year}
  {1967})}\BibitemShut {NoStop}%
\bibitem [{\citenamefont {Keldysh}\ and\ \citenamefont
  {Kozlov}(1968)}]{keldysh-kozlov.68}%
  \BibitemOpen
  \bibfield  {author} {\bibinfo {author} {\bibfnamefont {L.V.}\ \bibnamefont
  {Keldysh}}\ and\ \bibinfo {author} {\bibfnamefont {A.N.}\ \bibnamefont
  {Kozlov}},\ }\bibfield  {title} {\enquote {\bibinfo {title} {Collective
  properties of excitons in semiconductors},}\ }\href@noop {} {\bibfield
  {journal} {\bibinfo  {journal} {Sov. Phys. JETP}\ }\textbf {\bibinfo {volume}
  {27}},\ \bibinfo {pages} {521--528} (\bibinfo {year} {1968})}\BibitemShut
  {NoStop}%
\bibitem [{\citenamefont {Hanamura}\ and\ \citenamefont
  {Haug}(1977)}]{hanamura-haug.77}%
  \BibitemOpen
  \bibfield  {author} {\bibinfo {author} {\bibfnamefont {E.}~\bibnamefont
  {Hanamura}}\ and\ \bibinfo {author} {\bibfnamefont {H.}~\bibnamefont
  {Haug}},\ }\bibfield  {title} {\enquote {\bibinfo {title} {Condensation
  effects of excitons},}\ }\href@noop {} {\bibfield  {journal} {\bibinfo
  {journal} {Physics Reports}\ }\textbf {\bibinfo {volume} {33}},\ \bibinfo
  {pages} {209--284} (\bibinfo {year} {1977})}\BibitemShut {NoStop}%
\bibitem [{\citenamefont {Comte}\ and\ \citenamefont
  {Nozieres}(1982)}]{comte-nozieres.82}%
  \BibitemOpen
  \bibfield  {author} {\bibinfo {author} {\bibfnamefont {C.}~\bibnamefont
  {Comte}}\ and\ \bibinfo {author} {\bibfnamefont {P.}~\bibnamefont
  {Nozieres}},\ }\bibfield  {title} {\enquote {\bibinfo {title} {Exciton {Bose}
  condensation: the ground state of an electron-hole gas i. mean field
  description of a simplified model},}\ }\href@noop {} {\bibfield  {journal}
  {\bibinfo  {journal} {J. Physique}\ }\textbf {\bibinfo {volume} {43}},\
  \bibinfo {pages} {1069--1081} (\bibinfo {year} {1982})}\BibitemShut {NoStop}%
\bibitem [{\citenamefont {Jahnke}\ and\ \citenamefont
  {Henneberger}(1992)}]{jahnke-henneberger.92}%
  \BibitemOpen
  \bibfield  {author} {\bibinfo {author} {\bibfnamefont {F.}~\bibnamefont
  {Jahnke}}\ and\ \bibinfo {author} {\bibfnamefont {K.}~\bibnamefont
  {Henneberger}},\ }\bibfield  {title} {\enquote {\bibinfo {title}
  {{Light-induced effects in the interband absorption of semiconductors}},}\
  }\href@noop {} {\bibfield  {journal} {\bibinfo  {journal} {Phys. Rev. B}\
  }\textbf {\bibinfo {volume} {45}},\ \bibinfo {pages} {4077--4088} (\bibinfo
  {year} {1992})}\BibitemShut {NoStop}%
\bibitem [{\citenamefont {Byrnes}\ \emph {et~al.}(2010)\citenamefont {Byrnes},
  \citenamefont {Horikiri}, \citenamefont {Ishida},\ and\ \citenamefont
  {Yamamoto}}]{byrnes_bcs_2010}%
  \BibitemOpen
  \bibfield  {author} {\bibinfo {author} {\bibfnamefont {Tim}\ \bibnamefont
  {Byrnes}}, \bibinfo {author} {\bibfnamefont {Tomoyuki}\ \bibnamefont
  {Horikiri}}, \bibinfo {author} {\bibfnamefont {Natsuko}\ \bibnamefont
  {Ishida}}, \ and\ \bibinfo {author} {\bibfnamefont {Yoshihisa}\ \bibnamefont
  {Yamamoto}},\ }\bibfield  {title} {\enquote {\bibinfo {title} {{{BCS
  Wave}}-{{Function Approach}} to the {{BEC}}-{{BCS Crossover}} of
  {{Exciton}}-{{Polariton Condensates}}},}\ }\href {\doibase
  10.1103/PhysRevLett.105.186402} {\bibfield  {journal} {\bibinfo  {journal}
  {Physical Review Letters}\ }\textbf {\bibinfo {volume} {105}},\ \bibinfo
  {pages} {186402} (\bibinfo {year} {2010})}\BibitemShut {NoStop}%
\bibitem [{\citenamefont {Kamide}\ and\ \citenamefont
  {Ogawa}(2010)}]{kamide_what_2010}%
  \BibitemOpen
  \bibfield  {author} {\bibinfo {author} {\bibfnamefont {Kenji}\ \bibnamefont
  {Kamide}}\ and\ \bibinfo {author} {\bibfnamefont {Tetsuo}\ \bibnamefont
  {Ogawa}},\ }\bibfield  {title} {\enquote {\bibinfo {title} {What
  {{Determines}} the {{Wave Function}} of {{Electron}}-{{Hole Pairs}} in
  {{Polariton Condensates}}?}}\ }\href {\doibase
  10.1103/PhysRevLett.105.056401} {\bibfield  {journal} {\bibinfo  {journal}
  {Physical Review Letters}\ }\textbf {\bibinfo {volume} {105}},\ \bibinfo
  {pages} {056401} (\bibinfo {year} {2010})}\BibitemShut {NoStop}%
\bibitem [{\citenamefont {Keeling}\ \emph {et~al.}(2005)\citenamefont
  {Keeling}, \citenamefont {Eastham}, \citenamefont {Szymanska},\ and\
  \citenamefont {Littlewood}}]{keeling-etal.05}%
  \BibitemOpen
  \bibfield  {author} {\bibinfo {author} {\bibfnamefont {Jonathan}\
  \bibnamefont {Keeling}}, \bibinfo {author} {\bibfnamefont {P.~R.}\
  \bibnamefont {Eastham}}, \bibinfo {author} {\bibfnamefont {M.~H.}\
  \bibnamefont {Szymanska}}, \ and\ \bibinfo {author} {\bibfnamefont {P.~B.}\
  \bibnamefont {Littlewood}},\ }\bibfield  {title} {\enquote {\bibinfo {title}
  {{{BCS}}-{{BEC}} crossover in a system of microcavity polaritons},}\ }\href
  {\doibase 10.1103/PhysRevB.72.115320} {\bibfield  {journal} {\bibinfo
  {journal} {Physical Review B}\ }\textbf {\bibinfo {volume} {72}},\ \bibinfo
  {pages} {115320} (\bibinfo {year} {2005})}\BibitemShut {NoStop}%
\bibitem [{\citenamefont {Szymanska}\ \emph {et~al.}(2003)\citenamefont
  {Szymanska}, \citenamefont {Littlewood},\ and\ \citenamefont
  {Simons}}]{szymanska-etal.03}%
  \BibitemOpen
  \bibfield  {author} {\bibinfo {author} {\bibfnamefont {M.H.}\ \bibnamefont
  {Szymanska}}, \bibinfo {author} {\bibfnamefont {P.B.}\ \bibnamefont
  {Littlewood}}, \ and\ \bibinfo {author} {\bibfnamefont {B.D.}\ \bibnamefont
  {Simons}},\ }\bibfield  {title} {\enquote {\bibinfo {title} {{Polariton
  condensation and lasing in optical microcavities: The decoherence-driven
  crossover}},}\ }\href@noop {} {\bibfield  {journal} {\bibinfo  {journal}
  {Phys. Rev. A}\ }\textbf {\bibinfo {volume} {68}},\ \bibinfo {pages} {013818}
  (\bibinfo {year} {2003})}\BibitemShut {NoStop}%
\end{thebibliography}%

\end{document}


\begin{center}
{\large {\bf Supplementary Information}} \\
{\bf Signatures of a Bardeen-Cooper-Schrieffer Polariton Laser }\\
{\em J. Hu, Z. Wang, S. Hoon, H. Deng, S. Brodbeck, C. Schneider, \\
S. H\"ofling, N.H. Kwong, R. Binder } \\
\today%
\end{center}

\section{Additional information on the experiment}
\subsection{Simulation of the reflectance of the sub-wavelength grating and the electric field amplitude distribution in the cavity}

We show in this section the simulated reflectance and field distribution of the cavity structure. The simulations are done with a rigorous coupled-wave analysis (RCWA) program (\texttt{rcwa-1d}, developed by Pavel Kwiecien, \texttt{https://rcwa-1d.sourceforge.io/}).

\begin{figure}[htb]
\begin{center}
\includegraphics[angle=0, scale=0.5, trim= 0cm 0cm 0cm 0cm]{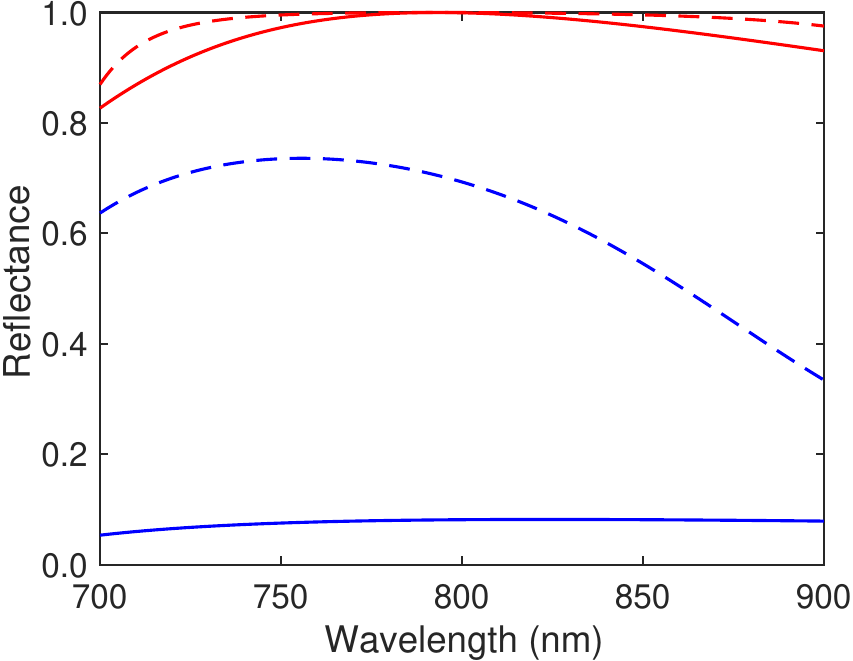}
\caption{\label{fig:FigS_mirror_R}
{\bf Simulated reflectance of the grating-based linear polarization-selective cavity mirror.}
Numerical simulation of the reflectance of the sub-wavelength grating (solid lines) and the top mirror (dashed lines) consisting of the grating and 2.5 pairs of DBR, at normal incidence, for TE-polarized light (red lines) and TM-polarized light (blue lines). The top mirror has very high reflectance for TE and low reflectance for TM.
}
\end{center}
\end{figure}

As shown in Fig.~\ref{fig:FigS_mirror_R}, the grating has high reflectance  at normal incidence for TE-polarized light near 800 nm, while the reflectance for TM-polarized light is low. In addition to the grating, there are 2.5 pairs of distributed Bragg reflector (DBR) layers on top of the cavity, which increase the reflectance for both polarizations. But the reflectance for TM-polarized light is still much lower than TE.

\begin{figure}[htb]
\begin{center}
\includegraphics[angle=0, scale=0.5, trim= 0cm 0cm 0cm 0cm]{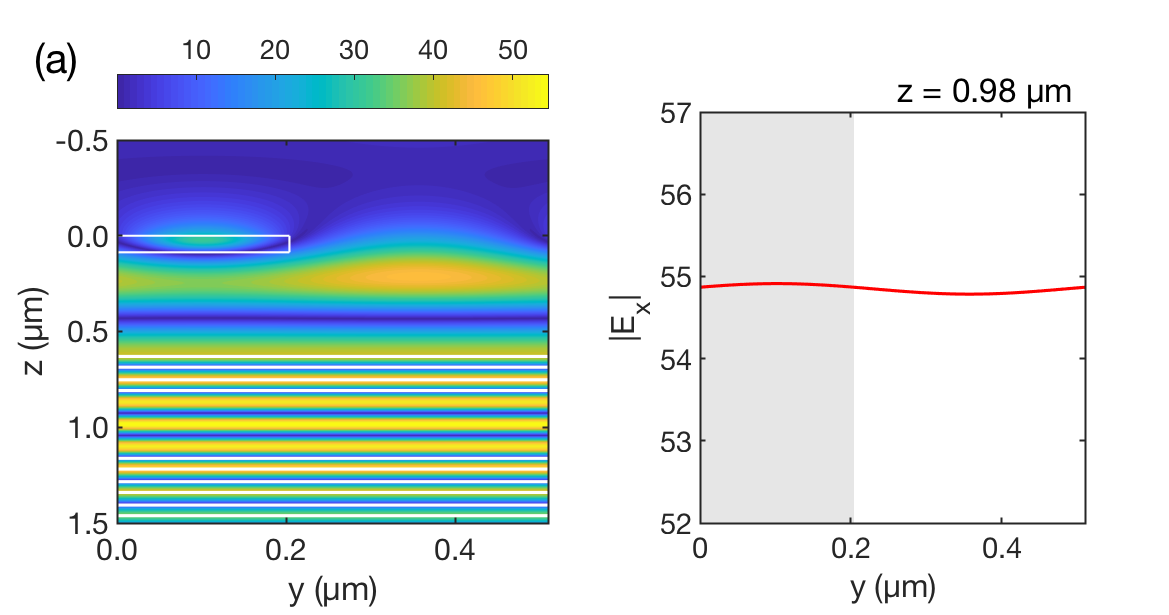}\\
\includegraphics[angle=0, scale=0.5, trim= 0cm 0cm 0cm 0cm]{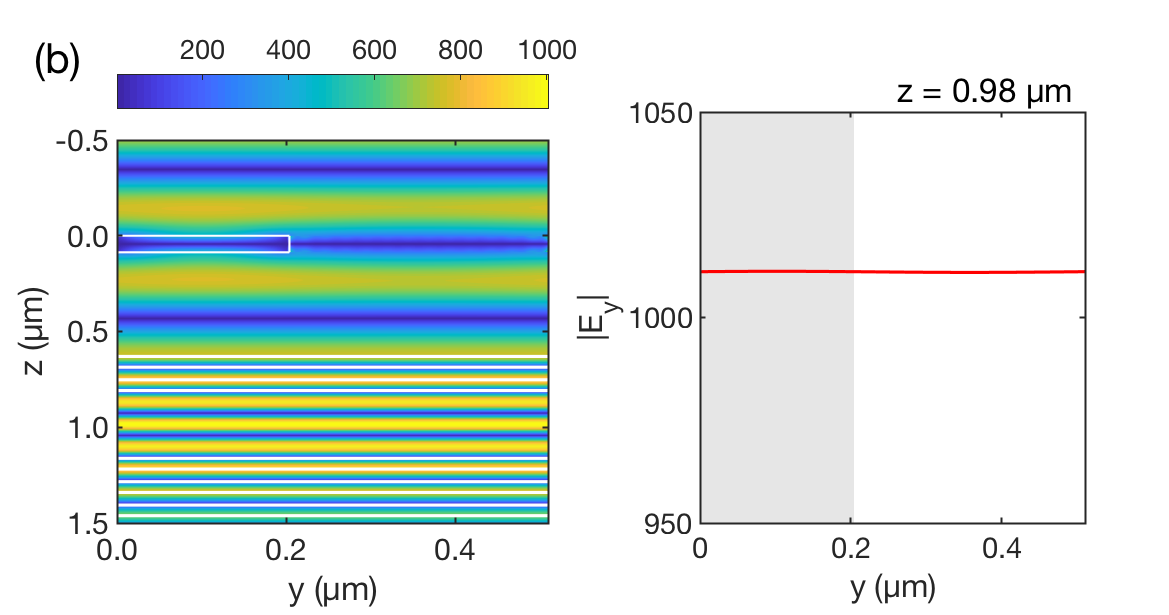}\\
\includegraphics[angle=0, scale=0.5, trim= 0cm 0cm 0cm 0cm]{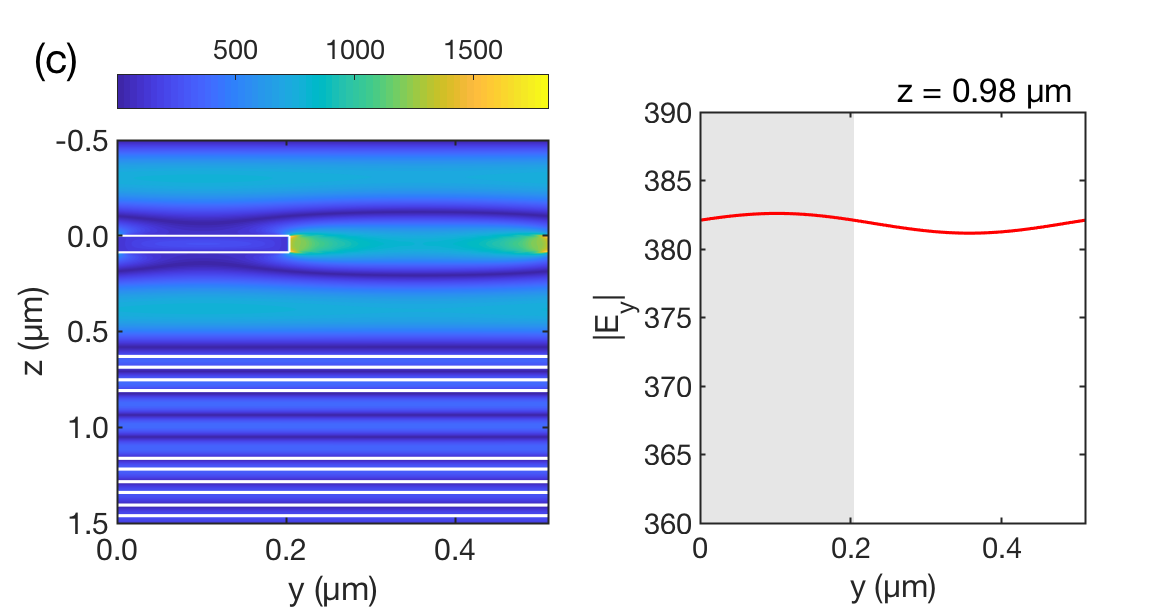}
\caption{\label{fig:FigS_field}
{\bf Numerical simulation of the electric field amplitude at normal incidence. }
(a) TE-polarized light at cavity resonance wavelength (799.45 nm). (b) TM-polarized light at cavity resonance wavelength (799.45 nm). (c) TM-polarized light at off-resonant pump wavelength (784 nm). In all three subfigures, the amplitude of the $x$-component (TE, parallel to grating bars) or $y$-component (TM, perpendicular to grating bars) of the electric field in $y$-$z$ plane for one grating period in $y$ is shown on the left side. The layer structure is outlined with white lines; the layers within the center $\frac{3}{2}\lambda$ of the cavity are omitted in the schematic. On the right side, the corresponding electric field amplitude along $y$ at the anti-node at the center of the cavity is shown. The part of the cavity directly under the grating bar is shaded in gray. As seen from these figures, the modulation of the field amplitude due to the grating is negligible.
}
\end{center}
\end{figure}

The field inside the cavity is strongly enhanced and uniform. In the near field of the grating, there is evanescent field periodic in $y$. However, separated by an air gap and 2.5 pairs of DBR, the field in the cavity becomes plane-wave-like with negligible amplitude fluctuation along $y$. Fig.~\ref{fig:FigS_field} shows examples of TE- and TM-polarized light at cavity resonance wavelength as well as TM-polarized light at the wavelength of the pump laser. In all three cases the field amplitude changes by less than $0.4\%$ along $y$. 

\subsection{Confirming equilibrium between the TE and TM populations in the reservoir}
\begin{figure}[htb]
\begin{center}
\begin{minipage}{0.5\textwidth}
\begin{center}
\includegraphics[width=0.95\textwidth]{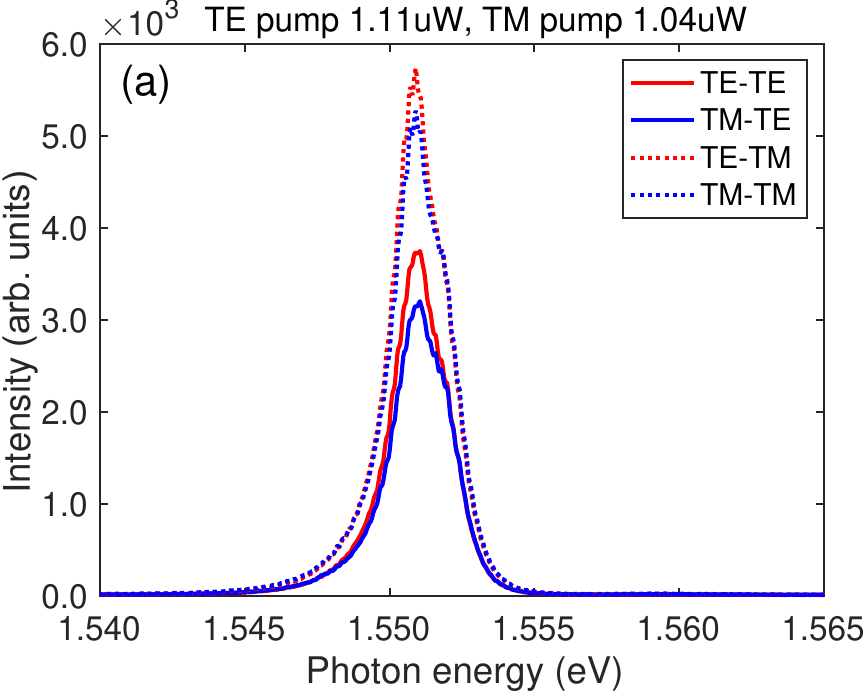}
\end{center}
\end{minipage}%
\begin{minipage}{0.5\textwidth}
\begin{center}
\includegraphics[width=0.95\textwidth]{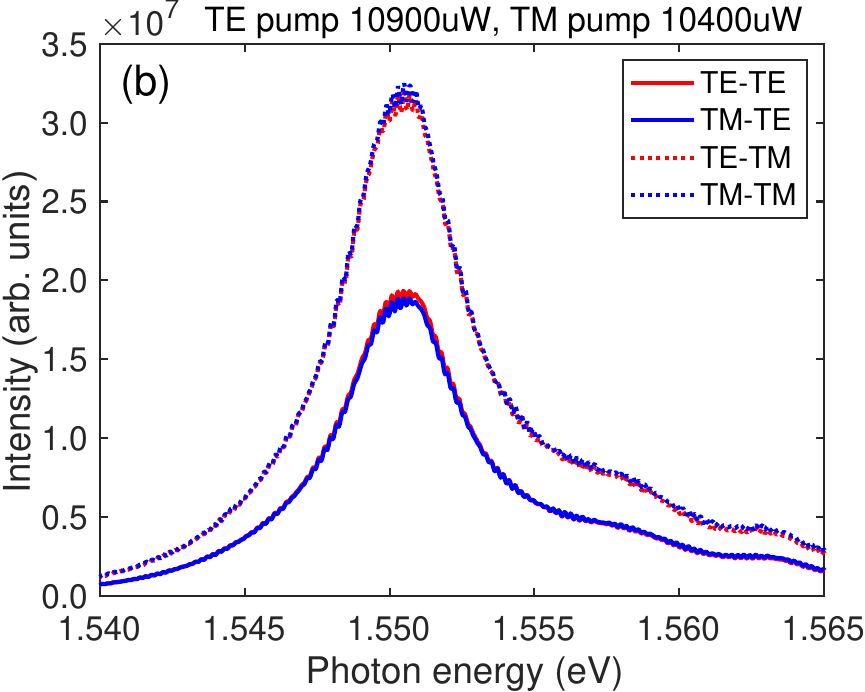}
\end{center}
\end{minipage}\\
\vspace{12pt}
\begin{minipage}{0.6\textwidth}
\begin{center}
\includegraphics[width=0.7\textwidth]{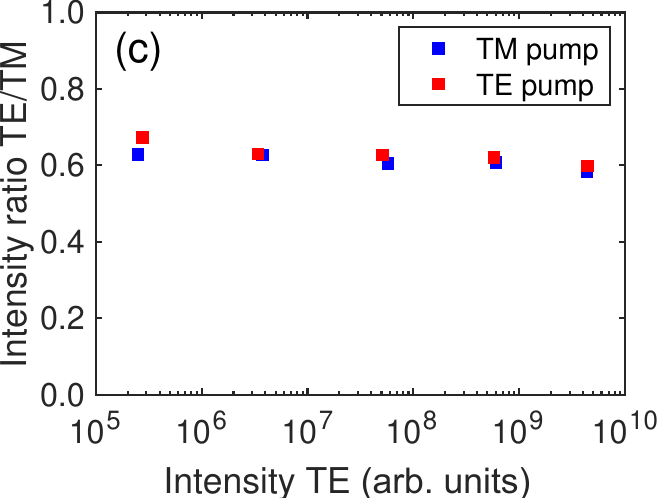}
\end{center}
\end{minipage}
\caption{\label{fig:FigS_outside_pol}
{\bf Emission in the unetched region of the sample with different polarization configurations.}
(a) and (b) show the emission spectra at different pump powers. The four configurations are: TE pump and TE collection (red solid line), TM pump and TE collection (blue solid line), TE pump and TM collection (red dotted line), TM pump and TM collection (blue dotted line). (c) The ratio between integrated intensities of TE and TM emission as a function of integrated intensity of TE emission. The ratio under TM pump is marked in blue squares while the ratio under TE pump is marked in red squares. (a) and (b) correspond to the pairs of data points in (c) near $3\times10^5$ and $4\times10^9$ on horizontal axis respectively. From these data, the polarization of the non-resonant pump laser does not significantly affect the emission intensity in the two orthogonal linear polarizations and therefore we ignore spin imbalance due to the pump laser.
}
\end{center}
\end{figure}

To verify that a TM-polarized non-resonant pump will not create an imbalance between the TE and TM population in the reservoir, we measure the TE- and TM-polarized emission spectra in the the unetched region of the sample (without the grating mirror), under TE- or TM-polarized pump. Fig.~\ref{fig:FigS_outside_pol}a and b shows the emission spectra of the four polarization configurations at pump powers of about $10^0~\mathrm{\mu W}$ and $10^4~\mathrm{\mu W}$ respectively. If there is no imbalance due to pump polarization, one would expect the emission spectra of the four configurations to be identical.
The spectra show that, firstly, indeed all four configurations yield spectra of similar lineshapes. Secondly, the TE (or TM) emission, indicative of TE (or TM) reservoir population, also has nearly the same intensities, regardless of the pump polarization (comparing lines of different color).
The small difference in the intensity is due to a slight difference in the TE and TM pump power.
Thirdly, we measure a higher intensity for the TM emission than for the TE emission by about the same ratio, again regardless of the pump polarization and pump power (comparing solid lines with dotted lines).
This is due to different collection efficiencies for TE and TM emission in the experiment setup. We also plot in Fig.~\ref{fig:FigS_outside_pol}c the ratio between the TE and TM intensities integrated in photon energy as a function of TE emission intensity. In this way, the preference of a certain emission polarization by the same pump polarization will be manifested in a difference of the ratio under the two pump polarizations at equal TE emission intensity. In the data, we still look for pairs of TE and TM pump powers that produce comparable TE emission intensities. The ratio is very close for each pair of them, with a difference smaller than $7.3\%$. There is an overall decreasing trend of this ratio as TE emission intensity increases, possibly because the collection efficiencies also changes with photon energy distribution. In summary, we verify that the pump does not create a significant imbalance between the two populations.

\subsection{Spatial profile of the reservoir}
\begin{figure}[htb]
\begin{center}
\includegraphics[angle=0, scale=0.6, trim= 0cm 0cm 0cm 0cm]{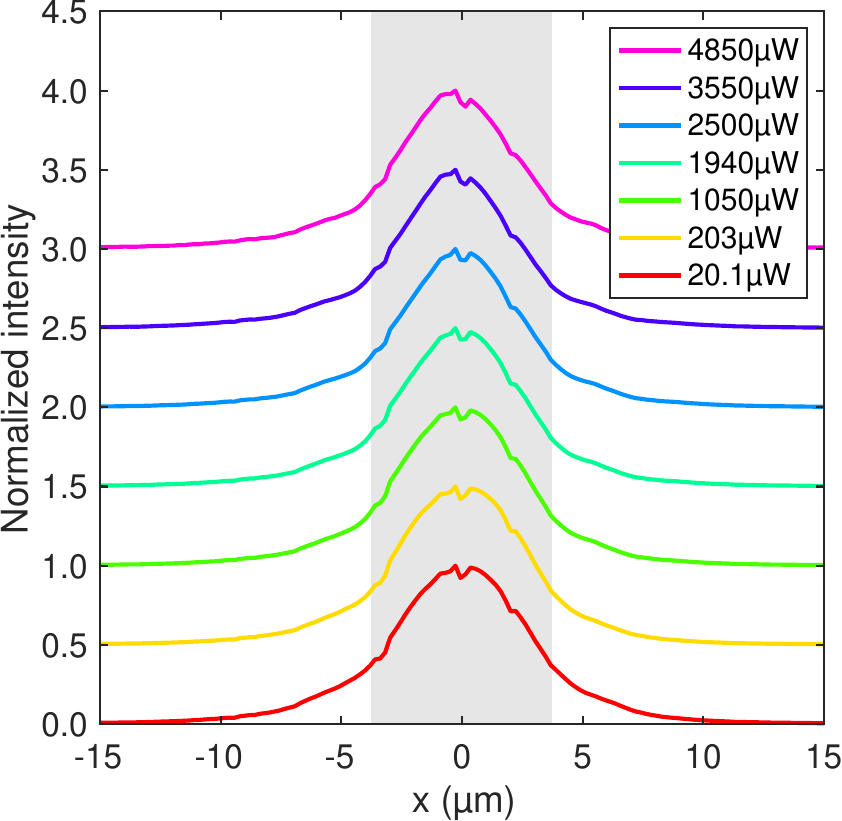}
\caption{\label{fig:FigS_TM_rx_profile}
{\bf Spatial profile of TM-polarized emission.} Spatial profile of the emission from a grating-based device along $x$-direction at $y=0$ in TM polarization. The profiles at different pump laser powers are displaced vertically for clarity. The grating dimension is shaded in gray. The emission remains mostly inside the grating region within the pump powers used.
}
\end{center}
\end{figure}

In the measurements, the pump laser has a spot size of about $3~\mathrm{\mu m}$ in diameter, which is smaller than the grating dimension of $7.5~\mathrm{\mu m} \times 7.5~\mathrm{\mu m}$. However, carrier diffusion leads to a larger spatial extent of carrier distribution, which can be observed from the TM-polarized emission. The emission profile of our device along $x$-direction at $y=0$ is shown in Fig.~\ref{fig:FigS_TM_rx_profile}. The intensity is integrated in photon energy from 1.546 eV to 1.572 eV. With the pump spot located at the grating center, the emission profile remains mostly inside the grating region and does not change much with increasing pump power.

\subsection{Comparison of the Fourier-space TM emission spectra of the BCS-like polariton laser and the photon laser}
\begin{figure}[htb]
\begin{center}
\includegraphics[angle=0, scale=0.6, trim= 0cm 0cm 0cm 0cm]{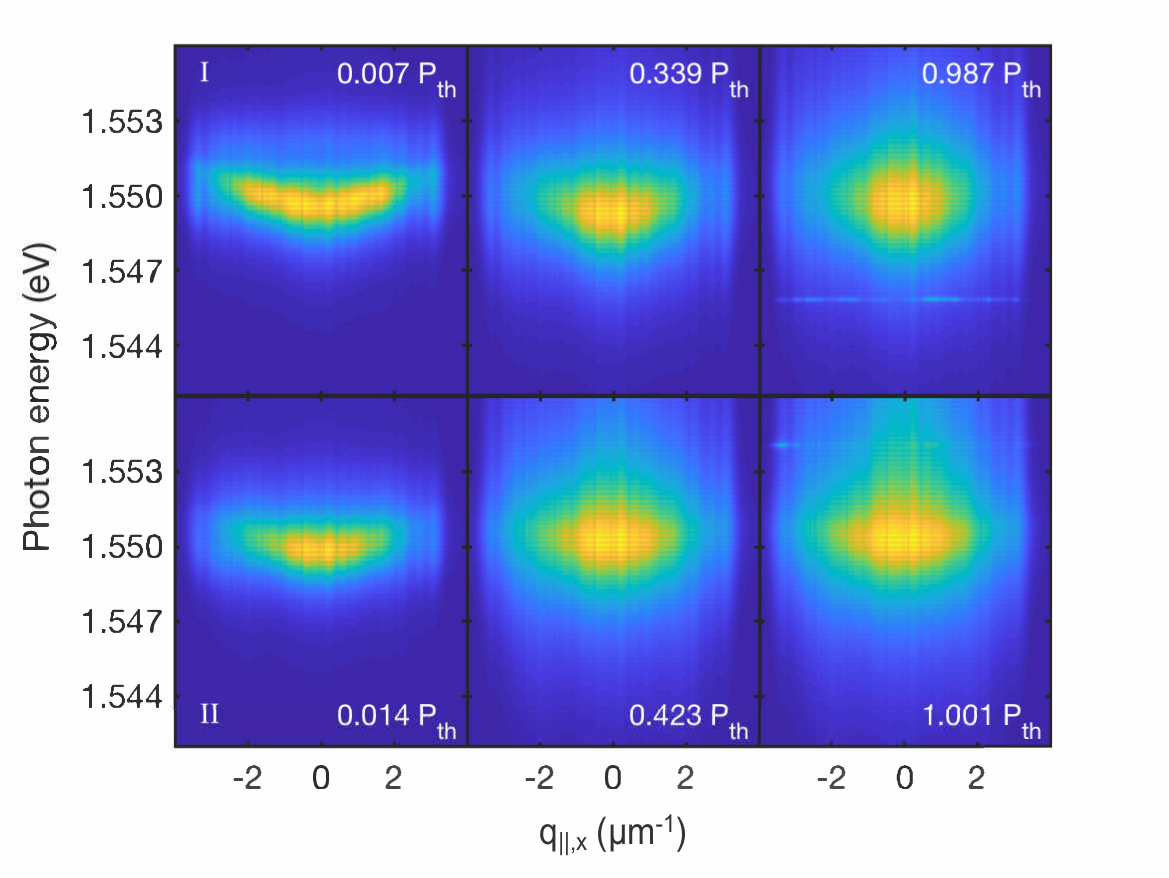}
\caption{\label{fig:FigS_kspace_image_TM}
{\bf Experiemental Fourier-space TM emission spectra.}
Fourier-space spectral images of TM-polarized emission of the polariton laser (upper row) and photon laser (lower row) shown Fig.~2 of the main text, at $q_{\parallel,y}\sim 0$ and at pump powers close to those shown in Fig.~2a of the main text.}
\end{center}
\end{figure}

Fourier-space TM-polarized emission spectra of the example devices shown in Fig.~2 of the main text are given in Fig.~\ref{fig:FigS_kspace_image_TM}. The pump laser powers here are chosen to be close to those in the TE-polarized spectra shown in Fig.~2 of the main text. For both devices, at low pump laser power, the TM-polarized spectra show only discrete exciton emission. As pump laser power increases, the emission linewidth broadens. Near TE lasing threshold, the spectra of both devices have an electron-hole continuum-like pattern, with a significant portion of emission at photon energies higher than the exciton energy at low pump power, signifying the Mott transition.

\subsection{Comparison of the real-space TE emission spectra of the BCS-like polariton laser and the photon laser}
\begin{figure}[htb]
\begin{center}
\includegraphics[angle=0, scale=0.6, trim= 0cm 0cm 0cm 0cm]{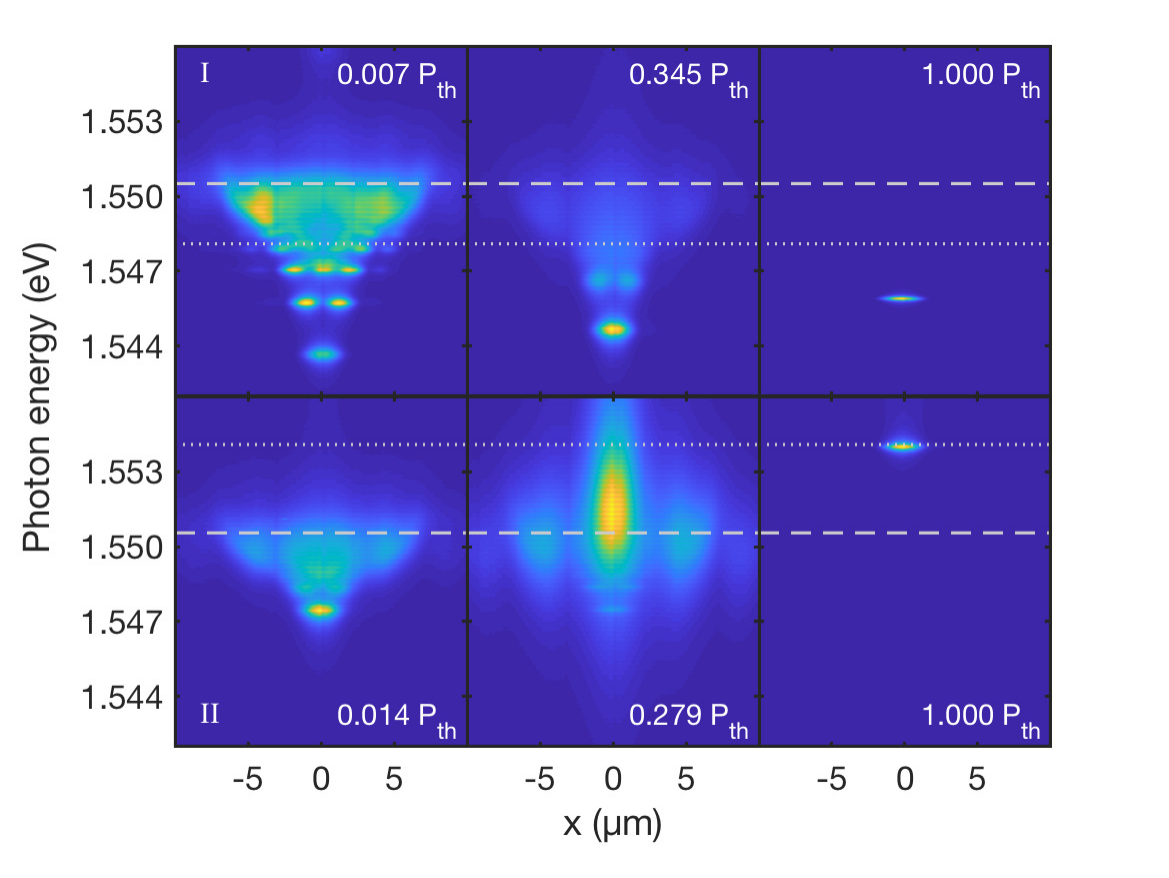}
\caption{\label{fig:FigS_rspace_image}
{\bf Experiemental real-space TE emission spectra.}
Real-space spectral images of TE-polarized emission of the polariton laser (upper row) and photon laser (lower row) shown in Fig.~2 of the main text. The grating of size $7.5~\mathrm{\mu m} \times 7.5~\mathrm{\mu m}$ is centered at the origin and the spectra are taken at $y\sim0$. The dashed lines mark the exciton resonances measured outside the grating region by reflection without a pump. The dotted lines mark the calculated empty cavity resonance at $q_{\parallel}=0$. The polariton ground state goes through a continuous blueshift towards lasing in a polariton laser but is significantly broadened before threshold in a photon laser.}
\end{center}
\end{figure}

Real-space TE-polarized emission spectra of the example devices in Fig.~2 of the main text are shown in the top and bottom rows of Fig.~\ref{fig:FigS_rspace_image}, respectively. The grating of size $7.5~\mathrm{\mu m} \times 7.5~\mathrm{\mu m}$ is centered at the origin and the spectra are taken at $y \sim 0$.

At low pump powers, discrete lower polariton states are visible in both devices inside the grating. With increasing pump power, in the polariton laser shown in the top row of Fig.~\ref{fig:FigS_rspace_image}, the polariton ground state retains its spatial profile and a narrow linewidth of a fraction of a meV, clearly separate from the excited states, and remains well below the cavity resonance even above the lasing threshold, suggesting the bound state is retained in the system and it evolves continuously from an exciton-polariton to e-h-polariton.

In contrast, in the photon laser shown in the bottom row (Fig.~\ref{fig:FigS_rspace_image}), the emission from inside the grating region broadens and blueshifts drastically even before reaching the lasing threshold (Fig.~\ref{fig:FigS_rspace_image}, bottom middle panel), spreading out over 3~meV and therefore indistinguishable from the excited states, characteristic of emission from the electron-hole continuum. Above the threshold power, lasing starts near the cavity resonance. These observations are consistent with Fourier-space spectra shown in the main text and in the supplemental animations.

\subsection{Devices with different parameters}

Devices at different positions on the sample can have different cavity-exciton detuning and normal mode splitting. On the sample from which we took data for Fig.~2 of the main text we have access to a series of devices with different parameters. At negative or small positive detunings, the emission has features of the BCS-like polariton laser described in the main text, namely a continuous blueshift to a lasing frequency that is significantly lower than the empty cavity resonance. At large positive detunings (and also smaller normal mode splitting), however, the lasing transition resembles a conventional photon laser, with a large blueshift towards a frequency close to the empty cavity resonance, and large linewidth broadening before threshold.

\begin{figure}[htbp]
\begin{center}
\begin{minipage}{0.45\textwidth}
\begin{center}
\includegraphics[angle=0, scale=0.5, trim= 0cm 0cm 0cm 0cm]{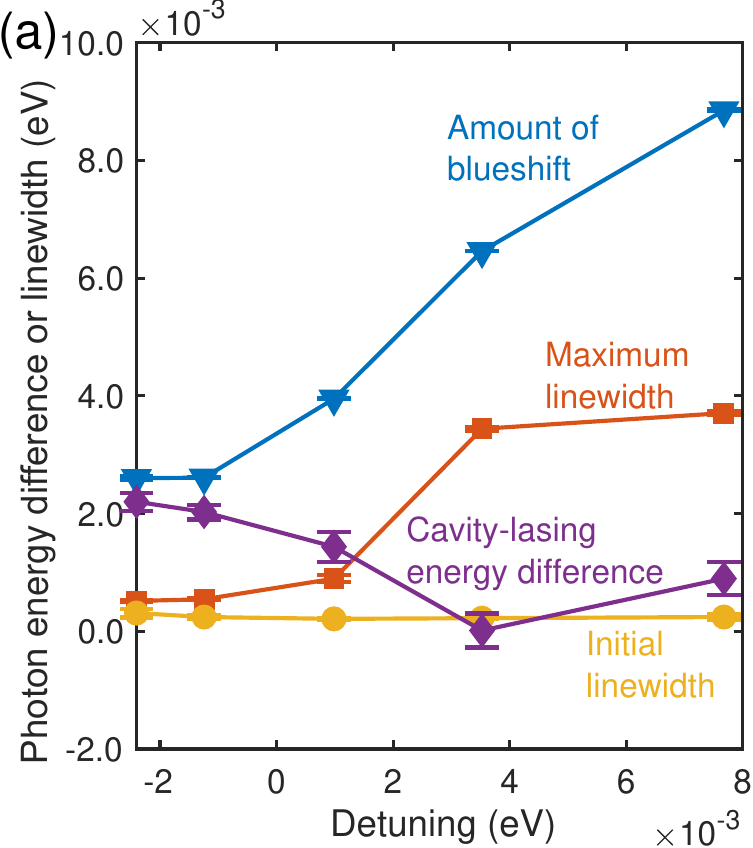}
\end{center}
\end{minipage}%
\begin{minipage}{0.45\textwidth}
\begin{center}
\includegraphics[angle=0, scale=0.5, trim= 1cm 0.5cm 0cm 0cm]{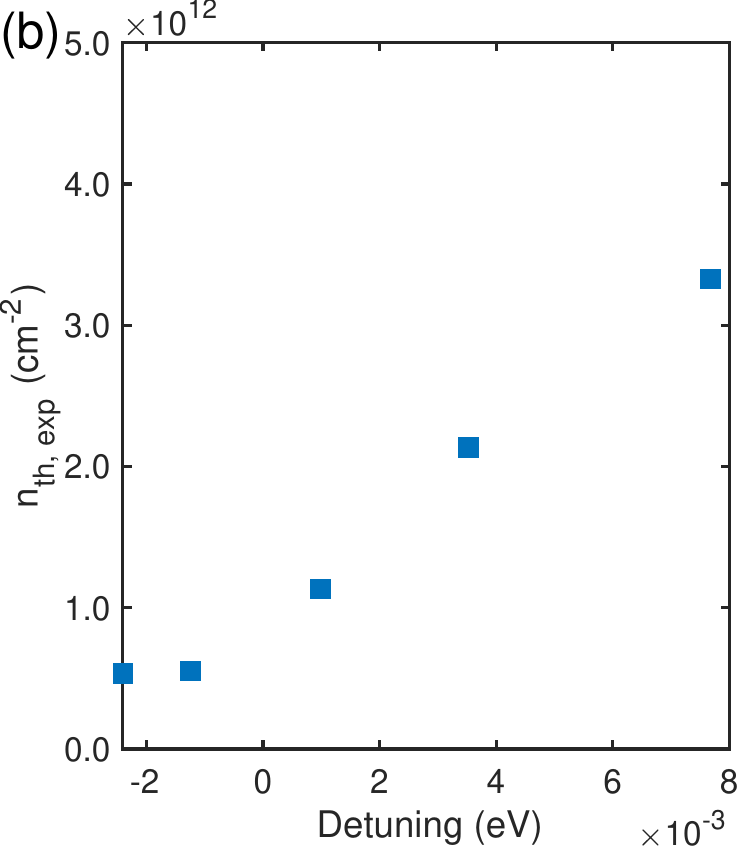}
\end{center}
\end{minipage}

\caption{\label{fig:detuning-freq}\label{fig:detuning-width}\label{fig:detuning-th-dens}
{\bf Emission features of devices with different parameters and estimated threshold density.}
(a) The photon energy difference between empty cavity resonance and laser emission at threshold pump power (purple diamonds), the linewidth of laser emission at low pump power (yellow circles), and the maximum linewidth before threshold (orange squares), compared to the amount of blueshift from low pump power to threshold pump power (blue downward-pointing triangles).
(b) Estimated threshold density. 
}
\end{center}
\end{figure}

In Fig.~\ref{fig:detuning-freq}a we show the photon energy difference between empty cavity resonance and laser emission at threshold pump power. At large positive detunings, the lasing frequency at threshold is close to the empty cavity.
The linewidth of the emission, both at low pump power, where the system is in the low density regime with bosonic exciton polaritons, and at the pump power before threshold, where the linewidth sees maximum broadening, are also compared to the amount of blueshift from low pump power to threshold pump power. For the BCS-like polariton laser, the linewidth remains close to that at low pump power, and only has small broadening compared to the amount of blueshift, whereas for those similar to conventional photon laser, the broadening is over one order of magnitude, and is comparable to the amount of blueshift. The latter can be considered as a discontinuous jump of the frequency from the polariton frequency to the cavity frequency. 
The lasers at larger positive detuning have higher estimated threshold density (Fig.~\ref{fig:detuning-th-dens}b).

\begin{figure}[htbp]
\begin{center}
\begin{minipage}{0.4\textwidth}
\begin{center}
\includegraphics[angle=0, scale=0.4, trim= 0cm 0cm 0cm 0cm]{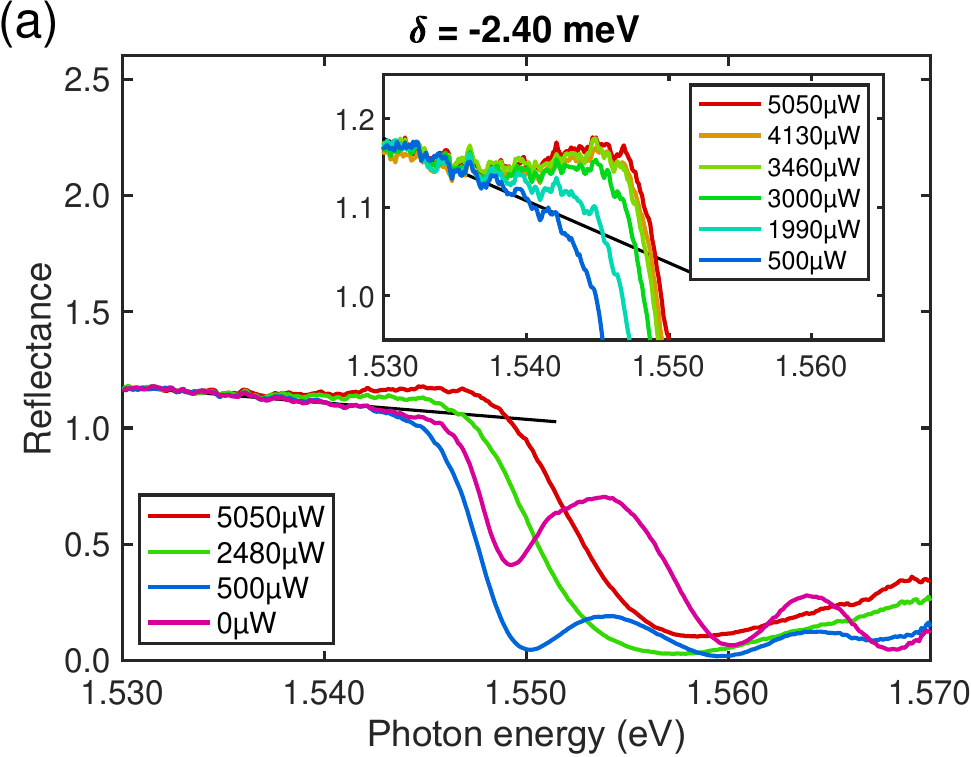}
\end{center}
\end{minipage}
\begin{minipage}{0.4\textwidth}
\begin{center}
\includegraphics[angle=0, scale=0.4, trim= 0cm 0cm 0cm 0cm]{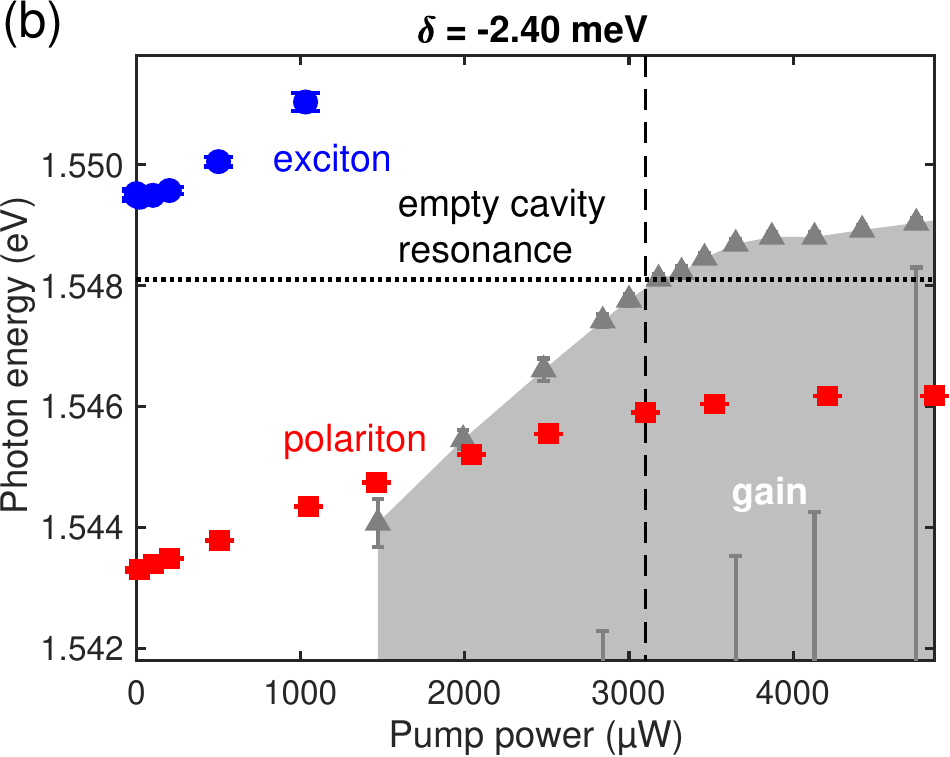}
\end{center}
\end{minipage}\vspace{0.5em}
\begin{minipage}{0.4\textwidth}
\begin{center}
\includegraphics[angle=0, scale=0.4, trim= 0cm 0cm 0cm 0cm]{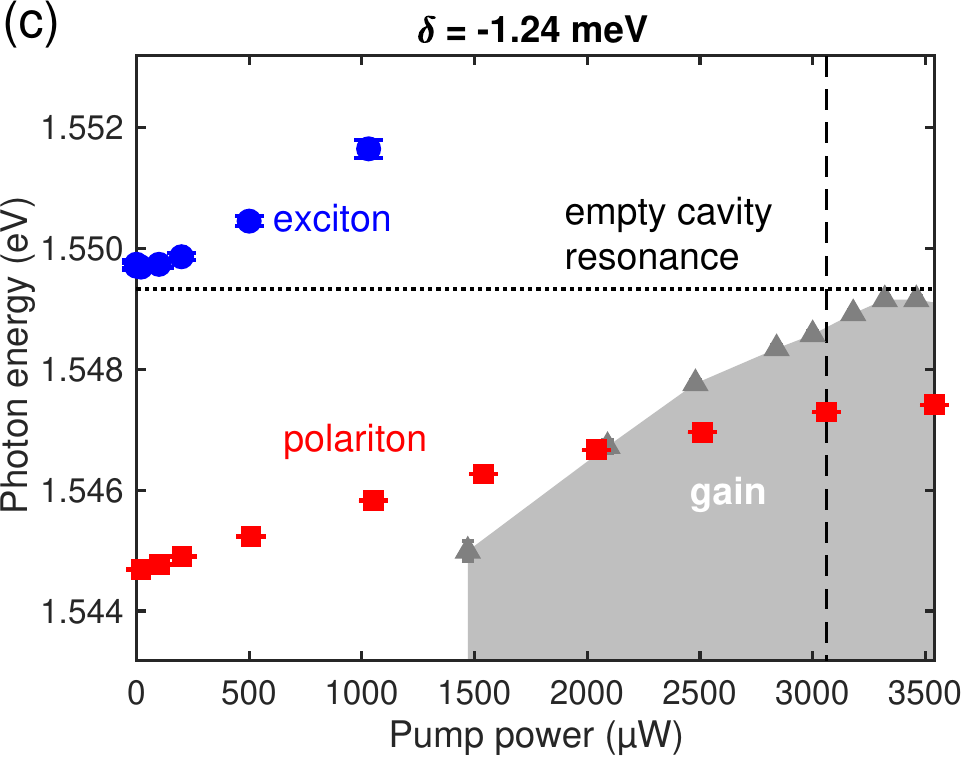}
\end{center}
\end{minipage}
\begin{minipage}{0.4\textwidth}
\begin{center}
\includegraphics[angle=0, scale=0.4, trim= 0cm 0cm 0cm 0cm]{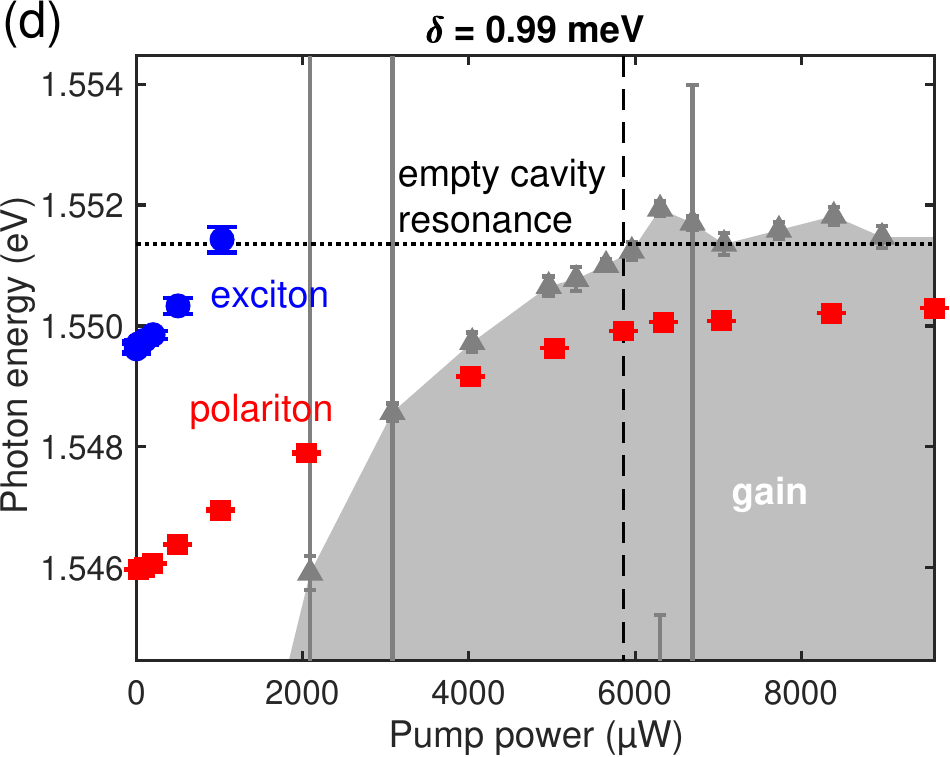}
\end{center}
\end{minipage}\vspace{0.5em}
\begin{minipage}{0.4\textwidth}
\begin{center}
\includegraphics[angle=0, scale=0.4, trim= 0cm 0cm 0cm 0cm]{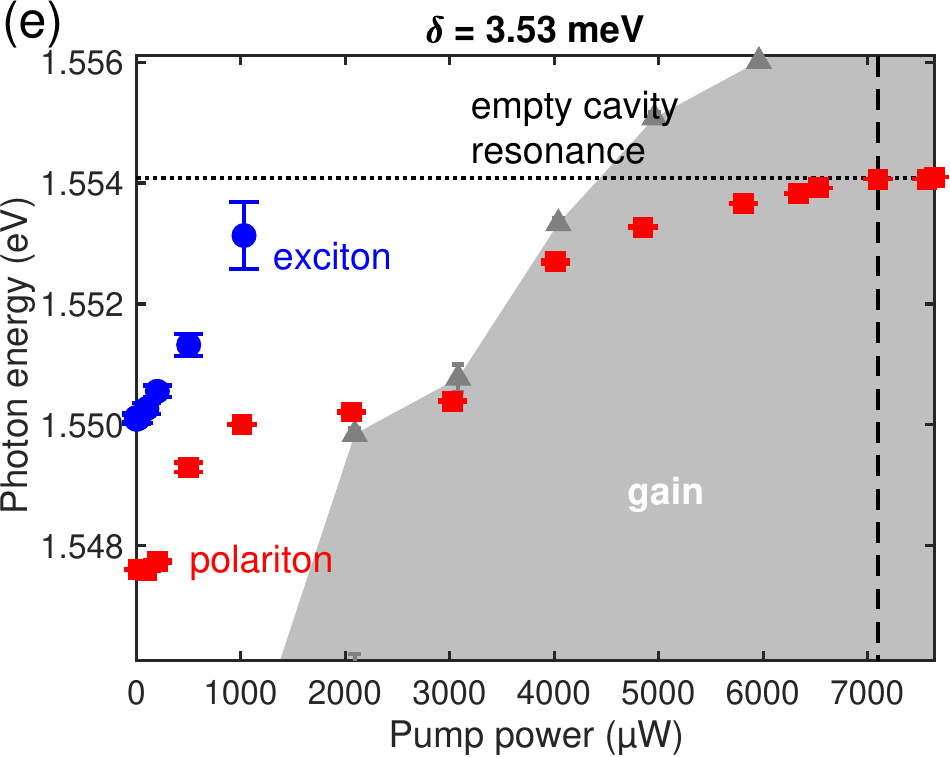}
\end{center}
\end{minipage}
\begin{minipage}{0.4\textwidth}
\begin{center}
\includegraphics[angle=0, scale=0.4, trim= 0cm 0cm 0cm 0cm]{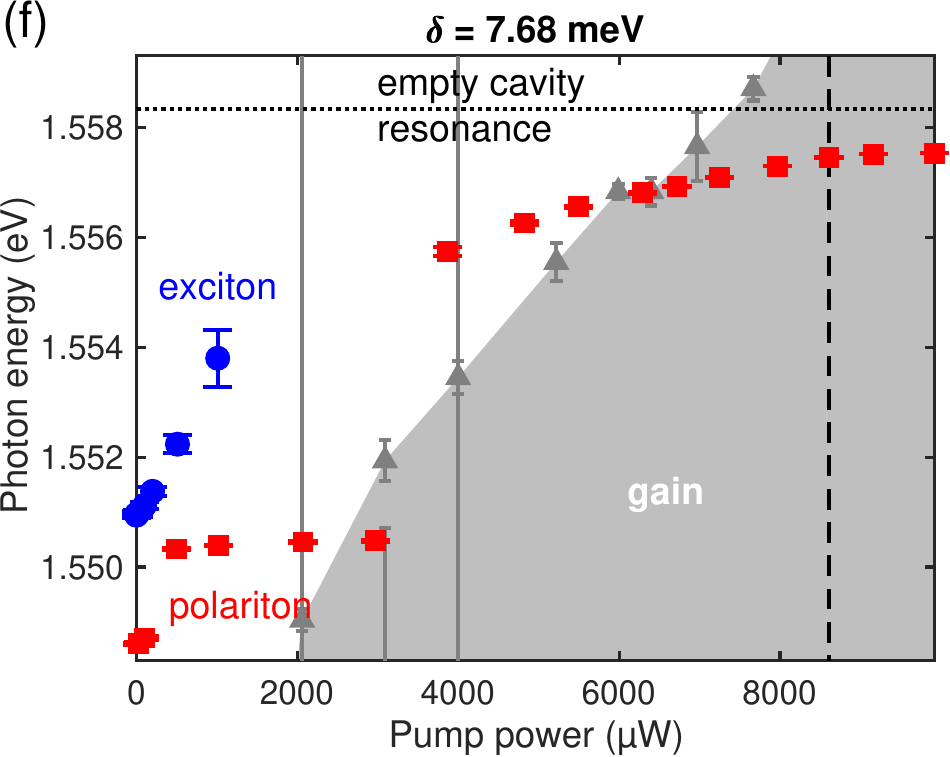}
\end{center}
\end{minipage}

\caption{\label{fig:detuning-summary}
{\bf Reflectance and gain of devices with different parameters.}
(a) TM reflectance spectra of the BCS-like polariton laser presented in Fig.~2 of the main text, at different pump powers. The reference level of unity is marked with tilted black lines. Inset: near the peak of gain.
(b-f) Pump power dependence of measured energetic positions of devices at different detunings, similar to Fig.~3b in the main text: TE polariton emission (red squares) and TM exciton absorption (blue circles) as a function of pump power, as well as empty cavity resonance (black dotted lines). The spectral bounds of gain are marked by the dark gray triangles. Error bars on the bounds are obtained by dividing the standard deviation of the reference of unity reflectance by the local slope at the boundary. Vertical dashed lines mark the threshold pump power. (b) and (e) are the polariton and photon lasing devices shown in Fig.~2 of the main text, respectively.}
\end{center}
\end{figure}

From reflectance measurement we can confirm that all these devices we measured have optical gain throughout lasing and therefore the lasing is due to the population inversion of the fermionic particles (Fig.~\ref{fig:detuning-summary}).
For this sample there is a non-zero slope in the reflectance in the range of wavelength which we use as a reference to determine the gain, because it is closer to a broad cavity resonance. In this case we take a linear fit to the slope at low pump power, instead of just the mean, as the reference (Fig.~\ref{fig:detuning-summary}a).

\section{Additional information on the theoretical analysis}
In the following we provide additional information regarding our theoretical results presented in the main paper.

\subsection{\protect Excitonic and polaritonic BCS model}

The original T=0K BCS state for superconductors, which follows from a HF
theory for Cooper pairs in an interacting Fermi gas is given, for example,
on p. 326-336 of Ref. \cite{fetter-walecka.71}. A central quantity in that
theory is the gap function, $\Delta (\mathbf{k})=\frac{1}{A}\sum\limits_{%
\mathbf{k}^{\prime }}V_{\mathbf{k-k}^{\prime }}^{c}u_{k^{\prime
}}v_{k^{\prime }}$, Eq. (37.25) of Ref. \cite{fetter-walecka.71}. The
extension of the BCS concept from superconductors to an electron-hole system in
semiconductors has been investigated extensively for about half a century.
To facilitate the discussion that puts our work into context, we will
brielfy review and summarize some of the established concepts.

As mentioned in the main text, there are three different approaches to
extend the superconductor BCS theory to a semiconductor: (i) the gap
function can be formed with an electric field that drives the interband
transitions, $\Delta (\mathbf{k})=a_{\mathbf{k}}^{c}E$ (using our notation),
(ii) the gap function can be formed with the interband polarization $\Delta (%
\mathbf{k})=\frac{1}{A}\sum\limits_{\mathbf{k}^{\prime }}V_{\mathbf{k-k}%
^{\prime }}^{c}P(\mathbf{k}^{\prime })$, in which case one can have an
excitonic BEC or BCS state, and (iii) the gap function is the sum $\Delta (%
\mathbf{k})=a_{\mathbf{k}}^{c}E+\frac{1}{A}\sum\limits_{\mathbf{k}^{\prime
}}V_{\mathbf{k-k}^{\prime }}^{c}P(\mathbf{k}^{\prime })$. Pioneering
publications and useful reviews include Ref.\ \cite{galitskii-etal.70} for approach
(i), Refs.\
\cite{jerome-etal.67,
keldysh-kozlov.68,
hanamura-haug.77,
comte-nozieres.82}
 for
(ii), and
\cite{jahnke-henneberger.92,byrnes_bcs_2010,kamide_what_2010} for (iii).

As pointed out in that literature one can define an analogous excitonic BCS
state that follows from the SBE
in the limit of vanishing light field
and upon taking only
the unscreened HF terms (all damping, relaxation and correlation terms being
omitted) and setting the time derivative to zero. Using the definition $P(%
\mathbf{k})=u_{k}v_{k}$ and $f(\mathbf{k})=v_{k}^{2}$ one recovers the
original BCS equations, that can be cast into an equation for the gap
function $\Delta (\mathbf{k})=\frac{1}{A}\sum\limits_{\mathbf{k}^{\prime
}}V_{\mathbf{k-k}^{\prime }}^{c}P(\mathbf{k}^{\prime })$. The requirement $%
u_{k}^{2}+v_{k}^{2}=1$ corresponds to the conservation of the Rabi vector
and can be written as
\begin{equation}
f(\mathbf{k})\left( 1-f(\mathbf{k})\right) =|P(\mathbf{k})|^{2} .
\label{equ:ftimes1-fequPsq}
\end{equation}
The extension of the concept of an excitonic BCS state to an ideal
polaritonic BCS state can then be achieved by combining Eqs.
(3), (22) and (27) of the main text,
again in the HF limit (removing all correlation or
damping terms).

In our cavity, the x and y-components of the polarization are treated separately. In particular, in the absence
of a weak y-polarized input field used to probe the TM response, only the x-polarization
 gives rise to a coherent light field (above threshold). A corresponding ideal T=0K BCS model is therefore restricted to the x-component of the polarization.
In HF approximation, the equation of motion for the x-component of the
interband polarization reads
\begin{eqnarray}
i\hbar \frac{\partial }{\partial t}P_{x}(\mathbf{k}) &=&\left( \frac{\hbar
^{2}k^{2}}{2m_{r}}+E_{g}+2\Sigma ^{HF}(\mathbf{k})-\hbar \omega _{0}-i\gamma_{BCS}\right) P_{x}(\mathbf{k})  \notag \\
&&-\left[ 1-2f(\mathbf{k})\right] \Omega _{eff}^{x}(\mathbf{k})
\label{equ:pdot-BCS-temporal}
\end{eqnarray}%
with
\begin{equation}
\Omega _{eff}^{x}(\mathbf{k})\ =\left( a_{\mathbf{k}}^{c}E_{x}+\frac{1}{A}%
\sum\limits_{\mathbf{k}^{\prime }}V_{\mathbf{k-k}^{\prime }}^{c}P_{x}(%
\mathbf{k}^{\prime })\right) .
\end{equation}%
The equation for the distribution functions is
\begin{equation}
\hbar \frac{\partial }{\partial t}f(\mathbf{k})=\func{Im}\left[ \Omega
_{eff}^{x}(\mathbf{k})^{\ast }P_{x}(\mathbf{k})\right]
\label{equ:fdot-BCS-temporal}
\end{equation}%
and that for the E-field is
\begin{equation}
i\hbar \frac{\partial }{\partial t}E_{x}=\left( \hbar \omega _{cav}-\hbar
\omega _{0}-i\gamma_{BCS}\right) E_{x}-\frac{N_{QW}}{A}\sum\limits_{\mathbf{%
k}}a_{\mathbf{k}}^{c\ast }P_{x}(\mathbf{k})  \label{equ:Edot-BCS-temporal}   .
\end{equation}%
The ideal polaritonic BCS solutions are obtained by setting the
phenomenological decay parameter $\gamma_{BCS}=0$, and solving for
non-trivial steady-state solutions:
\begin{equation}
\left( \frac{\hbar ^{2}k^{2}}{2m_{r}}+E_{g}+2\Sigma ^{HF}(\mathbf{k})-\hbar
\omega _{\mu }\right) P^{BCS}(\mathbf{k})-\left[ 1-2f^{BCS}(\mathbf{k})%
\right] \Omega ^{BCS}(\mathbf{k})=0  \label{equ:P_BCS_stationary}   ,
\end{equation}
\begin{equation}
\left( \hbar \omega _{cav}-\hbar \omega _{\mu }\right) E^{BCS}-\frac{N_{QW}}{%
A}\sum\limits_{\mathbf{k}}a_{\mathbf{k}}^{c\ast }P^{BCS}(\mathbf{k})=0
\end{equation}%
with
\begin{equation}
\Omega ^{BCS}(\mathbf{k})=\left( a_{\mathbf{k}}^{c}E^{BCS}+\frac{1}{A}%
\sum\limits_{\mathbf{k}^{\prime }}V_{\mathbf{k-k}^{\prime }}^{c}P^{BCS}(%
\mathbf{k}^{\prime })\right)   .
\end{equation}%
We denote the time-independent polaritonic BCS solutions by $P^{BCS}(\mathbf{%
k})$, $f^{BCS}(\mathbf{k})$, $E^{BCS}$ , and $\omega _{\mu }$. Here, $\omega
_{0}\equiv \omega _{\mu }$ plays the role of the chemical potential and is
obtained as part of the solution, rather than being an input parameter, if
an additional constraint is given. We use the density constraint
\begin{equation}
n^{BCS}=2\int \frac{d^{2}k}{(2\pi )^{2}}f^{BCS}(\mathbf{k})
\end{equation}%
as this additional constraint.
Since in our system we assume only the x-polarized light to form polaritons,
i.e. $E_{y},P_{y}\simeq 0$, the conservation law linking the distribution
function to the interband polarization becomes
\begin{equation}
2f^{BCS}(\mathbf{k})\left( 1-f^{BCS}(\mathbf{k})\right) =|P^{BCS}(\mathbf{k}%
)|^{2}  \label{equ:BCS-conservation-law-x}  .
\end{equation}%

We note that,
in a polaritonic BCS system, the polarization $P^{BCS}(\mathbf{k})$ does not
have to vanish at the transparency wavevector $\mathbf{k}_{tr}$, that
corresponds to the quasi-chemical potential. From Eq.\ (\ref%
{equ:P_BCS_stationary}) one sees that at the transparency wave vector, where
by definition $\left[ 1-2f^{BCS}(\mathbf{k}_{tr})\right] =0$, the equation
can be fulfilled for non-zero $P^{BCS}(\mathbf{k}_{tr})$ if $\ \hbar \omega
_{\mu }=\frac{\hbar ^{2}k_{tr}^{2}}{2m_{r}}+E_{g}+2\Sigma ^{HF}(\mathbf{k}%
_{tr})$. This is indeed the solution for $\omega _{\mu }$.

\subsection{\protect Excitation spectrum}

In the original BCS theory, the gap in the excitation spectrum opens up
when the gap function is non-zero, see for example Chapter 37 of Ref.\
\cite{fetter-walecka.71}. In the case of an excitonic (not polaritonic) BCS
theory, the gap function can be written as the integral
of the interband polarization and the e-h Coulomb potential,
\begin{equation}
\Delta _{BCS}(\mathbf{k})=\frac{1}{A}\sum\limits_{\mathbf{k}^{\prime }}V_{%
\mathbf{k-k}^{\prime }}^{c}P^{BCS}(\mathbf{k}^{\prime }) .
\end{equation}%
The excitation spectrum for electrons and holes, which we
assume to have identical spectra, is then given by
\begin{equation}
E^{xc}(\mathbf{k})=\sqrt{\xi ^{2}(\mathbf{k})+\Delta _{BCS}^{2}(\mathbf{k})}
\end{equation}%
where the single-particle energy without e-h interaction, measured relative
to the single-particle chemical potential is (again using $m_{e}=m_{h}\equiv
m$) is
\begin{equation}
\xi (\mathbf{k})=\frac{\hbar ^{2}k^{2}}{2m}+\Sigma ^{HF}(\mathbf{k})-\frac{1%
}{2}\left( \hbar \omega _{\mu }-E_{g}\right) .
\end{equation}%
The  BCS gap is the minimum of $E^{xc}(\mathbf{k})$, provided
that band minimum is below the single-particle chemical potential. If $%
\Sigma ^{HF}(\mathbf{k})$ is a monotonic function, this condition is $\Sigma
^{HF}(\mathbf{0})<\frac{1}{2}\left( \hbar \omega _{\mu }-E_{g}\right) $. In
order to discuss the BEC to BCS cross-over, it is helpful to formulate the
theory in terms of pair energies, rather than single-particle (electron or
hole) energies. This has been done in Ref.\ \cite{comte-nozieres.82}, where the HF
theory was formulated in terms of pair energies $\xi ^{pair}(\mathbf{k}%
)=2\xi (\mathbf{k})$, $E^{pair}(\mathbf{k})=2E^{xc}(\mathbf{k})$, $\Delta
_{pair}(\mathbf{k})=\Delta _{BCS}(\mathbf{k})$. This is useful because in
the low-density limit, in which $\Delta _{pair}(\mathbf{k})$ is negligible,
and the chemical potential is below the band minimum, more precisely $\hbar
\omega _{\mu }=\varepsilon _{1s}$, which is the 1s-exciton energy, the pair
gap energy obtained from minimizing $E^{pair}(\mathbf{k})$ is found at $%
\mathbf{k=0}$ (if the single-particle energy increases monotonically with $|%
\mathbf{k}|$, and its value is the exciton binding energy, $\varepsilon
_{1s}-E_{g}$). As mentioned in Ref.\ \cite{comte-nozieres.82} the exciton binding
energy is required to break up an exciton in an exciton condensate into an
e-h pair.

The extension of the BCS theory to the polariton case, as given in Refs.\
\cite{keeling-etal.05,byrnes_bcs_2010,kamide_what_2010}
 discusses the gap
function for the case of polaritons. It is found there that the fermionic
spectrum (i.e. for electrons and holes, not for bosonic
 states of the cavity field) is given by
\begin{equation}
E^{xc}(\mathbf{k})=\sqrt{\xi ^{2}(\mathbf{k})+\Omega _{BCS}^{2}(\mathbf{k})}
\label{equ:E-single-particle-BCS}
\end{equation}%
where $\Omega _{BCS}(\mathbf{k})$ is
\begin{equation}
\Omega _{BCS}(\mathbf{k})=\left( a_{\mathbf{k}}^{c}E^{BCS}+\frac{1}{A}%
\sum\limits_{\mathbf{k}^{\prime }}V_{\mathbf{k-k}^{\prime }}^{c}P^{BCS}(%
\mathbf{k}^{\prime })\right)  \label{equ:Omega-Rabi-BCS}
\end{equation}
 and we keep the notation $\Omega (\mathbf{k})$
(instead of $\Delta (\mathbf{k})$) for the gap function, which is also known
as the renormalized Rabi frequency. Hence, in a polaritonic BCS state a gap
in the excitation spectrum can emerge if either the interband
polarization, or the cavity field amplitude, or both are non-zero. Indeed,
in the limit where the photon number is much larger than the e-h pair number
(large photon fraction), which has been associated with a photon BEC in
Refs.\ \cite{byrnes_bcs_2010,kamide_what_2010}, the gap is dominated by $E^{BCS}$
and thus independent of the e-h Coulomb interaction. This means that, in
contrast to the exciton BCS theory (which is completely analogous to the
original BCS), the polaritonic BCS state does not require Coulombic e-h interaction
for the condensation to become possible. This is analogous to the Elesin
gaps \cite{galitskii-etal.70}, but instead of driving the system with an
external E-field, the E-field ``dresses'' the e-h pairs.

While the strict definition of BCS states (incl. a polariton BCS states)
assumes the system to be closed and in a quasi-thermal equilibrium at T=0K,
in reality all polariton systems are open and pumped systems which attain a
steady-state through compensation of pump and loss processes. This is also
true for photon lasers. If we formally adopt the definition of the
 states given in Eq.\ (\ref{equ:E-single-particle-BCS}) for
the pumped semiconductor microcavity system, we could define the
phenomenological quantity
\begin{equation}
\widetilde{\xi }(\mathbf{k})=\frac{\hbar ^{2}k^{2}}{2m}+\Sigma ^{HF}(\mathbf{%
k})+\func{Re}\Sigma ^{corr}(\mathbf{k})-\frac{1}{2}\left( \hbar \omega
_{0}-E_{g}\right) ,
\end{equation}%
the phenomenological open-system gap function (which omits the correlation
part of the e-h vertex)
\begin{equation}
\widetilde{\Omega }(\mathbf{k})=\left( a_{\mathbf{k}}^{c}E+\frac{1}{A}%
\sum\limits_{\mathbf{k}^{\prime }}V_{\mathbf{k-k}^{\prime }}^{c}P(\mathbf{k}%
^{\prime })\right)   \label{equ:Omega-Rabi-open} ,
\end{equation}%
and the phenomenological open-system excitation energies
\begin{equation}
\widetilde{E}(\mathbf{k})=\sqrt{\widetilde{\xi }^{2}(\mathbf{k})+|\widetilde{%
\Omega }(\mathbf{k})|^{2}}  \label{equ:E-single-particle-open} .
\end{equation}%
Minimizing $\widetilde{E}(\mathbf{k})$ would give a phenomenological quantity analogous to
the BCS gap, which can only be viewed as an estimate that will require future refinement or validation by a more
detailed analysis of the BCS gap for the case of open dissipative pumped systems with Coulombic electron and hole interactions
and correlations beyond the unscreened Hartree-Fock approximation.
Our present estimate should become quantitatively reliable if the system approaches the ideal system, as it
is designed to have the correct limit of the ideal system.

In Fig.~5c of the main paper, we show the pair gap,
\begin{equation}
E_{gap}^{pair} = 2 \min_{k} E^{xc}(\mathbf{k})
\end{equation}%
for the ideal system, and
\begin{equation}
\widetilde{E}_{gap}^{pair} = 2 \min_{k} \widetilde{E}^{xc}(\mathbf{k})
\end{equation}%
for the phenomenological estimate of pair gap energy in our experimental system. Showing the pair gap, with the factor of 2 relative to gap of the electron or the hole subsystem
facilitates comparison with the emission spectrum. For example, in the low density limit of the ideal T=0~K system, the pair gap corresponds to the lower polariton energy measured from the continuum band edge (18 meV in our case).

\begin{figure}
\begin{center}
\centerline{\includegraphics[angle=0, scale=1.0, trim= 0cm .4cm 0cm 0cm]{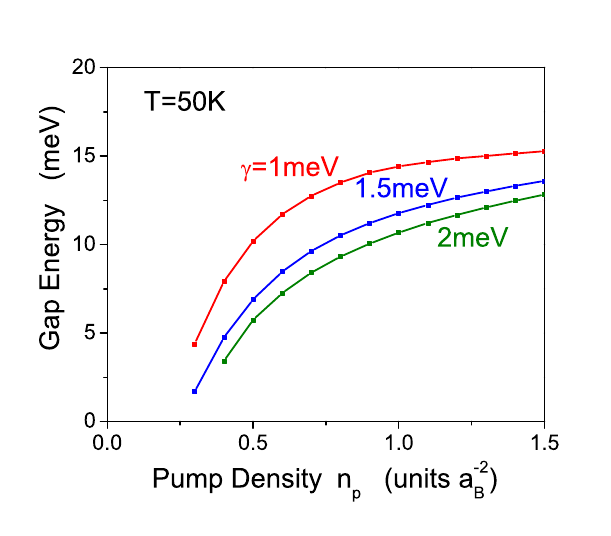}}
\caption{\label{s602-pairgap.fig}
{\bf  Estimated pair gap.}
Phenomenological estimate of the pair gap energy, $\widetilde{E}_{gap}^{pair}$,
for parameter values corresponding the experimental system at fixed temperature $T=50$K for three different dephasing rates. The blue curve ($\gamma$ = 1.5 meV) is the same
as in Fig.~6c of the main text.
}
\end{center}
\end{figure}

We note that it has been shown in Ref.\ \cite{szymanska-etal.03}
that losses (such as dephasing) reduce the actual gap determined by expressions of the form
(\ref{equ:E-single-particle-BCS}) and (\ref{equ:E-single-particle-open}), see for example Eq.\ (34) of  Ref.\ \cite{szymanska-etal.03}.
The trend of a gap reduction with increasing dephasing is also included our theory, see Fig.~\ref{s602-pairgap.fig}. However, as mentioned above,
future investigations aimed at a
 more
detailed analysis of the BCS gap for the case of open dissipative pumped system with Coulombic electron and hole interactions
and correlations beyond the unscreened Hartree-Fock approximation are desirable.

 We finally mention that in our system, in which we have only one
linear polarization substantially different from zero, the single-particle
energy contains a factor of $1/2$ under the square root, i.e. are of the form
\begin{equation}
E^{xc}(\mathbf{k})=\sqrt{\xi ^{2}(\mathbf{k})+|\Omega (\mathbf{k})|^{2}/2}
\end{equation}

\subsection{\protect Numerical implementation}

In the following, we discuss briefly numerical implementation techniques.
The generalized polariton SHF equations
(3), (22) and (27) of the main text,
are solved
numerically by integration of the first-order differential equations in time,
using a 4th-order Runge-Kutta algorithm.

The time-independent ideal-BCS equations are solved as a set of nonlinear
equations using a Broyden algorithm. All numerical solutions are limited to
s-wave solutions in which the wave vector dependent functions are only
functions of the magnitude of the wave vector, and the angle integration in
the Coulomb and correlation terms is performed numerically.

Unless otherwise noted, we use the following parameter values,
$N_{QW}=12$,
$a_B=1.4$nm,
$\varepsilon_b = 12.4$,
$E^{2D}_B=12$meV
where the exciton binding energy in 2D is related to the one in 3D via
$E^{2D}_B= 4 E^{3D}_B$ and
$E^{3D}_B = \frac{\hbar^2}{2 m_r a_B^2}$
determines the reduced mass $m_r$,
$\hbar \omega_{cav} = 1.551$eV,
$\gamma = 1.5$meV,
$\gamma_{pl} = 6$meV,
$\gamma_{p} = 0.2$meV,
$\gamma_{cav} = 0.2$meV,
$\gamma_{nr} = 10^{-4}$meV,
$r_{\rm cv} = 0.5$nm where $d_{cv} = - e r_{\rm cv}$,
$C_{pl} = 1$,
$k_{\max}=10a_{B}^{-1}$.
The material parameters, $E_G$, $E^{2D}_B$,  $\hbar \omega_{cav}$, $\gamma_{cav}$, $\gamma_{nr}$ and $r_{\rm cv}$ are chosen to correspond approximately to the experimental system.
We note that the value for the dephasing rate $\gamma$ is chosen so that the low-density exciton and polariton linewidth correspond approximately to the experimental linewidth. Hence, $\gamma$ is a phenomenological rate that accounts for both, homogeneous broadening due to scattering processes such as carrier-phonon scattering,
and inhomogeneous broadening, mainly due to well-width fluctuations. In our system, inhomogeneous broadening is the dominant source for the line broadening.
The value for  $\gamma_{pl}$ is chosen to correspond approximately to the typical value of the wavevector-dependent imaginary part of the correlation selfenergy,
Im$\{\Sigma _{a}^{corr}(\mathbf{k)} \}$.

\subsection{Supplemental numerical results.}

We show in the following additional theoretical results.

\begin{figure}
\begin{center}
\centerline{\includegraphics[angle=0, scale=1.0, trim= 0cm .4cm 0cm 0cm]{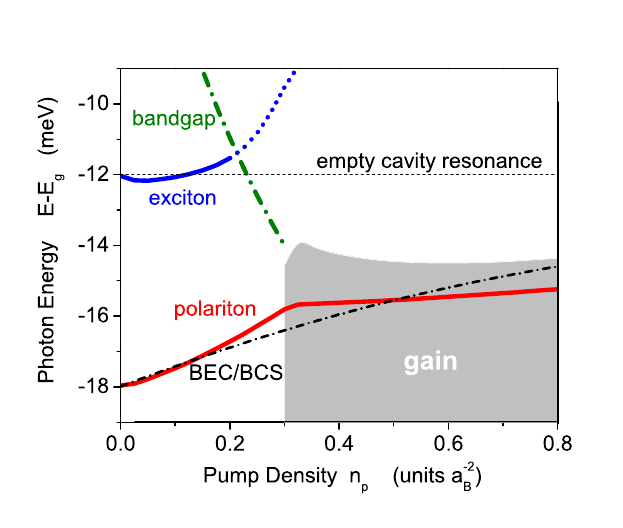}}
\caption{\label{r765vsD.fig}
{\bf Calculated energetic positions.}
Calculated energetic positions of polariton (TE) and exciton (TM) emission, polaritonic BCS quasi-chemical potential,
renormalized bandgap,  empty cavity resonance, and the TM gain region (grey shaded area) vs pump density $n_p$.
Temperature is $T=50$K.
}
\end{center}
\end{figure}

First, we show that the central message from Fig.~3c of the main text does not dependent on the specific assumption for the temperature. The assumption underlying that figure is that the effective bath temperature increases with pump power.
 But even without that assumption we obtain results that are in quite good agreement with the experiment. In Fig.~\ref{r765vsD.fig} we show results analogous to Fig.~3c of the main text but for fixed temperature of T=50K.
 All basic features seen in Fig.~3 of the main text are still reproduced, only some details of the various spectral positions are slightly modified and in less good agreement with the experiment.

\begin{figure}
\begin{center}
\centerline{\includegraphics[angle=0, scale=2.0, trim= 0cm 0.5cm 0cm 0cm]{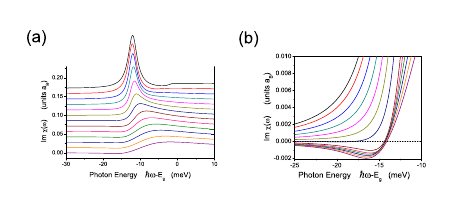}}
\caption{\label{s297absy.fig}
{\bf Calculated TM response spectra.}
(a) Calculated TM response spectra for various pump densities equidistantly spaced from  zero to $n_p = 0.6 a_B^{-2}$ (which is about twice the threshold pump density). Except for the highest pump density,
the spectra are shifted vertically for clarity. The parameters are the same as in Fig.~3c of the main paper.
(b) Same data as in (a) but zoomed in to the gain region and not vertically shifted.
}
\end{center}
\end{figure}

Second, we show the response spectra that determine the absorption/gain in the TM channel (imaginary part of the susceptibility) in Fig.~\ref{s297absy.fig}, which have the usual and well-known form of the transitions from exciton absorption to optical gain.

\begin{figure}
\begin{center}
\centerline{\includegraphics[angle=0, scale=0.8, trim= 0cm 0.5cm 0cm 0cm]{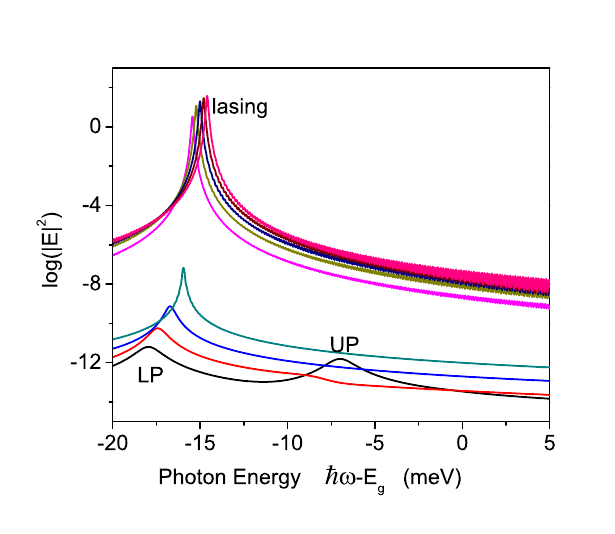}}
\caption{\label{r771LogEsq.fig}
{\bf Calculated TE spectra.}
Calculated spectra of the TE cavity field, shown here on a logarithmic scale,
for various pump densities equidistantly spaced from  zero to $n_p = 0.8 a_B^{-2}$. The parameters are the same as in Fig.~3c of the main text.
The spectra are not vertically shifted.
}
\end{center}
\end{figure}

Third, we show in Fig.~\ref{r771LogEsq.fig}
the spectra of the TE cavity mode, $|E_x(\omega)|^2$,
 for the parameters used in Fig.~3c of the main text.
 As mentioned in the main text, the only information we deduce from these spectra are spectral positions of the maxima, which, below threshold, correspond to the LP emission, and above threshold to the lasing frequency. No other information, such as lineshape, is contained in or taken from this plot.
  One can clearly see the difference between below and above threshold. The order of magnitude of the spectra below threshold is given by the classical fluctuation that we used in the numerical simulation. Hence, no information is contained in the order of magnitude below threshold.

\bibliography{references}

\clearpage

